\documentstyle[12pt]{book}

\begin{document}

\author{Eugene Perchik}
\title{Methodology of Syntheses of Knowledge: Overcoming Incorrectness of the
Problems of Mathematical Modeling }
\maketitle

\newpage

{\small J. Hadamard's ideas of correct formulation of problems of
mathematical physics as well as related Banach's theorem on the inverse
operator are analyzed. Modern techniques of numerical simulations are shown
to be in drastic contradiction to the concepts of J. Hadamard, S. Banach and
a number of other outstanding scientists in the sense that the priority is
given to the realization of inefficient algorithms, based on a belief that
ill-posed problems are adequate to real phenomena.}

{\small A new method of the solution of problems, traditionally associated
with Fredholm integral equations of the first kind, is developed. Its key
aspect is a constructive use of possibilities of the functional space }$l_{2}
${\small \ to ensure the conditions of correctness. A well-known phenomenon
of smoothing of information is taken into account by means of a special
composition that explicitly involves the sought function and is
infinitesimal in the space }$L_{2}${\small . A finite error in the
determination of the data by measurements can also be represented in the
same manner. By relatively simple transformations, the outlined class of
problems is reduced to the solution of Fredholm integral equations of the
second kind with properties most favorable for the numerical realization.}

{\small We demonstrate a reduction to Fredholm integral equations of the
first kind and, correspondingly, a possibility to extend the suggested
approach to wide classes of linear boundary-value and initial-boundary-value
problems characterized by variable coefficients, non-canonical domains, as
well as by other factors complicating their solution. Basically similar
techniques are developed for the solution of initial-boundary-value problems
for nonlinear differential equations of the evolutionary type. Also
considered are boundary-value problems for substantially nonlinear equations
and equations of a mixed type, the cases of nonlinearities in the boundary
conditions, the presence of a small factor by the highest-order derivative,
the inverse problem of the restoration of the variable coefficient of the
differential operator, and the problem of the Stefan type.}

{\small We put forward arguments that the determination of causal
relationships, based on the formulation restricted to a primitive renaming
of known and unknown functions of the corresponding direct problem, is
essentially illegitimate. The suggested approach may be considered as a
constructive realization of J. Hadamard's ideas of the existence of correct
formulations of physically meaningful problems.\bigskip }

{\small \bigskip Translated from Russian: www.pelbook.narod.ru (June 8,
2001).}

{\small E-mail: perchik@mail.com}

\newpage
\tableofcontents

\chapter{Introduction}

At the beginning, we should explain the title of the work and, in the first
place, the meaning of the employed notions. In this regard, we assume the
availability of information allowing us to formulate a mathematical model of
a certain phenomenon in a traditional way. Correspondingly, the
determination of unknown functions using the data of the problem is implied.
If the dependence of the solution to the problem on these data with respect
to the norm of the chosen space is continuous, such a problem, as a rule,
belongs to the domain of analysis or, in other words, its formulation is
direct.

However, the investigation of a concrete phenomenon in a variety of the
determining factors with the aim of obtaining, as a final result, of
qualitatively new information (the synthesis of knowledge) also envisages
the realization of problems in their inverse formulation,i.e., the
restoration of data using the hypothetically known solution: In other words,
the restoration of the cause using its consequence, which is usually
identified with the necessity of solving incorrectly formulated problems.

The purpose of the present investigation consists in the justification of
the illegitimacy of this statement and, on the contrary, in a constructive
development of J. Hadamard's ideas of the existence of correctly formulated
problems, adequately describing real processes and phenomena. Note that the
difference between these two notions in the context of the book is
unessential. However, the term ''process'' accentuates a time factor.

In the focus of the attention is a natural, to our mind, issue that, as an
example, can be explained by the evaluation of the integral 
\begin{equation}  \label{i1}
\left( A\psi \right) \left( x\right) \equiv \int\limits_0^1k\left( x,\xi
\right) \psi \left( \xi \right) d\xi =f\left( x\right) ,\quad x\in \left[ 0,1%
\right] ,
\end{equation}
which amounts to the determination of the function $f\left( x\right) $ using
given $k\left( x,\xi \right) $ and $\psi \left( x\right) $. This procedure
can be easily associated with a lot of physical, as well as other,
interpretations. Its realization, at least in the case of the bounded
integrand, does not pose any problem.

On the other hand, if the kernel $k\left( x,\xi \right) $ and the right-hand
side $f\left( x\right) $, evaluated according to Eq. (\ref{i1}), are given,
it is obvious that the corresponding function $\psi \left( x\right) $ is
objectively existent. Thus, the problem is whether it is legitimate to
restore this function by means of the solution of the Fredholm integral
equation of the first kind (\ref{i1}), just renaming the known and unknown
components, i.e., by assuming that the function $f\left( x\right) $ is given
and $\psi \left( x\right) $ is to be determined. And, generally speaking,
what is the basis to argue that mathematical formulations of the direct the
inverse problems can be absolutely identical?

By the way, it is very difficult to imagine the way of the formulation of
the inverse problem on physical grounds explicitly, without any relation to
the direct one. The fact of primitive renaming of the known and the unknown
function, without any additional corrections, itself causes a protest. Thus,
we put forward a thesis that a constructive approach to the formulation of
inverse problems should differ from the common one. This orientation caused
the presence in the title of the work the notion of methodology.

As a matter of fact, we hope to find reserves of the synthesis of the whole
complex of knowledge about the phenomenon by investigating it from different
sides, using formulations whose mathematical definitions are not identical.
Although the use of Eq. (\ref{i1}) for the determination of $f\left(
x\right) $ is fully justified, the problem of the restoration of the
function $\psi \left( x\right) $ does not need to coincide with it with
regard to its representation and, correspondingly, does not need to be
reducible to the Fredholm integral equation of the first kind, characterized
by instability of algorithms of numerical realization.

However, if there exists an alternative to the formulation of the inverse
problems by means of the above-mentioned renaming of known and unknown
components, it is logical to assume that corresponding formulations may
possess much more attractive properties in a computational sense. From this
point of view, the arguments of J. Hadamard acquire a rather concrete
meaning, stimulating a search of adequate, in the given situation, and, at
the same time, correct formulations of problems of mathematical physics. A
constructive realization of the outlined orientation turned out to be
possible in the context of the following considerations.

The reasons for the difficulties related to the solution of incorrectly
formulated and essentially mathematically senseless problems are, in
principle, well understood. As regards the Fredholm integral equation of the
first kind (\ref{i1}), the key role is played by the mismatch between the
function $f\left( x\right) $ and a hypothetically exact solution of the
corresponding direct problem (the result of integration) in the conditions
of approximations and different rounding of significant digits.

Correspondingly, considerable attention is paid to the so-called phenomenon
of smoothing of information in procedures of integration. At the same time,
the data of the problem, i.e., the free term $f\left( x\right) $ and also
the kernel $k\left( x,\xi \right) $, are usually determined experimentally,
which inevitably incurs a considerable error in Eq. (\ref{i1}). In this
regard, we should point out the dominance of the methodology of A. N.
Tikhonov that is based on objective incorrectness of the formulation of most
problems of mathematical simulation.

There appears a rather obvious, as it seems, question: Why not take into
account in practice the above-mentioned errors in the formulation of
problems, instead of merely bearing them in mind when identifying the
reasons for computational discrepancies? One can assume that an adequate
simulation of the error may contribute to a correct formulation of problems
that are considered to be inverse.

The adequacy implies, in the first place, the structure of the
representation of the error. In this regard, let us turn to the procedure of
integration (\ref{i1}). On the basis of general considerations, it is
logical to represent the loss of information about the function $\psi \left(
x\right) $ in the evaluation of $f\left( x\right) $ in the form 
\begin{equation}  \label{i2}
\left( \delta f\right) \left( x\right) =\psi \left( x\right) -\lambda
\int\limits_{-1}^1h\left( x,\xi \right) \psi \left( \xi \right) d\xi ,\quad
x\in \left[ 0,1\right] .
\end{equation}
Here, the function $\psi \left( x\right) $, $x\in [-1,0)$, the kernel $%
h\left( x,\xi \right) $ and the parameter $\lambda $ should satisfy the
requirement of the realization of the condition of the type $\left( \delta
f\right) \left( x\right) =0$ in $L_2\left( 0,1\right) $ for $\psi \left(
x\right) $, $x\in [0,1]$ from a rather representative class. As will be
shown, this requirement does not pose a considerable problem.

Thus, the outlined structure embodies the difference between the function $%
\psi \left( x\right) $, subject to integration, and a close expression that,
in turn, appears as a result of the execution of an analogous procedure. One
should note the absence of any a priori premises of self-sufficiency of (\ref
{i2}) in achieving the goal, namely, a correct formulation of the problem of
the restoration of the function $\psi \left( x\right) $ on the basis of the
data of (\ref{i1}). In other words, we rather put forward a hypothesis about
priority importance of a qualitative side of the model of smoothing of
information.

In light of the above, instead of the traditional incorrectly formulated
problem (\ref{i1}) for the determination of the function $\psi \left(
x\right) $, the following system of equations is employed: 
\[
\mu \left( A\psi \right) \left( x\right) =\mu f\left( x\right) +\left(
\delta f\right) \left( x\right) ;\quad \left( \delta f\right) \left(
x\right) =0,\quad x\in \left[ 0,1\right] , 
\]
where $\mu $ is a parameter, analogous to $\lambda $.

As a result of comparatively simple transformations, we managed to reduce
the considered problem to the solution of the Fredholm integral equation of
the second kind with more favorable properties for numerical realization. We
assume the absence of singularities as well as oscillations of the kernel,
not caused by $k\left( x,\xi \right) $, i.e., those that are enforced by the
employed algorithm.

Aside from explicit presence of the sought function and, in general, the
structure of the representation of the error (\ref{i2}), a necessary
condition turned out to be the completeness of the kernel 
\[
h\left( x,\xi \right) =\frac{1-r^2}{1-2r\cos \left[ 2\pi \left( x-\xi
\right) +r^2\right] },\quad 0<\left| r\right| <1, 
\]
whereas the simplification of the evaluation proved to be possible due to
its symmetry $h\left( x\right) \equiv h\left( x\pm 1\right) $ and dependence
on the difference of the arguments.

However, a broad class of linear boundary-value and initial-boundary-value
problems of mathematical physics can be rather elementarily reduced to the
Fredholm integral equations of the first kind. As a result, the
above-mentioned procedure can be considered as an efficient method of their
numerical realization. This statement is illustrated by the following
example: 
\begin{equation}  \label{i3}
\partial _x^2u+\partial _y^2u=-1,\quad u=0,\quad x=y=0;\quad x=y=1.
\end{equation}
Form the notation $\partial _x^2u=\psi $, in view of (\ref{i3}), it follows: 
\begin{equation}  \label{i4}
u\left( x,y\right) =\left[ \int\limits_0^x\left( x-\xi \right)
-x\int\limits_0^x\left( 1-\xi \right) \right] \psi \left( \xi ,y\right) d\xi
;
\end{equation}
\begin{equation}  \label{i5}
u\left( x,y\right) =\frac 12y\left( 1-y\right) -\left[ \int\limits_0^y\left(
y-\eta \right) -y\int\limits_0^y\left( 1-\eta \right) \right] \psi \left(
x,\eta \right) d\eta .
\end{equation}
The elimination of the function $u\left( x,y\right) $ leads to a
two-dimensional Fredholm integral equation of the first kind. In the case of
conditions on the contour of a more general type, one can use, for example
the following relation: $\partial _x^2u+u=\psi $. The outlined scheme is
practically indifferent to the type and order of differential operators, the
presence of variable coefficients, the configuration of the boundary of the
domain of the function and some other factors that usually complicate the
realization of numerical algorithms.

It should be noted that the dimensionality of the boundary-value problem,
regarding its further reduction to the Fredholm integral equation of the
first kind, is of no principal importance, because this procedure is, in
fact, performed with respect to one variable while the rest play the role of
particular parameters. By the way, in this context, the conventional
division of problems of mathematical simulation into direct and inverse
becomes rather conditional.

Nonlinear boundary-value and initial-boundary-value problems are, in
general, analogously reduced to nonlinear integral equations of the second
kind that are specially adopted for an effective realization of iterative
algorithms.

Perhaps, the motivation of the proposed investigation could be of certain
interest. At the beginning, the author was confused by the absence in the
specialized literature of a clear statement of the universality of the
presented method of the reduction of problems of mathematical physics to the
Fredholm integral equations of the first kind. Given the complexity of the
solution of incorrect problems, the factor of the elimination of boundary
and initial conditions as well as a placement of the whole lot of initial
data in the kernels of derived equations seems to be rather attractive.
Indeed, a conventional classification of problems according to the
complexity of their numerical realization, in fact, vanishes, and the
construction of an effective algorithm of the solution of the Fredholm
integral equation of the first kind appears in the foreground.

At the same time, a specific degeneracy of the relations for the new unknown
function $\psi $, generating the representation (\ref{i4}), (\ref{i5}), is
worth noting. There appears a natural question: Is there a possibility to
reduce boundary-value problems to equations with more acceptable properties
(in the sense of a numerical realization) by means of relations between the
functions $u$ and $\psi $ of a more general type. For example, setting 
\[
u\left( x,y\right) =\left[ \int\limits_0^xK\left( x,y,\xi \right)
-x\int\limits_0^xK\left( 1,y,\xi \right) \right] \psi \left( \xi ,y\right)
d\xi , 
\]
where 
\[
K\left( x,y,x\right) \equiv 0,\quad x,y\in \left[ 0,1\right] , 
\]
can one complete the definition of the kernel $K\left( x,y,\xi \right) $ to
reduce the problem (\ref{i3}) to the Fredholm integral equation of the
second kind with respect to the function $\psi $? However, an attempt to do
something like this did not succeed.

Analogously, no results were yielded by attempts to reduce the considered
problems to integral equations of the first kind $B\psi =A\psi +f$, with the
operator $B$ having a suitable inversion form [by analogy with (\ref{i1}), $%
A $ and $f$ characterize the data of a concrete problem]. At the same time,
this did not mean that one could not realize, in principle, the algorithm of
a derivation by means of transformations of the Fredholm integral equation
of the second kind.

The use of the model of smoothing of information by means of integration of
the type (\ref{i2}) proved to be constructive. In this regard, a profound
elaboration of methodological aspects of correct formulation of problems of
mathematical physics proved to be necessary.

In section 2, we analyze J. Hadamard's arguments concerning the issue of
correct formulation of problems for partial differential equations. Both
related and alternative positions on this issue of known specialists are
also illuminated. Different interpretations of contextually close Banach's
inverse operator theorem, one of the three main principles of functional
analysis, are given.

Section 3 contains a critical review of methods of the solution of incorrect
problems, based, in the first place, on the concepts of A. N. Tikhonov and
V. M. Fridman. We express considerations on the reasons for a crisis in the
field of mathematical simulation, related to an orientation at the use of
incorrect formulations that reduce to a numerical realization of ill-defined
algebraic systems.

The material of section 4, in a sense refracts principle difficulties,
accompanying the solution of incorrectly formulated problems, by the prism
of fundamental concepts of J. Hadamard and S. Banach. In this context, we
formulate general premises for correct formulation of problems of
mathematical simulations, traditionally associated with the Fredholm
integral equations of the first kind.

Section 5 is devoted to the presentation of an original approach to the
solution of a canonical equation of the above-mentioned type and is basic
constructively. Here we investigate issues related to the simulation of the
error and to the transformation of the initial formulation. We give an
algorithm of the reduction of the problem to the Fredholm integral equation
of the second kind. and also consider corresponding aspects of
methodological character.

In section 6, we concretely demonstrate the universality of the technique of
the reduction of linear boundary-value and initial-boundary-value problems
to Fredholm integral equations of the first kind with respect to one of
high-order derivatives. We also demonstrate an extension to them of the
algorithm of the previous section.

Section 7 develops the outlined orientation involving into the sphere of
effective realization sufficiently nontrivial applications (including
factors of nonlinearity, singular perturbations and some other). The
presentation of the material has the form of sketches.

In section Conclusions, we summarize the main points of the proposed
investigation from the position of priority importance of correct
formulation of problems of mathematical physics for their numerical
realization.

Mathematical techniques employed in the presentation of the material is
comparatively simple: basics of the classical theory of integral equations;
elements of functional analysis; general principles of formulations of
problems of mathematical physics and methods of their solution.

The literature to each section is given in reference order. We refer mainly
to original publications and those ones that treat the subject in most
detail. Sections and subsections (chapters and paragraphs of the literature
sources) are denoted, respectively, as section 1, section 1.1, sections 1.1,
1.2.

The author has published about 30 papers on the subject of this
investigation in leading scientific journals and conference proceedings. All
these publications are written by the author on his own. They clearly state
the general orientation of the investigation and the means of its practical
realization. However, only by turning to the methodology of correct
formulation of problems of mathematical physics, the author managed to
provide the justification of the developed approach.

The concepts, described below, were formed at different stages under the
influence of well-known specialists, in particular: Profs. S. M. Belonosov,
P. M. Varvak, E. S. Wentzel, N. A. Virchenko, Yu. V. Gandel, Yu. A. Danilov,
V. A. Dobrushkin, M. D. Dolberg, N. T. Zhadrasinov, P. A. Zhilin, B. Y.
Kantor, V. V. Kopasenko, K. A. Lurie, S. G. Mikhlin, N. F. Morozov, M. V.
Paukshto, A. S. Sakharov, E. A. Simson, V. C. Ryabenkii, L. A. Filshtinskii,
N. P. Fleishman, D. G. Khlebnikov, A. V. Cherkaev, I. Y. Chudinovich, I. D.
Chueshov, V. A. Shterbina, as well as Corresponding Member of The Academy of
Sciences of the USSR A. I. Lurie, Academician of the National Academy of
Sciences of Ukraine V. L. Rvachev, Prof. E. B. Koreneva, and Senior
Researcher I. Z. Reutfarb.

Prof. S. N. Kan stimulated the author's interest in the development of
approximate methods of the solution of problems of engineering mechanics.
The author acknowledges the influence of Prof. L. P. Vinokurov, who
supported the reorientation of the author's interest beyond the framework of
the outlined Dr. Sc. Thesis. Of great importance were discussions with
Profs. B. G. Korenev and P. I. Perlin. For a long period, the author enjoyed
discussions with Prof. G. I. Drinfeld, who gave valuable advices.

The author's investigations were supported by his employers: The Department
of Engineering Mechanics of Kharkov Civil-Engineering Institute (Prof. V. P.
Sachko, Rector; Prof. A. V. Borisov, Head of the Department; A. S.
Shevchenko, Head of ONIL; Profs. V. A. Voblykh, N. M. Grigoriants, V. I.
Pustynnikov, N. V. Zinchenko); North-Eastern Research Center of the National
Academy of Sciences of Ukraine (P. T. Bubenko, Ph.D., Director; Prof. V. A.
Fedynskii, Scientific Secretary; A. L. Razumovskii. Head of the Department);
Joint-Stock Company ''Research-Tecnological Institute for Transcription,
Translation and Replication'' (S. I. Chernyshov, President; V. M.
Zamolotskii, Vice-Presedent; S. I. Dotsenko, Director); Joint-Stock Company
''Interregional Electric-Energetic Association'' (A. L. Livshits, Ph.D.,
President; V. G. Katsman, General Director); Joint-Stock Company ''Ukrainian
Information Systems'' (Prof. E. A. Simson, President; V. B. Liubetskaya,
Head of the Board); Joint-Stock Company ''AVEC'' (A. B. Feldman, President;
G. K. Bronshpak, President Adviser; S. V. Ingultsov, Head of the Analytical
Department.

An important role in the development of the ideas of causal relations was
played by applied studies of interdisciplinary problems carried out in close
collaboration with the following specialists: Profs. N. N. Vnukova, V. M,
Moskvin; I.G. Kiuila, Dr. Sc.; Profs. L. B. Gerjula, A. S. Gugel, E. A.
Moldavskii, V. A. Filshtinskii,V. A. Frydynskii; V. E. Vasiliev, Ph.D.; S.
I. Dotsenko, Ph.D.; V. M. Zamolotskii, Ph.D.; Director of Ecological Center
''Region'' V. V. Ganichev; Vice-Director of PEO ''Kharkovenergo'' S. T.
Prishtepov.

Of great support were long-term friendly relations with A. V. Dolgarev, A.
I. Korobov, A. Mangubi, A. L. Razumovskii, A. S. Shevchenko, I. G.
Shepelenko, V. V. Shestakov, and K. Z. Yagudin.

The author thanks Sergey V. Kuplevakhsky for translating the manuscript into
English and Ilya Stepanov for computer processing of the text.

While preparing the present version, we have corrected typographic errors
noticed in www.pelbook.narod.ru, added four footnotes, two literature
references in sections 2, 3 and Appendix.

\chapter{The issue of the correct formulation of problems of mathematical
physics}

\section{Hadamard's definition of correctness}

J. Hadamard has defined two conditions that should be satisfied by a
correctly formulated boundary-value (initial-boundary-value) problem for
partial differential equations: existence and uniqueness of the solution ( 
\cite{1.1}, p. 12).\footnote{%
For the first time, the concept of correct formulation was put forward by J.
Hadamard in his article of 1902.} At the same time, the third condition of
Hadamard's definition of correctness is well-known that concerns continuous
dependence on the data of the problem. Indeed, he paid serious attention to
the investigation of this issue with regard to Cauchy-Kovalevskaya's theorem
concerned with the solution of the differential equation 
\begin{equation}  \label{1.1}
\partial _t^ku=f\left( t,x_1,x_2,\ldots ,x_n,\partial _tu,\partial
_{x_1}u,\partial _{x_2}u,\ldots ,\partial _{x_n}^ku\right)
\end{equation}
(a system of analogous equations), where $f$ is an analytical function of
its arguments in the vicinity of the origin of coordinates, with initial
conditions 
\[
u\left( 0,x_1,x_2,\ldots ,x_n\right) =\varphi _0\left( x_1,x_2,\ldots
,x_n\right) ; 
\]
\begin{equation}  \label{1.2}
\ldots
\end{equation}

\[
\partial _t^{k-1}u\left( 0,x_1,x_2,\ldots ,x_n\right) =\varphi _{k-1}\left(
x_1,x_2,\ldots ,x_n\right) . 
\]

As is pointed out by J. Hadamard, the consideration of the problem (\ref{1.1}%
), (\ref{1.2}), named after Cauchy, raises three questions (\cite{1.1}, p.
17):

1) Does it admit a solution?

2) Is the solution unique? (In general, is the problem correctly formulated?)

3) How the solution can be actually derived?

Cauchy-Kovalevskaya's theorem (in its authors' interpretation) states that,
except for some special cases, the above-mentioned problem admits a unique
solution that is analytical at the origin of coordinates. Moreover, the
functions $\varphi _0$,..., $\varphi _{k-1}$ can be not only analytical but
regular, i.e., continuous together with their derivatives up to a certain
order. A possibility of a uniform approximation of $\varphi _0$,..., $%
\varphi _{k-1}$ by Taylor series expansions in powers of $x_1$, ..., $x_n$ ,
retaining all operations on analytical functions (including differentiation
up to a corresponding order), is implied.

However, such an approach was strongly criticized by J. Hadamard. In his
opinion, the question is not how such an approximation affects the initial
data, but rather what is an effect on the solution? He emphasized the
non-equivalence of the notion of small perturbation for given Cauchy's
problem and of the solution to this problem (\cite{1.1}, p. 39). In this
regard, J. Hadamard presented his prominent example of a solution of the
differential equation 
\begin{equation}  \label{1.3}
\partial _t^2u+\partial _x^2u=0,
\end{equation}
subject to the conditions 
\begin{equation}  \label{1.4}
u\left( x,0\right) =0;\quad \partial _tu\left( 0,x\right) =\alpha _n\sin
\left( nx\right) ,
\end{equation}
where $\alpha _n$ is a rapidly decreasing function of $n$.

The expression on the right-hand side of (\ref{1.4}) can be arbitrarily
small. Nevertheless, the problem admits the solution 
\begin{equation}  \label{1.5}
u\left( x,t\right) =\frac{\alpha _n}n\sin \left( nx\right) \sinh \left(
nt\right) .
\end{equation}
For $\alpha _n=1/n$ or $1/n^\mu $, or $e^{-\sqrt{n}}$, this solution is
rather large for any nonzero $t$, because of the prevailing growth of $%
e^{nt} $ and, correspondingly, of $\sinh \left( nt\right) $. Thus, the
function (\ref{1.5}) does not depend continuously on the initial data and,
as a result, the problem (\ref{1.3}), (\ref{1.4}) is incorrect.

Concerning the regularity of the right-hand side of (\ref{1.2}), J. Hadamard
remarked: ''...actually, one of the most curious facts of the theory is that
equations, seemingly very close to each other, behave in a completely
different way'' (\cite{1.1}, p. 29).

A large number of investigations devoted to the issue of the correct
formulation of Cauchy's problems. The authors of these investigations
concerned themselves with specification of corresponding classes of
differential equations and with minimization of requirements imposed on the
initial data (see \cite{1.2}). However, we are mostly interested in the
actual character of the dependence of the solution on the data of the
problem and, in this regard, the classic J. Hadamard's statement that ''an
analytical problem is always correctly formulated in the above-mentioned
sense, when there exists a mechanical or physical interpretation of the
question'' (\cite{1.1}, p. 38).

As was pointed out by V. Y. Arsenin and A. N. Tikhonov \cite{1.3}, the
latter questioned the legitimacy of studies of incorrect problems, specified
by the authors as the following: the solution of integral equations of the
first kind; differentiation of approximately known functions; numerical
summation of Fourier series whose coefficients are approximately known in
the metric $l_2$; analytical continuation of functions; the solution of
inverse problems of gravimetry and of ill-defined systems of linear
algebraic equations; minimization of functionals for divergent sequences of
coordinate elements; some problems of linear programming and of optimal
control; the design of optimal systems and, in particular, the synthesis of
aerials. It is emphasized that this list is by no means complete, because
incorrect problems appear in investigations of a broad spectrum of problems
of physics and engineering.

In his talk at the meeting of the Moscow Mathematical Society devoted to J.
Hadamard's memory, G. E. Shilov said the following \cite{1.4}: ''Our time
has brought about corrections in Hadamard's instructions, because it turned
out that incorrect, according to Hadamard, problems could have meaning (as,
e.g., the problem of restoration of a potential from scattering data).
However, the studies of correct problems, proclaimed by Hadamard, was a
cementing means for the formation of the whole theory'' (functional analysis
is implied). This quotation is borrowed from a biographical sketch by E. M.
Polishtuk and T. O. Shaposhnikova \cite{1.5}, where it is also pointed out
that in the course of time J. Hadamard's opinion about the importance for
practice of exclusively correct problems was understood in a less absolute
sense.

At the same time, rather sharp statements were made:

'' And what is more, Hadamard put forward a statement that incorrect
problems had no sense at all. Since (as can be seen from a modern point of
view) most applied problems, represented by equations of the first kind, are
incorrect, this statement of the outstanding scientist, apparently, strongly
slowed down in 1920-1950's the development of the theory, methods and
practice of the solution of problems of this class'' (\cite{1.6}, p. 12).

''Until quite recently, it was thought that incorrect problems had no
physical sense and that it was unreasonable to solve them. However, there
are many important applied problems of physics, engineering, geology,
astronomy, mechanics, etc., whose mathematical description is adequate
although they are incorrect, which poses an actual problem of the
development of efficient methods of their solution'' (\cite{1.6}, p. 225).%
\footnote{%
In the context of what follows, we draw attention to the ''adequate
description''.}

''From the results of this work [of A. N. Tikhonov] followed a limitation of
the well-known notion of J. Hadamard \cite{1.1} of a correctly formulated
problem of mathematical physics, which was of indisputable methodological
interest, and inconsistency of Hadamard's thesis, wide-spread among
investigators, that any incorrectly formulated problem of mathematical
physics was unphysical.'' (\cite{1.7}, p. 3).

''For a long time, activities related to the analysis and solution of
problems called incorrect used to be relegated (by famous mathematicians
too) to the domain of metaphysics'' (\cite{1.8}, p. 126). ''A prevailing
number of mathematicians (including Hadamard) expressed.their attitude
towards this problem in the following way: If a certain problem does not
meet the requirements of correctness, it is of no practical interest and,
hence, does not need to be solved'' (\cite{1.8}, p.127) (I. G.
Preobrazhenskii, the author of the section ''Incorrect problems of
mathematical physics'').

Note that the latter paper most distinctively reveals the style that causes
a principal objection. Thus, A. Poincar\'e is accused of inconsistency of
methodological views on the nature of causal relationship (\cite{1.9})
(''The Last Thoughts''). Indeed, the text does not contain any evidence that
he makes a fetish of the problem of restoration of the cause from the
effect. On this basis, a conclusion is made about the great scientist's
misunderstanding of the essence of instability of computation procedures
inherent to incorrect problems and, in particular, to integral equations of
the first kind.\footnote{%
In particular, the exact statement reads: ''However, one must remember that
vagueness of philosophical positions of some scientists in the West, even
rather renown, results in the fact that, based on correct starting points,
they draw rather inaccurate conclusions, repeating old mistakes of, for
example, A. Poincar\'e, who writes: 'If two organisms are identical, or
simply similar, this similarity could not occur by chance, and we can assert
that they lived under the same conditions...' In other words, the fact of
possible incorrectness of the inverse problem is completely ignored.''
However, one would hardly mention Poincar\'e's mistakes... if modern
''spontaneous supporters of the principle of determinism did not repeat
them'' ( [8], p. 134).}

The adequacy of employed models to considered concrete processes is not even
touched on. Thus, a quite legitimate question arises: How does one know that
A. Poincar\'e, if necessary, could not find a way of a mathematically
correct formulation of the same physical problems? Anyway, is there any
contradiction in general arguments for the existence of such a possibility,
including the aspects of its constructive realization?

By the way, exactly A. Poincar\'e repeatedly mention J. Hadamard while
establishing a relationship between the correct formulation of problems and
a practical realization of employed models. We draw attention to an
expressive thesis: ''If a physical problem reduces to an analytical one,
such as (\ref{1.3}), (\ref{1.4}), it will seem to us that it is governed by
a pure occasion (according to Poincar\'e, it means that determinism is
violated) and it does not obey any law'' (\cite{1.1}, p. 43).

In light of the above, the arguments of I. Prigogine and I. Stengers \cite
{1.10} are of interest: ''...one can speak of a 'physical law' of some
phenomenon only in the case when this phenomenon is 'coarse' with respect to
a limiting transition from a description with a finite accuracy to that with
an infinite accuracy and thus inaccessible to any observer, whoever he may
be'' (p. 9). ''Scientist in a hundred different ways expressed their
astonishment that a correct formulation of the question allows them to solve
any puzzle suggested by nature'' (p. 44).

Thus, underlined are: first, methodological importance of correct
formulation of problems of mathematical physics; second, a leading role of
the employed procedures and, finally, substantial influence of the quality
of their realization on the degree of complexity of obtaining the final
result. In other words, one and the same problem can be better or worse
formulated. The above-mentioned statement, which is, essentially, the same
hypothesis of J. Hadamard, as a matter of fact, implies a possibility of a
''good'' (correct) formulation of any meaningful problem and, consequently,
can be interpreted as having a global orientation.

In this regard, one can establish an obvious relationship to D. Hilbert's
comments on his 20th problems that suggested a possibility of correct
formulation of arbitrary boundary-value problems of mathematical physics by
means of special requirements on boundary values of corresponding functions
(a type of continuity or piecewise differentiability up to a certain order)
and, by necessity, by giving an extended interpretation to the notion of the
solution (\cite{1.11}, pp. 54-55).

For the first time, the three conditions of the correctness of problems of
mathematical physics were clearly pointed out by D. Hilbert and R. Courant ( 
\cite{1.12}, pp. 199-200): existence, uniqueness and continuous dependence
of the solution on the data of the problem. Concerning the last, they say:
''...it has crucial importance and is by no means trivial... A mathematical
problem can be considered adequate to the description of real phenomena only
in the case when a change of given data in sufficiently narrow limits is
matched by an alike small, i.e. restricted by predetermined limits, change
of the solution''.

V. A. Steklov's position is quite analogous (\cite{1.13}, p. 62): ''...if
differential equations with the above-mentioned initial and boundary
conditions are not constructed on erroneous grounds, are not in direct
contradiction to the reality, they must yield for each problem a unique and
completely definite solution...''. Along the same lines, I. G. Petrovskii
writes (\cite{1.14}, p. 87): 'The above-mentioned arguments for the correct
formulation of Cauchy's problem show that other boundary-value problems for
partial differential equations are of interest for natural science only in
the case when there is, in a sense, continuous dependence of the solution on
boundary conditions''.

S. L. Sobolev is less categorical (\cite{1.15}, p. 38): ''The solution to an
incorrectly formulated problem in most cases has no practical value''.

Of considerable interest is the opinion of V. S, Vladimirov (\cite{1.16}, p.
69): ''The issue of finding correct formulations of problems of mathematical
physics and methods of their solution (exact or approximate) is the main
content of the subject of equations of mathematical physics''.

V. V. Novozhilov, in fact, drew attention to the potential of variation of
the formulation of the considered problem with the aim of the simplification
of the procedure of its numerical realization (\cite{1.17}, p. 352): ''The
absence in the term ''a mathematical model'' of the indication of its
inevitable approximate character leaves way for a formal mathematical
approach to models, disregarding those concrete problems for whose solution
they were intended, which is, unfortunately, wide-spread at present''.

\section{J. Hadamard's hypothesis and incorrectness of ''real'' problems}

Thus, J. Hadamard and other outstanding scientists thought that any
physically interpretable problem could be correctly formulated. However, a
quite opposite point of view dominates in modern publications. Indeed, a
visibly larger part of practically important problems considered therein are
incorrect. However, is the actual methodology of mathematical formulation of
these problems and, correspondingly, the results of its refraction with
respect to realities adequate?

Here we will not elaborate on something like general principles of the
construction of differential equations, and, generally speaking, it is
reasonable at the beginning to restrict the question to the following: What
arguments allow one to conclude that an incorrectly formulated problem
adequately describes an observable phenomenon or a potentially real process?
In this regard, let us turn to the procedure of the solution of the Fredholm
integral equation of the first kind 
\begin{equation}  \label{1.6}
\int\limits_0^1k\left( x,\xi \right) \psi \left( \xi \right) d\xi =f\left(
x\right) ,\quad x\in \left[ 0,1\right] ,
\end{equation}
which is a classical incorrect problem: the kernel $k\left( x,\xi \right) $%
and the free term $f\left( x\right) $ are given; the function $\psi \left(
x\right) $ is to be restored.

For definitiveness, let the kernel $k$ be symmetric and closed, i.e. $%
k\left( x,\xi \right) \equiv k\left( \xi ,x\right) $ and its eigenfunctions $%
\bar \psi _n\left( x\right) $, being nontrivial solutions of the integral
equation 
\[
\psi \left( x\right) =\lambda \int\limits_0^1k\left( x,\xi \right) \psi
\left( \xi \right) d\xi =f\left( x\right) ,\quad x\in \left[ 0,1\right] 
\]
with characteristic numbers $\lambda =\lambda _n$, $n=1,2,\ldots $, form a
complete in $L_2\left( 0,1\right) $orthogonal system of elements. Besides,
the function $f\left( x\right) \in L_2\left( 0,1\right) $. In this case,
according to Picard's theorem, the solution to Eq. (\ref{1.6}) exists and is
unique under the condition (see, e.g., \cite{1.18}) 
\begin{equation}  \label{1.7}
\sum_{n=1}^\infty \alpha _n^2\lambda _n^2<\infty ,\quad \alpha
_n=\int\limits_0^1f\left( x\right) \bar \psi _n\left( x\right) dx.
\end{equation}

If we assume that all the above-mentioned conditions are fulfilled, there is
still the third condition of correctness that, as is known (\cite{1.19}, p.
69), is not satisfied by Eq. (\ref{1.6}). Numerous literature references
clearly illustrate an inadequately strong influence on the solution of small
perturbations of the data of the problem, in the first place of $f\left(
x\right) $. As a rule, this function is determined experimentally and
mismatch the kernel $k\left( x,\xi \right) $ with respect to smoothness.
Thus, Eq. (\ref{1.6}), strictly speaking, looses sense. At the same time, a
possibility of an equivalent description of realistic problems by means of
integral equations of the first kind is indisputably admitted at present,
which is confirmed by their colossal list \cite{1.6}.

Let us specify Eq. (\ref{1.6}): 
\begin{equation}  \label{1.8}
k\left( x,\xi \right) =\left\{ 
\begin{array}{c}
\left( 1-x\right) \xi ,\quad 0\leq \xi \leq x; \\ 
x\left( 1-\xi \right) ,\quad x\leq \xi \leq 1;
\end{array}
\right. \quad f\left( x\right) =\frac 1{\left( m\pi \right) ^2}\sin \left(
m\pi x\right) ,
\end{equation}
where $m$ is an integer.

For this choice, $\bar \psi _n\left( x\right) =\sin \left( n\pi x\right) $; $%
\lambda $$_n=\left( n\pi \right) ^2$, $n=1,2,\ldots $ . The use of Mercer's
theorem \cite{1.18}, according to which 
\[
k\left( x,\xi \right) =\sum_{n=1}^\infty \frac{\bar \psi _n\left( x\right)
\bar \psi _n\left( \xi \right) }{\lambda _n}, 
\]
and a representation of $\psi \left( x\right) $ as a series expansion in
terms of $\bar \psi _n\left( x\right) $ with undetermined coefficients
allows one to find the solution to Eq. (\ref{1.6}): 
\begin{equation}  \label{1.9}
\psi \left( x\right) =\sin \left( m\pi x\right) .
\end{equation}

However, the procedure of calculations turned out to be so simple owing to a
special choice of the data of the problem. If this is not the case or in the
case of the solution of Eq. (\ref{1.6}) with the kernel and the free term (%
\ref{1.8}) by means of one of numerical methods, the complexity of the
realization of a an approximation of sufficiently high order is practically
identical to the most general situation, characterized by an error in the
determination of $f\left( x\right) $ and $k\left( x,\xi \right) $.\footnote{%
This complexity implies an ill definition of systems of linear algebraic
equations obtained as a result of some sort of discretization. The methods
allowing one to overcome this difficulty are discussed in section 3.} As a
matter of fact, even if the data are objectively compatible, the
incorrectness of Eq. (\ref{1.6}) is incurred by an error in the
approximation of $f$ and $k$ and also by rounding off the digits by the
computer.

The key factor of the incorrectness of Eq. (\ref{1.6}) follows from a
comparison of the free term (\ref{1.8}) with the solution (\ref{1.9}).
Indeed, by increasing $m$, the function $f\left( x\right) $ can be made
arbitrarily small without any change of the amplitude value of $\psi \left(
x\right) $. Correspondingly, any error in the calculations with $f\left(
x\right) $ is projected onto $\psi \left( x\right) $ with the factor $m^2$.
The mechanism of this effect related to the smoothing of information about
the sought function by the procedure of integration will be repeatedly
discussed in what follows.

However, let us return to the question of the relation of an incorrect
formulation to the reality. In this regard, we draw attention to the
following. By considering (\ref{1.6}) as the Fredholm integral equation of
the first kind (the function $\psi $ is to be determined), we mean the
solution of the inverse problem (I). However, equation (\ref{1.6}) can be
used for the solution of the corresponding direct problem (D): the
determination of the function $f\left( x\right) $ from the data $k\left(
x,\xi \right) $ and $\psi \left( x\right) $. This procedure is correct and
thus is radically simple than the problem I. It is sufficient to note the
absence of any principal difference between an essentially numerical
realization and the integration of expressions (\ref{1.8}), (\ref{1.9}) in
the analytical form.

Here we want to draw attention to an issue that seems to be of substantial
importance. The problem D, as a rule, is transparent: in its categories, we
quite naturally model current processes and phenomena by, which should be
emphasized, explicit means of linear superposition. Correspondingly, if, for
instance, $k\left( x,\xi \right) $ is a characteristic of the media and $%
\psi \left( x\right) $ is intensity of external influence, a resulting
effect in this or that subject sphere is to be elementarily summed up.

The situation is diametrically different for the problem I. One could hardly
point out any realistic process (phenomenon) for which it could be
formulated in mathematical terms directly on the basis of the subject
sphere. In other words, without any relation to the problem D, which
commonly implies a transformation of the latter into the problem I by means
of mechanical renaming of known and unknown components.

In our opinion, the methodology, which states the adequacy of the problem I,
obtained by the above-mentioned renaming of the components, to the realities
on the basis of a quality information about a concrete problem D, is
profoundly erroneous. Correspondingly, the statements of experts who reject
J. Hadamard's hypothesis of the existence of correct formulations of
problems of mathematical physics should be considered unjustified.

Let us turn to the problem D that describes some realistic process (\ref{1.6}%
).For this process, the determination of $\psi \left( x\right) $ from the
data $k\left( x,\xi \right) $ and $f\left( x\right) $, i.e. the formulation
of the corresponding inverse problem that will be denoted as I$^{^{\prime }}$%
, is, of course, reasonable. Suppose that in this case J. Hadamard's
hypothesis is correct, and, hence, the problem I$^{^{\prime }}$ is correct.
However, the problem I', the solution of the Fredholm integral equation of
the first kind (\ref{1.6}), is incorrect by definition.

The conclusion is obvious: Mathematical formulations of the problems I and I$%
^{^{\prime }}$ are non-identical. As a result, the formulation of the
problem I$^{^{\prime }}$ cannot be restricted to readdressing the status of
the unknown variable between the functions $f$ and $\psi $ in the problem D.
Note in this regard that constructive methodology of the correct formulation
of the problem I$^{^{\prime }}$ is the basis of the subject and the main
objective of the present investigation.

The above arguments seem to be rather convincing, however, at this stage of
our consideration, we can neither prove the correctness of J. Hadamard's
hypothesis nor illustrate its constructive character with respect to
separate classes of problems. One should also bear in mind that, using
special methods, the solution of the incorrect problem I, as a rule, can be
obtained with accuracy that is considered to be practically acceptable. In
this regard, the question arises: Should one aim at the correct formulation I%
$^{^{\prime }}$, if the algorithm of the calculation of the function $\psi
\left( x\right) $ in the formulation of the problem I in some way realizes
its regularization? This implies a well-known deformation of the formulation
I with the use of a small parameter that yields the property of correct
solvability.

Thus, can the algorithm to a full extent, including the efficiency of
numerical realization, level off the principal difficulties inherent to the
incorrectness of the problem I in the form (\ref{1.6})? It is clear that the
answer is definitely negative: Otherwise, the deep-rooted differentiation of
problems in correctly and incorrectly formulated ones would make no sense.

Furthermore, the indicated difference is of exceptional importance, because
correctness of the formulation is a criterion of a qualitative level,
whereas the efficiency of a method of the solution of the Fredholm integral
equation of the first kind can be estimated only in terms of quantitative
factors of a palliative property. The latter is caused by a direct
relationship between a degree of regularization and the deformation
(distortion) of the problem I.

What is, however, the actual difference in the interpretation of the
formulations I and I$^{^{\prime }}$? The answer to this question is the
basic concept of the suggested approach. At this stage, we only note that a
transformation of the formulation I into the formulation I' will be mostly
qualitative and will be realized by means of zero in $L_2\left( 0,1\right) $
perturbation of Eq. (\ref{1.6}), modeling the phenomenon of smoothing of
information.

\section{Banach's theorem on the inverse operator}

Let us quote (\cite{1.5}, p. 175): ''First, Hadamard defined the correctness
of the problem by the conditions of solvability and uniqueness and strongly
insisted on continuous dependence of the solution on the initial data only
in the consideration of Cauchy's problem. In the book ''The theory of
partial differential equations'', published in Peking a year after his
death, he wrote: 'This third condition that we introduced in ''Lectures on
Cauchy's problem...'' but did not consider as part of well formulated
problems, was added, quite justified, by Hilbert and Courant \cite{1.12}.
Here, we accept their point of view.'''

E. M. Polishuk and T. O. Shaposhnikova made the following comment on this
text \cite{1.5}, pp. 175-176]: ''From a mathematical point of view, the
question of the necessity of the requirement of the continuity of the
solution with respect to the data seems to be rather delicate. As a matter
of fact, according Banach's well-know theorem on closed graph, unique
solvability of a linear problem leads to boundedness of the inverse operator
and, thus, continuous dependence of the solution on the right-hand sides.''
It is pointed out that variations of the coefficients of differential
equations and of the boundary of the considered domain can also influence
the solution of the problem; hence, the use of the three conditions of the
correctness is preferable.

At the same time, Banach's theorem on the inverse operator (\cite{1.20}, pp.
34-36), being a consequence of the above-mentioned one, is more closely
related to the considered issue. Its formulation, given by A. I. Kolmogorov
and S. V. Fomin, is the following (\cite{1.21}, pp. 259-260): Let $A$ be a
linear bounded operator that maps a Banach space $B_1$ in a one-to-one
fashion onto a Banach space $B_2$. Then the inverse operator $A^{-1}$ is
unique.

In addition, L. A. Lyusternik and V. I. Sobolev (\cite{1.22}, pp. 159-161)
emphasized that a one-to-one mapping of the whole Banach space $B_1$ onto
the whole Banach space $B_2$ is implied. Besides, a situation is discussed
when ''...an operator, being the inverse of a bounded operator, although
linear, turn out to be defined not on the whole space $B_2$ but only on a
certain linear manifold and unbounded on this manifold''.

A somewhat different interpretation of the same theorem in (\cite{1.23}, p.
60) is reduced to the following: If a linear bounded operator $A$ that maps
a Banach space $B_1$ onto a Banach space $B_2$ has an inverse $A^{-1}$, then 
$A^{-1}$ is bounded. It is pointed out that this statement becomes invalid
if one gives up the requirement of completeness of one of the spaces. There
is also a clarification: The existence and uniqueness of the solution of the
equation $A\psi =f$ with an arbitrary right-hand side from $B_2$ leads to
continuous dependence of the solution $\psi =A^{-1}f$ on $f$.

L. V. Kantorovich and G. P. Akilov made an addition concerning a mapping
under the specified conditions onto a closed subspace of the Banach space $%
B_2$ (\cite{1.24}, p. 454). As a matter of fact, a closed subspace of a
Banach space is itself a Banach space.

S. G. Mikhlin gave a proof of the theorem (\cite{1.25}, p. 507): For the
linear problem $A\psi =f$ to be correct in a pair of Banach spaces $B_1$, $%
B_2$, it is necessary and sufficient that the operator $A^{-1}$ exist, be
bounded and map the whole space $B_2$ onto $B_1$ . At the same time, the
author made a clear division between the category of the existence and
uniqueness of the solution of the boundary-value problem and its correctness
as a whole, which implies, as a result, continuous dependence on the data
(the third condition according to Hadamard). In this regard, note a typical
definition: ''A boundary-value problem is called correct in a pair of Banach
spaces $B_1$, $B_2$ if its solution is unique in $B_1$ and exists for any
data from $B_2$, and if an arbitrarily small change of the solution in the
norm $B_1$ corresponds to a sufficiently small change of the initial data in
the norm $B_2$'' (p. 204).

It is emphasized that the considered problem may turn out to be correct in
one pair of spaces and incorrect in another one. Besides, the incorrectness
of the Fredholm integral equation of the first kind (\ref{1.6}) is
established by contradiction: If the problem is correct, there exists a
bounded operator $A^{-1}$ and, hence, the identical operator $I=A^{-1}A$ is
compact in an infinite dimensional space, which contradicts the fundamentals
of the general theory \cite{1.23}. S. G. Mikhlin also quite encouragingly
pointed out the approach of an approximate solution of incorrect problems,
headed by A. N. Tikhonov.

In an analogous, as to its content, course \cite{1.26}, S. G. Mikhlin
reiterated the above-mentioned formulations. However, A. N. Tikhonov is not
mentioned at all, whereas the discussion of Eq. (\ref{1.6}) found a rather
interesting continuation. (p. 171). It is shown that the problem of its
solution becomes correct a pair of spaces $B_1$, $B_2$ is replaced with such
one that the operator $A$ is no longer compact. The general considerations
are illustrated by the following example. Let $k\left( x,\xi \right) $ and $%
f\left( x\right) $ satisfy the conditions of section 2.2, including (\ref
{1.7}). Then, according to Picard's theorem, the solution to Eq. (\ref{1.6})
has the form 
\begin{equation}  \label{1.10}
\psi \left( x\right) =\sum_{n=1}^\infty a_n\lambda _n\bar \psi _n\left(
x\right) ,\quad x\in \left[ 0,1\right] .
\end{equation}

It turns out that if one retains $L_2\left( 0,1\right) $ as $B_2$ and for $%
B_1$ also takes a Hilbert space of functions normalized according to (\ref
{1.7}), i.e. $l_2$, the solution of Eq. (\ref{1.6}) becomes a correctly
formulated problem, and, correspondingly, the operator $A^{-1}$ is bounded.
Indeed, the value domain of the operator $A$ in $l_2$ is not closed (\cite
{1.27}, p. 34).

A decade later, S. G. Mikhlin, in fact, gave up the investigations related
to the issue of correctness \cite{1.28}: ''The author adheres to the
classical point of view, according to which the problem being solved by
mathematical methods should be considered as correctly formulated. Of
course, there are other opinions (p. 7)... Thus, we neglect the so-called
incorrigible errors related to the formulation of the above-mentioned
problem as a problem of natural science or of social studies (measurement
errors, insufficient accuracy of basic hypotheses, etc.)'' (p. 17).

M. M. Lavrentiev and L. Y. Saveliev characterized investigations of the
issue of the solvability of Eq. (\ref{1.6}) on the basis of considerations
of the type of \cite{1.26} as trivial, because it is difficult to imagine
that for experimentally determined $f\left( x\right) $ the corresponding
error may prove to be small in the norm of the space $l_2$ (\cite{1.29}, p.
217). At the same time, it is pointed out that, generally speaking, for any
operator equation, one can choose pairs of spaces such that the problem of
its solution will be correct.

G. M. Vainikko and A. Y. Veretennikov draw attention to the complexity of
the description of such spaces. And what is more, even the Volterra integral
equation of the first kind 
\[
\int\limits_0^xk\left( x,\xi \right) \psi \left( \xi \right) d\xi =f\left(
x\right) ,\quad x\in \left[ 0,1\right] , 
\]
which admits the regularization 
\[
\psi \left( x\right) +\int\limits_0^x\partial _xk\left( x,\xi \right) \psi
\left( \xi \right) d\xi =f^{\prime }\left( x\right) ,\quad x\in \left[ 0,1%
\right] 
\]
and is elementarily solvable by quadratures, for reasons of the norm for $%
\psi \left( x\right) $, as a rule, has to be considered as an incorrectly
formulated problem (\cite{1.30}, p. 6).

As regards the pair of spaces that realize the conditions of the correct
formulation, an original remark of K. I. Babenko is of interest (\cite{1.31}%
, p. 304): ''Hadamard's well-known example (\ref{1.3}), (\ref{1.4}) that
yields the solution of Cauchy's problem of the type (\ref{1.5}) by no means
tells of the absence of continuous dependence on the initial data, as it is
usually interpreted. It rather tells of the fact that small changes of the
initial data may result in leaving the totality of the initial data for
which the solution of Cauchy's problem exists.'' By the way, R. Richtmyer
demonstrated the correctness of the procedure of a numerical realization of
a rather complicated problem of the above-mentioned type with the
representation of sought functions by two-dimensional power series and with
the use of special methods of suppression of errors of arithmetical
operations (\cite{1.32}, section 17.B).

In the context of the present consideration, two theorems, given by V. A.
Trenogin, are of interest (\cite{1.33}, p. 225):

Let $E_1$ and $E_2$ be infinite dimensional normalized spaces, with $E_2$
being complete. If $A$ is a compact linear operator from $E_1$ to $E_2$,
different from finite dimensional, its value domain $R\left( A\right) $ is
not a closed manifold in $E_2$.

Let $A$ be a compact operator from an infinite dimensional normalized space $%
E_1$ to a normalized space $E_2$, with the inverse operator $A^{-1}$
existing on $R\left( A\right) $. Then $A^{-1}$ is bounded on $R\left(
A\right) $.

\section{The premises of the realization of the conditions of correctness}

Let us assume that $f=f_{*}\left( x\right) $ is the result of exact
integration of the function $\psi \left( x\right) \in L_2\left( 0,1\right) $
and of the symmetric kernel $k\left( x,\xi \right) $ by means of the formula
(\ref{1.6}). In this case, $f_{*}\in l_2$. Moreover, the corresponding set,
which is important, is not compact. This leads to the boundedness of the
operator $A^{-1}$ from $l_2$ to $L_2\left( 0,1\right) $ defining the
solution of the equation 
\begin{equation}  \label{1.11}
\left( A\psi \right) \left( x\right) =f_{*}\left( x\right) ,\quad x\in \left[
0,1\right] ,
\end{equation}
as mentioned above. There is no need to explain that such a situation would
be, in principle, rather favorable, in view of the fact that Hadamard's
third condition of correctness, directly following from the existence and
uniqueness of the solution, is fulfilled automatically.

The reason is the stability, by definition, of the inverse procedure of the
restoration of $\psi \left( x\right) $ with respect to small, in a sense,
perturbations of the data of the problem. Correspondingly, the solution to
Eq. (\ref{1.11}) can be obtained by the use of the formula (\ref{1.10}), and
the error of the evaluation of the coefficients $a_n$, following (\ref{1.7})
with $f=f_{*}\left( x\right) $, and of the values of $\lambda _n$ will
adequately influence $\psi \left( x\right) $ in $L_2\left( 0,1\right) $.

However, in reality, the coefficients $a_n$ are determined with an error
equivalent to $\left( \delta f\right) \left( x\right) $, which is inherent
to the definition of the free term in (\ref{1.6}). That is, in reality, the
right-hand side of Eq. (\ref{1.11}) has the form 
\begin{equation}  \label{1.12}
\tilde f\left( x\right) =f_{*}\left( x\right) +\left( \delta f\right) \left(
x\right) .
\end{equation}

As a matter of fact, a necessary condition of the correctness of Eq. (\ref
{1.11}), namely $\tilde f\in l_2$, proves to be illusive, because it is
impossible to obtain any quantitative information about $\left( \delta
f\right) \left( x\right) $.

Note that specific''diffusion'' of the space $l_2$ is caused by the
representation of the normalizing functional (\ref{1.7}) as an infinite
series, whose terms are determined on the basis of spectral characteristics
of the kernel $k\left( x,\xi \right) $. In this sense, such a space as $%
L_2\left( 0,1\right) $ is much more explicit for the function $f\left(
x\right) $. However, the use of it incurs rather negative consequences.

Indeed, in this case $R\left( A\right) $ does not belong to the closed space 
$L_2\left( 0,1\right) $, the operators $A$ and $A^{-1}$ become,
respectively, compact and unbounded. As a consequence, an investigation and
a numerical realization of the original equation (\ref{1.6}), in fact, turn
out to be beyond the sphere of the application of Banach's fundamental
theorem on the inverse operator. Isn't it a too high price to pay for
seemingly ephemeral clarity in the formulation of the problem?

We draw attention to a known point of view that a choice of appropriate
spaces for the solutions to problems of mathematical physics should be done
on the basis of practical applications, which can hardly be disputed. As the
same time, a wide-spread opinion that, for example, a sociologist should
formulate a problem to be solved by mathematical methods with a
specification of appropriate spaces for its data. This, as a rule, admits
variety, which is a prerequisite for an increase in the efficiency of
procedures of numerical realization.

Are there any prospects to overcome the above-mentioned complexity in mating
the free term of the Fredholm integral equation of the first kind (\ref{1.6}%
) with the adequate space $l_2$? In this regard, let us turn to Eq. (\ref
{1.11}) that by virtue of (\ref{1.12}) takes the form 
\begin{equation}  \label{1.13}
\left( A\psi \right) \left( x\right) =\tilde f\left( x\right) -\left( \delta
f\right) \left( x\right) ,\quad x\in \left[ 0,1\right] .
\end{equation}

It seems that there is a chance of a reduction of the given function $\tilde
f\left( x\right) $ to $f_{*}\in l_2$ by means of adaptive simulations of the
error $\left( \delta f\right) \left( x\right) $. Indeed, $\delta f$ and,
essentially, the smoothing of information by the procedure of integration
can be naturally reflected by the difference between the explicit form of
the sought function $\psi \left( x\right) $ and an integral over this
function whose kernel would not impose any additional restrictions on the
formulation of the problem.\footnote{%
A realization of this approach is a key aspect of the constructive part of
the present consideration (see section 5).} As a development of analogous
considerations, there appears the condition 
\begin{equation}  \label{1.14}
\left\| \delta f\right\| _{L_2\left( 0,1\right) }=0.
\end{equation}
Thus, instead of traditional restoration of the function $\psi \left(
x\right) $ directly from the solution of the Fredholm equation of the first
kind, we suggest to employ a nonessential, at the first sight, perturbation
(zero in $L_2$), which is adequate to the problem (\ref{1.13}), (\ref{1.14}%
). As will be shown below, as a result of comparatively simple
transformations, the determination of $\psi \left( x\right) $ will be
reduced to the solution of the Fredholm integral equation of the second
kind, which is rather preferable from the point of view of the efficiency of
procedures of a numerical realization.

In conclusion, we want to point out that known methods of the solution of
equations of the type (\ref{1.6}) are widely used when there is a cardinal
mismatch between $R\left( A\right) $ in $L_2\left( 0,1\right) $, or in some
other space, and the functions $\tilde f\left( x\right) $ that are
determined by means of measurements. From this point of view, the
fulfillment of the condition (\ref{1.14}) can hardly be interpreted in a
direct way. Nevertheless, the above-indicated approach still holds by
interpreting, figuratively, an approximation of the considered problem to
the correctly formulated one. Thus, the convergence of the series (\ref{1.7}%
) must be implied with regard to a corresponding variation of the kernel $%
k\left( x,\xi \right) $. As far as algorithms of the solution of resulting
Fredholm integral equations of the second kind are concerned, their
stability is practically not affected by the error in the determination of $%
\tilde f\left( x\right) \in L_2\left( 0,1\right) $ (see, e.g., \cite{1.19}).

\chapter{The existing approaches to the solution of incorrect problems}

\section{A. N. Tikhonov's methodology}

The consideration of this subsection is based on the material of the
monograph by A. N. Tikhonov and V. Y. Arsenin \cite{2.1} that is, literally,
pieced by the concept of the adequacy of incorrect formulations and, in
particular, of integral equations of the first kind to problems of
mathematical physics. As an illustration, we show that the solution to the
Fredholm integral equation of the first kind 
\begin{equation}  \label{2.1}
\left( A\psi \right) \left( x\right) \equiv \int\limits_a^bk\left( x,\xi
\right) \psi \left( \xi \right) d\xi =f\left( x\right) ,\quad x\in \left[ a,b%
\right] ,
\end{equation}
with $k\left( x,\xi \right) $ and $\partial _xk\left( x,\xi \right) $ being
continuous with respect to $x$, can undergo arbitrarily considerable changes
both in the metric $C$ and $L_2$ for small in $L_2\left( a,b\right) $
variations of the right-hand side in the form 
\[
N\int\limits_a^bk\left( x,\xi \right) \sin \left( \omega \xi \right) d\xi . 
\]

The situation with the perturbation of the kernel $k\left( x,\xi \right) $
is, in fact, analogous. In this regard, the authors pose the question: What
should be understood as the solution of Eq. (\ref{2.1}) when $k$ and $f$ are
known approximately? In their opinion, a problem of this type should be
considered as ''underspecified'', and, correspondingly, a choice of possible
solutions should be made taking into account ''usually'' available
additional qualitative or quantitative information about the function $\psi
\left( x\right) $. In this regard, we draw attention to N. G.
Preobrazhenskii's considerations concerning a system of linear algebraic
equations, obtained by the discretization of (\ref{2.1}) (\cite{2.2}, p.
130):

''An analysis shows that choosing sufficiently high order of an
approximation, we transform [the above-mentioned problem] into an
arbitrarily ill-defined one... Under these conditions, it is necessary to
add to the algorithm some a priori nontrivial information, only by the use
of which we can expect to filter out veiling false variants and single out
the solution, closest to the sought one. any purely mathematical tricks that
do not employ additional a priori data are equivalent to an attempt to
construct an informational perpetuum mobile producing information from
nothing.''

The so-called method of the selection of the solution to incorrectly
formulated problems is based on a priori quantitative information. It is
shown that if a compactum $M$ of a metric space $E_1$ is mapped in a
one-to-one and continuous manner onto a set $F$ of a metric space $E_2$, the
inverse map $F$ onto $M$ is also continuous. Correspondingly, an assumption
that the solution, in particular, to Eq. (\ref{2.1}) belongs to the
compactum $M$ allows us to consider the operator $A^{-1}$ to be continuous
on the set $F=AM$. A practical realization is reduced to an approximation of 
$M$ by a series with parameters that change within certain limits (for $M$
to represent a closed set of a finite dimensional space) and should be
determined from the condition of the minimum of the error of closure of (\ref
{2.1}). Note the absence of any more or less general recommendation with
respect to the choice of $M$.

In light of the above, M. M. Lavrentiev has formulated the notion of
correctness according to Tikhonov for an equation of the type (\ref{2.1}),
with the functions $\psi $ and $f$ belonging to Banach spaces $B_1$ and $B_2$%
, respectively \cite{2.3}:

1) It is a priori known that the solution $\psi _{*}$ to the considered
equation exists and belongs to a set $M$ of the space $B_1$.

2) The solution $\psi _{*}$ is unique on the set $M$.

3) The operator $A^{-1}$ is continuous on the set $AM$ of the space $B_2$.

If $M$ is a compactum (this case is called ''usual'') the last condition
becomes a consequence of the first two conditions.

Those problems in which the operator $A^{-1}$ is unbounded on the set $AE_1$
and the set of possible solutions $E_1$ is not a compactum are called
substantially incorrect. For such problems, A. N. Tikhonov has put forward
an idea of a regularizing operator $G$, in a sense close to $A^{-1}$, whose
value domain for the map from $E_2$ into $E_1$ admits matching to the
right-hand side of (\ref{2.1}), known approximately. Moreover, $G$ must
contain a regularization parameter$\alpha $ that depends on the accuracy of
the initial information.

The operator $G\left( f,\alpha \right) $ is called a regularizing operator
for Eq. (\ref{2.1}) if it possesses the following properties:

1) It is defined for any $\alpha >0$ and $f\in E_2$.

2) For $A\psi _{*}=f_{*}$, where $\psi _{*}$ and $f_{*}$ are corresponding
exact expressions, there exists such $\alpha \left( \delta \right) $ that
for any $0<\epsilon \leq \rho _{E_1}\left( \psi _{_{*}},\psi _\alpha \right) 
$ there is $\delta \left( \epsilon \right) \geq \rho _{E_2}\left( \psi
_{_{*}},\psi _\alpha \right) $. Here, $\psi _\alpha =G\left( f,\alpha
\right) $.

It is implied that there is a possibility of a choice of $\alpha \left(
\delta \right) $ such that for $\delta \rightarrow 0$ the regularized
solution $\psi _\alpha \rightarrow \psi _{*}$, i.e., $\epsilon \rightarrow 0$%
. At the same time, it is pointed out that the construction of the
dependence $\alpha \left( \delta \right) $, for which the operator $G\left(
f,\alpha \left( \delta \right) \right) $ is a regularizing one, is
algorithmically complicated for classes of practically important problems.
There are a lot of publications of A. N. Tikhonov's followers devoted to the
resolution of this difficulties, which will be discussed below.

In \cite{2.1}, the construction of $G\left( f,\alpha \right) $ is carried
out by the use of techniques of calculus of variations that reduce the
evaluation of $\psi \left( x\right) $ to the minimization of the functional 
\begin{equation}  \label{2.2}
\Phi ^\alpha \left[ f,\psi \right] =\rho _{E_1}^2\left( A\psi ,f\right)
^2+\alpha \Omega \left[ \psi \right] .
\end{equation}

For Eq. (\ref{2.2}), its stabilizing component is recommended to be taken in
the form 
\begin{equation}  \label{2.3}
\Omega \left[ \psi \right] =\int\limits_a^b\left\{ p_0\left( x\right) \psi
^2\left( x\right) +p_1\left( x\right) \left[ \psi ^{\prime }\left( x\right) %
\right] ^2\right\} dx,
\end{equation}
where $p_0$, $p_1\geq 0$ are given functions.

In the case of a symmetric kernel $k\left( x,\xi \right) $, the procedure of
the minimization is equivalent to the solution of the integrodifferential
equation 
\begin{equation}  \label{2.4}
\alpha \left\{ p_0\left( x\right) \psi \left( x\right) -\left[ p_1\left(
x\right) \psi ^{\prime }\left( x\right) \right] ^{\prime }\right\} +\left(
A\psi \right) \left( x\right) =f\left( x\right) ,\quad x\in \left[ a,b\right]
,
\end{equation}
under the conditions 
\begin{equation}  \label{2.5}
\left. p_1\left( x\right) \psi ^{\prime }\left( x\right) \upsilon \left(
x\right) \right| _a^b=0.
\end{equation}

Here, $\upsilon \left( x\right) $ is an arbitrary variation of $\psi \left(
x\right) $ in the class of admissible functions.

In the opinion of the authors of \cite{2.4}, an overwhelming majority of
inverse problems are incorrectly formulated, and attempts to solve them, in
view of their great practical importance, were being undertaken for a long
period. ''But only as a result... of the appearance of fundamental
publications of academician A. N. Tikhonov, the modern theory of the
solution of inverse problems, based on the notion of a regularizing
algorithm, was constructed'' (p. 7). In what follows, the authors
demonstrate the efficiency of the procedure of a numerical realization of
the Fredholm integral equations of the first kind, related to the
interpretation of astrophysical observations, by means of the selection of
the compactum of possible solutions in the class of monotonically bounded
functions.

As is pointed out by O. A. Liskovets \cite{2.5}, ''...the correctness
according to Tikhonov is achieved at the expense of the reduction of the
admissible manifold of solutions to the class of correctness'' (p. 13). The
following quotation from the above-mentioned monograph is also of
considerable interest: ''In contrast to a previously prevailing opinion that
all the problems describing physical reality are correct, according to the
modern point of view any realistic problem can be regularized, i.e., it has
at least one regularizer'' (p. 14).

Let us refer to V. A. Morozov's conclusion (\cite{2.6}, p. 9): ''A. N.
Tikhonov's method of regularization turned out to be simple in practice,
because it did not require actual knowledge of the compactum $M$ that
contained the sought solution to Eq. (\ref{2.1})... The main difficulty of
the application of this method consists in the formulation of algorithmic
principles of the selection of the parameter of regularization $\alpha $''.
According to his own monograph (\cite{2.7}, p. 4), ''the importance of A. N.
Tikhonov's paper \cite{2.8} can hardly be overestimated. It served as
impetus for a number of publications by other investigators in different
fields of mathematical analysis and natural science: spectroscopy, electron
microscopy, identification and automatic regulation, gravimetry, optics,
nuclear physics, plasma physics, meteorology, automation of scientific
research and some other spheres of science and engineering''.

V. V. Voevodin's considerations \cite{2.9} are rather typical: ''The success
of the application of the regularization method to the solutions of unstable
systems of algebraic equations is explained to a large extent by the fact
that A. N. Tikhonov and his followers did not restrict themselves to an
investigation of separate fragments of this complicated problem but
considered the whole complex related issues. This, in the first place,
concerns a clear formulation of the problem itself, the construction of a
stable with respect to perturbation of the input data algorithm of its
solution, the development of an efficient numerical method, estimates of a
deviation of the actually evaluated object from the sought one taking into
account a perturbation of the input data and errors of rounding''.

A quotation from the preface to the collected volume by A. A. Samarsky and
A. G. Sveshnikov \cite{2.9} reads: ''A clarification of Andrey Nikolaevich
Tikhonov of the role of incorrect problems in classic mathematics and its
applications (inverse problems) is of fundamental importance for the who;e
modern mathematics. He proposed a principally new approach to this class of
problems and developed methods of the construction of their stable solutions
based on the principle of regularization''.

\section{A brief review of the development of the outlined concepts}

The results of investigations devoted to the determination of the
regularization parameter $\alpha $ are summarized in \cite{2.10}. Based on
the assumption that errors in the determination of the free term $f\left(
x\right) $ and the kernel $k\left( x,\xi \right) $ of Eq. (\ref{2.1}) are
known, one uses different methods of the minimization of the error of
closure of the type 
\[
\left\| \tilde A\psi _\alpha -f\right\| _{E_2}=\mu \left\| \delta f\right\|
_{E_2},\quad \mu \in \left( 0,1\right) . 
\]
The evaluation of the parameter $\alpha $ as a root of the corresponding
equation does not pose any problem. However, a choice of $\mu $ is, in fact,
related to considerable uncertainty. The main obstacle is that a reliable
estimate of the error caused by incompatibility of the concretely considered
equation $\tilde A\psi =f$ is rather questionable.

Considerable efforts were undertaken to reduce the volume of information
necessary for the evaluation of the parameter $\alpha $. A noticeable step
in this direction was made by A. N. Tikhonov and V. B. Glasko who suggested
a criterion of the minimization of the functional $\left\| \alpha d\psi
_\alpha /d\alpha \right\| $ with respect to $\alpha >0$ \cite{2.11} (see
also \cite{2.1}, section 2.7). However, its theoretical justification proved
to be possible only for rather narrow classes of problems. A number of
methods of the determination of $\alpha $ is related to the use of solutions
to Eq. (\ref{2.1}) for a special form of the expressions $f\left( x\right) $.

In \cite{2.10}, the status of the studies of estimates of the accuracy of
methods of the solution of the integral equation (\ref{2.1}) is also
discussed. If $\psi \left( x\right) $ belongs to a compactum, any serious
complications, as a rule, do not arise, and the main interest is focused on
the algorithm of regularization. If $p_1\equiv 0$ in (\ref{2.3}) and the
parameter $\alpha $ is finite, Eq. (\ref{2.4}) becomes a Fredholm integral
equation of the second kind, to which, under the assumption that the error
in the determination of $k\left( x,\xi \right) $ and $f\left( x\right) $ is
known, the whole general theory of approximate methods of L. V. Kantorovich
applies (\cite{2.12}, section 14.1).

At the same time, as shown by V. A.Vinokurov \cite{2.13}, when a priori
information about the solution to Eq. (\ref{2.1}) is missing, the estimate
of the error of the evaluation of $\psi \left( x\right) $ by means of
regularization is impossible in principle. Justified is only a formulation
of the question of the convergence of the procedure of computation or of a
possibility of the regularization of the corresponding problem.

In this regard, we note the arguments of A. B. Bakushinskii and A. V.
Goncharskii (\cite{2.14}, p. 13): ''Unfortunately, in the general case, it
is impossible to estimate the measure of closeness of $G\left( f,\alpha
\right) $ to $A^{-1}\left( f_{*}\right) $ without additional information
about the solution to Eq. (\ref{2.1}). This is a characteristic feature of
incorrect problems. In the general case, a regularization algorithm ensures
only asymptotic convergence of an approximate solution to the exact one for $%
\delta \rightarrow 0$''.

The name of M. M. Lavrentiev is associated, in fact, with a particular case
of a practical realization of A. N. Tikhonov's method consisting in the
reduction of the problem (\ref{2.4}), (\ref{2.5}) to the solution of the
Fredholm integral equation of the second kind 
\begin{equation}  \label{2.6}
\alpha \psi \left( x\right) +\int\limits_a^bk\left( x,\xi \right) \psi
\left( \xi \right) d\xi =f\left( x\right) ,\quad x\in \left[ a,b\right] ,
\end{equation}
where $\alpha >0$ is a small parameter.

It is shown that $\left\| \psi _\alpha -\psi _{*}\right\| \rightarrow 0$ for 
$\delta \rightarrow 0$, $\gamma \rightarrow 0$ and $\left( \delta +\gamma
\right) /\alpha \left( \delta ,\gamma \right) \rightarrow 0$. Here, $\gamma $
is an error in the determination of the kernel $k\left( x,\xi \right) $,
analogous to $\delta $ (see section 3.1).

V. K. Ivanov's method \cite{2.15} allows one to find the so-called
quasisolution minimizing the error of closure of (\ref{2.1}) for a class of
functions $\psi \left( x\right) \in M_R$, where $M_R\in E_1$ is a compactum
whose specification, generally speaking, are analogous to that employed in
Picard's theorem (see section 2.2). The quasisolution to (\ref{2.1}) on such
a compactum has the form 
\begin{equation}  \label{2.7}
\psi \left( x\right) =\sum_{n=1}^\infty c_n\left( \lambda +\lambda _n\right)
\bar \psi _n\left( x\right) ,\quad x\in \left[ a,b\right] .
\end{equation}
Here 
\[
c_n=\int\limits_a^bf\left( x\right) \bar \psi _n\left( x\right) dx; 
\]
$\lambda _n$ and $\bar \psi _n\left( x\right) $ are, respectively,
characteristic numbers and eigenfunctions of the kernel $k\left( x,\xi
\right) $; the parameter $\lambda =0$ and represents a positive root of the
equation 
\begin{equation}  \label{2.8}
\sum_{n=1}^\infty \left( \frac{c_n\lambda \lambda _n}{\lambda +\lambda _n}%
\right) ^2=R^2
\end{equation}
under the conditions, respectively, 
\begin{equation}  \label{2.9}
\sum_{n=1}^\infty c_n^2\lambda _n^2\leq R^2;\quad \sum_{n=1}^\infty
c_n^2\lambda _n^2>R^2.
\end{equation}

Special methods of regularization are developed for the situation when there
is considerable volume of information of statistical character (spectral
densities, mathematical expectations, etc.) about the solution to an
equation of the type (\ref{2.1}). Thus, V. N. Vapnik \cite{2.16} rather
constructively employed the specifics of problems concerned with recognition
of images, related to nonuniqueness and, as a result, to extreme behavior of
the sought functions. We point out a definition in the above-mentioned
monograph (p. 8) that, apparently, was implied by many authors but did not
receive such a clear formulation.

''The problem of the restoration of dependencies from empirical data was
and, probably, will always be central in applied analysis. This problem is
nothing but mathematical interpretation of one of the main problems of
natural science: How to find the existent regularity from random facts.''

\section{V. M. Fridman's approach}

Let $k\left( x,\xi \right) $ be symmetric, positive definite kernel and Eq. (%
\ref{2.1}) be solvable. Then, as shown by V. M. Fridman \cite{2.17}, a
sequence of functions determined by iteration 
\begin{equation}  \label{2.10}
\psi _{n+1}\left( x\right) =\psi _n\left( x\right) +\lambda \left[ f\left(
x\right) +\int\limits_a^bk\left( x,\xi \right) \psi _n\left( \xi \right)
d\xi \right] ,\quad n=0,1,\ldots ,
\end{equation}
converges in $L_2\left( a,b\right) $ to the solution of Eq. (\ref{2.1}) for
an arbitrary choice of the initial approximation $\psi _0\left( x\right) \in
L_2\left( a,b\right) $ and $0<\lambda <2\lambda _1$, where $\lambda _1$ is
the smallest characteristic number of the kernel $k\left( x,\xi \right) $.

M. A. Krasnoselskii \cite{2.18} extended this result to an arbitrary
solvable equation of the type (\ref{2.1}) with a linear bounded operator $A$
in a Hilbert space $H$. A theorem on the convergence of successive
approximations 
\begin{equation}  \label{2.11}
\psi _{n+1}=\left( I-\nu A_1\right) \psi _n+\nu f_1
\end{equation}
to the solution is proved. Here, $A_1=A^{*}A$; $f_1\left( x\right) =\left(
A^{*}f\right) \left( x\right) $; $I$ is the identity operator; $A^{*}$ is
the conjugate operator to $A$; $0<\nu <2/\left\| A_1\right\| $; $\psi
_0\left( x\right) \in H$.

Note that in the case of the integral operator (\ref{2.1}) 
\[
A_{1\bullet }=\int\limits_a^bk_1\left( x,\xi \right) _{\bullet }d\xi , 
\]
where 
\[
k_1\left( x,\xi \right) =\int\limits_a^bk\left( \zeta ,x\right) k\left(
\zeta ,\xi \right) d\zeta . 
\]

A number of procedures are known that improve convergence of iterations
according to Fridman (see \cite{2.10}). For example, under the conditions
that are specified with respect to the procedure (\ref{2.10}), 
\begin{equation}  \label{2.12}
\psi _{n+1}\left( x\right) =\frac 1{m+1}\sum_{n=0}^m\varphi _n\left(
x\right) ,
\end{equation}
where $\varphi _0\left( x\right) \in L_2\left( a,b\right) $; 
\[
\varphi _n\left( x\right) =\varphi _{n-1}\left( x\right) +f\left( x\right)
-\int\limits_a^bk\left( x,\xi \right) \varphi _{n-1}\left( \xi \right) d\xi
,\quad n=1,2,\ldots . 
\]

G. M. Vainikko and A. Y. Veretennikov \cite{2.19} studied an iteration
algorithm of an implicit type: 
\begin{equation}  \label{2.13}
\alpha \psi _{n+1}\left( x\right) +\int\limits_a^bk\left( x,\xi \right) \psi
_{n+1}\left( \xi \right) d\xi =\alpha \psi _n\left( x\right) +f\left(
x\right) ,\quad n=0,1,\ldots ,
\end{equation}
where $\psi _0\left( x\right) \in L_2\left( a,b\right) $; the parameter $%
\alpha $$>0$.

Note that in contrast to the regularization of the type (\ref{2.6}), based
on the smallness of $\alpha $, the considered approach is characterized by
multiple iteration with, on the contrary, sufficiently large value of this
parameter. Moreover, one of the merits of the procedures (\ref{2.10})-(\ref
{2.13}) is a possibility of a constructive application of an a posteriori
estimate of the error to accomplish the iteration.

In the simplest case, one finds the number $n$ for which for the first time 
\[
\left\| \psi _{n+1}-\psi _n\right\| _{L_2\left( a,b\right) }\leq c_1\delta
+c_2\gamma , 
\]
where $\delta $ and $\gamma $ are errors in the determination of $f\left(
x\right) $ and $k\left( x,\xi \right) $, respectively; $c_1$, $c_2$ are
constants meeting a number of requirements to ensure the stability of the
procedures of computation.

The influence of errors, small in a probabilistic sense, on the convergence
of successive approximations is also investigated \cite{2.19}.

The authors of \cite{2.20} gave arguments for usefulness of the combination
of the regularization of the equation of the type (\ref{2.1}), whose
parameter is the number of iterations, with algorithms of the saddle-point
type. This approach has its origin in the publication by V. M. Fridman \cite
{2.21} and is realized, in particular, according to the scheme 
\begin{equation}  \label{2.14}
\psi _{n+1}=\psi _n-\beta _nA^{*}\left( A\psi _n-f\right) ,
\end{equation}
where 
\[
\beta _n=\frac{\left\| A^{*}\left( A\psi _n-f\right) \right\| ^2}{\left\|
AA^{*}\left( A\psi _n-f\right) \right\| ^2}, 
\]
which is adequate to the choice of the step of the descent from the
condition of the minimum of the error of closure 
\[
\Delta _{n+1}=\left\| A\psi _{n+1}-f\right\| ;\quad \left\| _{\bullet
}\right\| =\left\| _{\bullet }\right\| _{L_2\left( a,b\right) }. 
\]

\section{Inverse problems for differential equations of mathematical physics}

Modern viewpoints on this subject are characterized in the monograph by O.
M. Alifanov, E. A. Artyukhin and S. V. Rumyantsev \cite{2.20}. In the
procedure of mathematical formulation of the problems, structural and
parametric identification is emphasized, which implies, respectively, a
qualitative description of the considered processes by means of differential
operators and allotting quantitative information to the model.
Interpretation in terms of causality is also given. The cause includes
boundary and initial conditions with their parameters, coefficients of the
differential equations and also the domain of the problem. The effect
reflects the status of the investigated object and represents, mostly,
fields of physical quantities of different types.

The restoration of the cause from the information about physical fields is
considered as an inverse problem. A key consideration is as follows (p. 19):
''A violation of a natural causal relation that takes place in the
formulation of the inverse problem can lead to its mathematical
incorrectness, such as, in most cases, instability of the solution.
Therefore, inverse problems constitute a typical example of incorrectly
formulated problems''.

According to the sought function, the following types of inverse problems of
the identification of physical processes for partial differential equations
are singled out:

1) Retrospective problems: the determination of the prehistory of a certain
state of the problem.

2) Boundary problems: the restoration of boundary conditions or of the
parameters contained therein.

3) Coefficient problems: the restoration of the coefficients of the
equations.

4) Geometrical problems: the determination of geometrical characteristics of
the contour of the domain or of the coordinates of points inside.

A principal difference between inverse problems of identification and those
of regulation is pointed out, concerning the width of classes of possible
solutions. Whereas in the former case their increase leads to complications
in the numerical realization, in the latter case, on the contrary, this is a
favorable factor. By the way, the algorithmic means \cite{2.20} are almost
completely based on the methods of the solution of integral equations of the
first kind, to which the considered problems of heat exchange are reduced.

In the formulation of inverse problems of mathematical physics, the proof of
corresponding theorems of existence and uniqueness is of crucial importance.
In this regard, a general approach, outlined schematically by A. L. Buchgeim 
\cite{2.22}, is of interest. Thus, the following formulation is considered: 
\begin{equation}  \label{2.15}
Pu=f;\quad Qf=g,
\end{equation}
where $P$ is an operator of the direct problem; $Q$ is an ''information''
operator describing the law of the change of the right-hand side; $g$ is
given, whereas $u$ and $f$ are the sought elements of corresponding
functional spaces.

The application of the operator $Q$ to the first equation (\ref{2.15})
yields $QPu=g$, which is equivalent to 
\[
PQu=\left[ P,Q\right] u+g, 
\]
where $\left[ P,Q\right] =PQ-QP$ is the commutator of the operators $P$ and $%
Q$. The meaning of the commutation lies in the fact that, as a rule, there
is no information, except for (\ref{2.15}), about the function $f$.
Therefore, it is easier to study the operator on the solution of the direct
problem $u$ that satisfies some manifold of boundary conditions. It is
important that in typical applications the operator $Q$ does not ''spoil''
the part of boundary conditions that reflects the domain of the operator $P$%
. As a result, one gets a specific factorization of the inverse problem (\ref
{2.15}) as a product of two direct problems, induced by the operators $P$
and $Q$ under the condition that the commutator is, in a sense,
''subordinate'' to them.

In the trivial case $\left[ P,Q\right] =0$, the initial problem decomposes
into two simple ones: $Pv=g$; $Qu=v$. For the description of properties of
the employed operators, a priori estimates are used.

Of interest is also a quotation from the introduction to the monograph by R.
Lattes and J.-L. Lions \cite{2.23}: ''In this book, we suggest a method of
quasiinversion, intended for the numerical solution of some classes of
incorrect, according to Hadamard, boundary value problems. Practical and
theoretical importance of such problems is being more and more realized by
investigators''. Moreover: The main idea of the method of quasiinversion
(universal in numerical analysis!) consists in an appropriate change of
operators entering the problem. This change is done by the introduction of
additional differential terms that are

i) sufficiently ''small'' (they can be set equal to zero);

ii) ''degenerate on the boundary'' (to prevent, for example, the appearance
of complicated boundary conditions and of such conditions that may contain
unknown, sought variables)''.

In particular, the incorrect formulation of the problem of thermal
conductivity 
\begin{equation}  \label{2.16}
\partial _tu-\partial _x^2u=0,
\end{equation}
\[
u\left( 0,t\right) =u\left( 1,t\right) =0;\quad u\left( x,T\right) =\kappa
\left( x\right) , 
\]
where $\kappa \left( x\right) $ is an unknown function, can be replaced by
the following, with a small parameter $\epsilon $: 
\begin{equation}  \label{2.17}
\partial _tu-\partial _x^2u-\epsilon \partial _x^4u=0;
\end{equation}
\[
u=\partial _x^2u=0,\quad x=0;\quad x=1;\quad u\left( x,T\right) =\kappa
\left( x\right) . 
\]

The authors point out (p. 36): ''In a numerical realization, it is natural
to choose $\epsilon $ as the smallest possible one. However, in problems of
the considered type, one should expect numerical instability for $\epsilon
\rightarrow 0$. Therefore one can expect at most that for any problem there
exists a certain optimal value of $\epsilon $ equal to $\epsilon _0$''. The
absence of convergence ''in a usual sense'' of the solution of the problem (%
\ref{2.17}) to the exact one for $\epsilon \rightarrow 0$ was pointed out by
A. N. Tikhonov and V. Y. Arsenin (\cite{2.1}, p. 52).

\section{Alternative viewpoints and developments}

In Y. I. Liubich's opinion, any more or less general theory of integral
equations of the first kind is absent, and only in some cases it is possible
to use special methods. An example is given by known Abel's equation (\cite
{2.24}, p. 83).

K. I. Babenko's remark (\cite{2.25}, p. 310) is rather typical: ''Although
from the point of view of the loss of information algorithms are not
estimated, it seems to us that this is an important characteristic and it
should be taken into account''. In what follows, the lack of optimality of
the traditional approach to a numerical realization of incorrectly
formulated problems is concretely demonstrated.

A profound analysis of methodological aspects of this sphere is given by R.
P. Fedorenko (\cite{2.26}, sections 40, 41). In particular, he failed to
establish the value of the regularization parameter $\alpha $ by minimizing
the functional (\ref{2.2}), because for small values the sought function
began to oscillate, whereas with it increase the value of $\Phi ^\alpha $
considerably exceeded the admissible one. It is found that the reason is the
inadequacy of the theory \cite{2.1} to problems of control, characterized by
discontinuity of solutions.

In the context of investigations of the problem (\ref{2.16}), R. P.
Fedorenko brought up the following consideration: '' All the methods of the
solution of incorrect problems more or less consist in preventing the
appearance in the sought solution of higher harmonics with large or even
simply finite coefficients. But what is ''high frequency''? Beginning with
what number $n$ should we consider the function $\sin \left( n\pi x\right) $
redundant, only spoiling the solution? This, of cause, depends on $T$''. It
is implied that a hypothetically known solution of the corresponding direct
problem can be expanded into a Fourier series 
\[
u\left( x\right) =u\left( x,0\right) =\sum_{n=1}^\infty \sin \left( n\pi
x\right) . 
\]

It is shown that the use in \cite{2.23} of the value $T=0.1$ and the errors
in $L_2\left( 0,1\right) $ of the satisfaction of the last condition (\ref
{2.16}), with $\delta $ of order $10^{-3}$, imposes the restriction $n=2$.
In this context, the method of P. Lattes and G.-L. Lions came under
criticism. This authors, while solving the problem (\ref{2.17}) on a grid
with a step of $\Delta x=0.02$, obtained an absolutely unacceptable
component $u_0$, namely, $10^8\sin \left( 6\pi x\right) $. This occurred for 
$\delta $ at the level of $0.05$, under the conditions when $\left| \kappa
\left( x\right) \right| \leq 1\ldots $ .

Note also the remark \cite{2.26} that, aside the fact of the boundedness of
the regularizing operator $G$ (see section 3.1), its norm $\left\| G\right\| 
$ is an exceptionally important characteristic whose value directly
influences a relation between the accuracy of the given function $\kappa $
and the solution $u_0=G\kappa $. \footnote{%
By the way, in most specialized publication this issue is not accentuated.}

Indeed, let us consider Eq. (\ref{2.6}), written in the canonical form 
\begin{equation}  \label{2.18}
\psi \left( x\right) =-\frac 1\alpha \int\limits_a^bk\left( x,\xi \right)
\psi \left( \xi \right) d\xi +\frac 1\alpha f\left( x\right) ,\quad x\in 
\left[ a,b\right] .
\end{equation}

Let $a=0$, $b=1$, and the kernel $k\left( x,\xi \right) $ is determined by
the expression (\ref{1.8}). In this case, its characteristic numbers and
eigenfunctions are $\lambda _n=-\left( n\pi \right) ^2$, $\bar \psi _n\left(
x\right) =\sin \left( n\pi x\right) $, $n=1,2,\ldots $, respectively. Thus,
on the basis of general theory (see, e.g., \cite{2.27}), for $\alpha
^{-1}\neq \lambda _n$, the solution of Eq. (\ref{2.18}) is 
\[
\psi \left( x\right) =\frac 1\alpha f\left( x\right) -\frac 1\alpha
\sum_{n=1}^\infty \frac{c_n}{1+\alpha \left( n\pi \right) ^2}\sin \left(
n\pi x\right) , 
\]
where 
\[
c_n=\int\limits_0^1f\left( x\right) \sin \left( n\pi x\right) dx. 
\]

It is not difficult to notice that for small values of $\alpha $ the error
in the determination of the function $f\left( x\right) $ can considerably
distort $\psi \left( x\right) $.

In a constructive aspect, R. P. Fedorenko recommends to use traditional
formulations of inverse problems of differential or variational character
with an application of additional conditions that rationally restrict
classes of possible solutions. As the main factor to achieve the desired
efficiency, a comprehensive analysis of qualitative peculiarities of
solutions to the considered problems, involving elements of numerical
simulations, is suggested.

What are the values of the regularization parameter $\alpha $, typical of
computational practice? The authors of \cite{2.28} point out that for
problems of restoration of time-dependent density of thermal flux on the
surface from the results of temperature measurements at internal points of
the samples the corresponding range is rather representative: $%
10^{-7}-10^{-4}$. The editors of the above-mentioned book have a different
point of view: ''One can give a lot of examples of solutions to inverse
problems thermal conductivity, when the range of acceptable values of $%
\alpha $ turn out to be rather narrow'' (p. 141).

The main technique of a numerical realization \cite{2.28} is interpreted by
the authors as a complement to the method of least squares by a procedure
that smooths oscillations of the solutions in high order approximations. In
this regard, they point out a relationship between Tikhonov's regularization
and algorithms of singular expansions and ridge regression (or damping) that
are widely used for the suppression of the instability of the method of
least squares \cite{2.29}.

In a number of publications, one can see an orientation towards
regularization of Eq. (\ref{2.1}) without the distortion of the original
operator along the lines of (\ref{2.4}) or (\ref{2.6}). Thus, A. P. Petrov 
\cite{2.30} suggested a formulation of the problem with $f\left( x\right)
\in R\left( A\right) $ by means of the representation $f=A\psi +\omega $,
where $\omega $ is a random process reflecting errors of the data and of the
calculations. At the same time, the author failed to use his formally
achieved correctness to construct an efficient algorithm of a numerical
realization. The reason, in our opinion, is non-constructive structure of $%
\omega $ from the point of view of adaptive compensation of the error of
closure of the satisfaction of (\ref{2.1}).

A. V. Khovanskii \cite{2.31} put forward arguments for the regularization of
the algorithm of the solution of Eq. (\ref{2.1}), not the operator $A$
(which is the basis of the theory of \cite{2.1}). The following quotation is
of interest: ''Tikhonov's regularization contains in an inseparable form two
completely different notions, accuracy and stability, and there is a
transformation of one into another. Nevertheless, there exists for a long
time an idea of the predetermination of the operator \cite{2.32}, although
only in the context of conjugate gradients and in a multiplicative form''.

In fact, this implies the use of Eq. (\ref{2.6}) with $\alpha =\alpha _0$,
where $\alpha _0$ is the minimal value allowing us to obtain information
about the unknown function $\psi \left( x\right) $ of indirect character.
With its use, one can determine the regularization parameter $\alpha $
matching the error of the free term $f\left( x\right) $.

However, the method of conjugate gradients is, in fact, Fridman's iterations
of the type (\ref{2.14}). Note that nonlinearities contained therein
facilitate the smoothing of a well-known slow-down of the convergence of the
procedure (\ref{2.10}) with approaching the solution to Eq. (\ref{2.1}).
This effect was demonstrated by A. D. Myshkis \cite{2.33} with the help of
the representation of the components of (\ref{2.10}) by series in terms of
the eigenfunctions of the kernel $k\left( x,\xi \right) $. This leads to the
relations 
\[
c_{n+1,m}=\left( 1-\lambda /\lambda _m\right) c_{n,m}+\lambda f_m,\quad
m=1,2,\ldots , 
\]
where $c_{n,m}$ and $f_m$ are coefficients of the above-mentioned expansion
of $\psi _n\left( x\right) $ and $f\left( x\right) $, respectively.

When the number of the terms in the representation of the solution
increases, which seemingly had to improve the accuracy, the coefficient of
convergence $\left( 1-\lambda /\lambda _m\right) $ approaches unity and, as
a result of the accumulation of errors, the iterations become
counterproductive.

Note an effective method of the suppression of instability of the algorithm
of the solution of the Fredholm integral equation of the second kind 
\begin{equation}  \label{2.19}
\psi \left( x\right) =\lambda \int\limits_a^bk\left( x,\xi \right) \psi
\left( \xi \right) d\xi +f\left( x\right) ,\quad x\in \left[ a,b\right] ,
\end{equation}
''positioned on the spectrum'', i.e., in the case $\lambda =\lambda _n$,
developed by P. I. Perlin (\cite{2.34}, pp. 105-107).

This problem is incorrect both with respect to the uniqueness of the
solution and as a result of the degeneracy of the system of linear algebraic
equations obtained by discretization. Nevertheless, a perturbation of the
right-hand side of $f\left( x\right) $ by a zero (within the limits of the
accuracy of calculations) component 
\[
-\bar \psi _n^{*}\left( x\right) \int\limits_a^bf\left( x\right) \bar \psi
_n^{*}\left( x\right) dx, 
\]
where $\bar \psi _n^{*}\left( x\right) $ is a normalized eigenfunction of
the conjugate kernel $k\left( \xi ,x\right) $, allows one to improve
radically the situation.

The essence lies in the fact that, theoretically, the solution to Eq. (\ref
{2.19}) is expanded in a power series of $\lambda $. Provided that
computational procedures, matching this situation, are identical, one can
compensate for the errors.

\section{A comparison between the main concepts of A. N. Tikhonov and V. M.
Fridman}

A. N. Tikhonov's original suggestion (1943) admitting of the consideration
of incorrectly formulated problems by an a priori restriction on the class
of possible solutions is a kind of refraction of general methodology of
investigations of the issues of existence and uniqueness into the sphere of
numerical analysis.\footnote{%
There is a translation of the first edition of [1]: A. N. Tikhonov and V. Y.
Arsenin, Solutions of Ill-Posed Problems (Winston, Washington, 1977). See
also V. A. Morozov, Solutions of Incorrectly Posed Problems (Springer, New
York, 1984). An English translation of the article of V. M. Fridman [17] is
given in Appendix.} Note that A. N. Tikhonov's proof of the well-known
theorem on the uniqueness of the solution of the inverse problem of thermal
conductivity in an infinite $n$-dimensional domain under an additional
condition of the type $\left| \partial _x^nu\right| \leq M$ dates back to
1935. A clear illustration of these considerations is provided by the
algorithm of the search for a quasisolution (\ref{2.7})-(\ref{2.9}) that
artificially subjects the data of Eq. (\ref{2.1}) to conditions of the type
of those that figure in Picard's theorem.

Behind A. N. Tikhonov's method of regularization (1963), there is a global
idea of a limiting transition to the exact solution with respect to a small
parameter of the problem, which is unambiguously pointed out in (\cite{2.1},
p. 56): ''Note that regularizing operators, dependent on a parameter, have
been employed in mathematics since Newton's times. Thus, the classic problem
of an approximate calculation of the derivative $u^{\prime }\left( x\right) $
by means of approximate (in the metric $C$) values $u\left( x\right) $ can
be solved with the help of the operator 
\[
G\left( u,\alpha \right) =\frac{u\left( x+\alpha \right) -u\left( \alpha
\right) }{\alpha }". 
\]

Then, instead of the exact value of the function $u\left( x\right) $, an
approximate one $u_\delta \left( x\right) =u\left( x\right) +\Delta u\left(
x\right) $ with $\left| \Delta u\left( x\right) \right| \leq \delta $ is
substituted. On the basis of these calculations, one makes the statement:
''If $\alpha =\delta /\eta \left( \delta \right) $, where $\eta \left(
\delta \right) \rightarrow 0$ for $\delta \rightarrow 0$, then $2\delta
/\alpha $$=2\eta \left( \delta \right) \rightarrow 0$ for $\delta
\rightarrow 0$. Thus, for $\alpha =\alpha _1\left( \delta \right) =\delta
/\eta \left( \delta \right) $, $G\left( u_\delta ,\alpha _1\left( \delta
\right) \right) \rightarrow u^{\prime }\left( x\right) $''.

It should be noted that, using the methodology of a small parameter, A. N.
Tikhonov obtained fundamental results in the field of investigations of
differential equations with a singular perturbation of the type 
\[
\epsilon \dot u=f\left( u,v,t\right) ;\quad \dot v=g\left( u,v,t\right) , 
\]
where $\epsilon $ is a small parameter; $f\left( u,v,t\right) $ is a
nonlinear function (1948)-(1952)\footnote{%
See the review by A. B. Vasileva [35].}.

The solution of the system of equations does not depend continuously on the
parameter $\epsilon $. Proceeding to the limit $\epsilon \rightarrow 0$
creates a new object of investigations with completely different properties.
In the first place, it implies the issue of the so-called violation of the
stability of the root of the equation $f\left( u,v,t\right) =0$.
Nevertheless, A. N. Tikhonov managed to develop a rather constructive theory
that served as a basis for a number of productive approaches of both
fundamental and applied character. The importance of A. N. Tikhonov's
achievements in the sphere of system analysis is analyzed in detail by N. N.
Moiseev (\cite{2.36}, section 5).

However, properties of the integral equation (\ref{2.6}) for $\alpha =0$
also change radically. In this regard, generally speaking, a certain analogy
emerges. One can suggest that A. N. Tikhonov undertook an attempt to use the
techniques of his theory of singular perturbations for the solution of
incorrectly formulated problems.

This suggestion is supported by the following quotation from the monograph
by S. A. Lomov (\cite{2.37}, p. 12): ''Now it is becoming clear how to
isolate in singularly perturbed differential equations small terms that can
be neglected. It turned out that one needed additional information about the
solution to do this.''

Note J. Hadamard's remark that an extension of methods of the theory of
ordinary differential equations to problems of mathematical physics should
be done with great care (\cite{2.38}, p. 38). At the same time, at the turn
of the 1950s, the theory of singular perturbations became an efficient tool
in investigations of complicated problems of partial differential equations
(publications by M. I. Vishik and L. A. Liusternik, O.A. Olejnik, K. O.
Fiedrichs, and others). By the way, explaining the conceptual basis of their
method of quasiinversion, R. Lattes and G.-L.\ Lions (\cite{2.23}, p. 11)%
\footnote{%
Ideological closeness of quasiinversion and Tikhonov regularisation was
pointed out by M. M. Lavrentiev [23, p. 5].} refer to these authors and A.
N. Tikhonov.

Simultaneously, they pointed out that A. N. Tikhonov's priority publication
on the method of regularization \cite{2.8} (see also \cite{2.39}) was
preceded by D. L. Phillips' article \cite{2.40}, whose results with respect
to integral equations were analogous. In the monograph by F. Natterer \cite
{2.41} this regularization figures as Tikhonov-Phillips' method. V. A.
Morozov estimated the achievements of the latter author in a much more
restrained manner (\cite{2.6}, p. 10): ''Some recommendations on the use of
this method are contained in the publications by L. V. Kantorovich \cite
{2.42} and D. L. Phillips \cite{2.40}. There is no theoretical justification
of this approach in the above-mentioned publications''.

The chronological reference to the most important results in the field of
the construction of stable algorithms for the solution of integral equations
of the first kind (\cite{2.10}, p. 234) gives the following information:
''1962, Phillips's publication \cite{2.40}, where he suggested a variational
method of conditional minimization of the functional (with the use of
restrictions on the smoothness of the solution) and put forward the idea ...
of a choice of the regularization parameter $\alpha $''.

Turning to V. M. Fridman's achievements, note that it is rather difficult to
evaluate the premises that form the basis of the iteration procedure (\ref
{2.10}). At the first sight, such a computational method has a lot of
analogs. However, its adequacy, in a sense, to the object of investigation,
the incorrect problem for the Fredholm integral equation of the first kind,
turned out to be rather unexpected.

Later on, with the aim to improve convergence, V. M. Fridman also employed
the nonlinear algorithm (\ref{2.14}). In our opinion, different ways of the
determination of the number of the final iteration and of the increase of
the rate of global convergence (see \cite{2.14,2.19,2.20}), despite their
actuality for practical application, should be interpreted as a technical
complement to V. M. Fridman's methodology.

Nowadays, the algorithm of conjugated gradients is considered to be nearly
the most efficient one for the solution of large ill-defined sparse systems
of linear algebraic equations, obtained by the reduction of, apparently, of
most problems of numerical simulations \cite{2.32,2.43,2.44}. As is pointed
out by J. Ortega \cite{2.32}, this method was proposed by M. P. Hestens and
E. L. Stiefel (1952). However, for certain reasons, it was not employed for
a long time. It attracted considerable interest at the turn of the 1970s,
when one realized the actual sphere of its applications, the potential of
the above-mentioned predetermination and adaptivity with respect to
paralleling of computational operations in combination with the architecture
of modern computers.

Thus, the priority of the method of conjugated gradients ensured its
refraction to a class of problems of linear algebra, characterized by the
instability of the numerical realization, that is, in fact, incorrectly
formulated. In this regard, we emphasize that V. M. Fridman's ''methods of
the saddle-point type'' \cite{2.21} can be interpreted as somewhat
simplified representatives of the family of the methods of conjugated
gradients (\cite{2.20}, section 2.1; \cite{2.43}, section 7.1). It seems
that V. M. Fridman, who was the first to use systematically iterations for
the solution of incorrectly formulated problems, essentially foresaw the
development of computational mathematics that followed.

In light of above, the position of M. A. Krasnoselskii and the co-authors is
worth noting \cite{2.18}. The role of V. M. Fridman in the development of
the iteration procedure (\ref{2.11}), which is an analog of (\ref{2.10}), is
described as follows: ''A transition to the equation $\left[ \psi =\left(
I-\nu A_1\right) \psi +\nu f_1\right] $ was pointed out for some cases by I.
P. Natanson \cite{2.45}. For Fredholm integral equations of the second kind,
it was already employed by G. Wiarda \cite{2.46}. For integral equations of
the first kind, it was, essentially, employed in the publication by V. M.
Fridman \cite{2.17}'' (p. 73). There is no comment on a qualitative
difference between the objects of the investigation.

The nontriviality of V. M. Fridman's approach is noted in the remark of I.
P. Natanson \cite{2.45}: ''Our method does not apply to the solution of the
integral equation of the first kind. This could be expected, because the use
of the method implies complete arbitrariness of the free term of the
equation $A\psi =f$, whereas Eq. (\ref{2.1}) is solvable not for all $%
f\left( x\right) $''. In what follows, the author gives an extended proof of
the degeneracy of the corresponding discrete problem.

The gradient algorithm of V. M. Fridman \cite{2.21} is mentioned by the
authors of \cite{2.18} exclusively in the context of the equation $A\psi =f$%
, where both the operators $A$ and $A^{-1}$ are bounded (p. 115). We quote
the abstract to V. M. Fridman's paper \cite{2.21}: ''We present a new proof
of the convergence of methods of the saddle-point type for a linear operator
equation. We do not assume, unlike L. V. Kantorovich \cite{2.47}, M. A.
Krasnoselskii and S. G. Krejn \cite{2.48}, that zero is an isolated point of
the spectrum of the operator''.\footnote{%
This is equivalent to the boundedness of the operator $A^{-1}$.}

\section{Ill-defined finite-dimensional problems and issues of discretization%
}

In this subsection, $A\psi =f$ denotes a system of linear algebraic
equations. The conditionality number of the matrix $A$ (see, e.g., \cite
{2.49}) 
\[
cond\left( A\right) =\max\limits_{\psi }\frac{\left\| A\psi \right\| }{%
\left\| \psi \right\| }/\min_{\psi }\frac{\left\| A\psi \right\| }{\left\|
\psi \right\| }, 
\]
where $\psi $ is a manifold of vectors of the Euclidean space, represents a
raising coefficient between a relative error of the data and the solution.
At the same time, cond$\left( A\right) $ characterizes the measure of
closeness of $A$ to a degenerate matrix, for which the solution of the
corresponding system of algebraic equations does not exist or is nonunique.

An algorithm of the solution of a degenerate system of linear algebraic
equations, based on the method of least squares, is presented in the book by
A. N. Malyshev \cite{2.50}. First, the matrix $A$ is transformed to a
two-diagonal one by means of a special transformation, and one finds its
eigenvalues that are subdivided into two groups, $\sigma _1$, $\sigma _2$%
,..., $\sigma _n$ and $\sigma _{n+1}$,..., such that $\sigma _n/\left(
\sigma _n-\sigma _{n+1}\right) $ is not very large. Then, with the help of a
rather laborious procedure of the exhaustion of the second group of the
eigenvalues, one constructs a matrix $A_n$ that is stably invertible
beginning with a certain value $n$. The accuracy of the thus obtained
generalized solution $\tilde \psi $ is determined by the error of closure $%
\left\| A\tilde \psi -f\right\| /\left\| A\right\| $, using heuristic
considerations.

It seems that in a methodological sense this scheme reminds of B. K.
Ivanov's algorithm \cite{2.15} that reflects computational relations (\ref
{2.7})-(\ref{2.9}).

L. Hageman and D. Young \cite{2.43} studied the approach of
predetermination, employed for the solution of systems of linear algebraic
equations, close to degenerate ones, to accelerate by the method of
conjugated gradients iterations of the type 
\[
\psi _{n+1}=P\psi _n+g, 
\]
where $P=I-Q^{-1}A$; $g=Q^{-1}f$. It is assumed that this procedure can be
symmetrized in the sense that there exists a non-degenerate matrix $W$ such
that the matrix $W\left( I-P\right) W^{-1}$ is symmetric and positive
definite.

By use of $W$, the initial problem can be reduced to the solution of much
better defined systems of algebraic equations $B\varphi =q$, where 
\[
B=W\left( I-P\right) W^{-1};\quad \varphi =W\psi ;\quad q=Wg. 
\]

Formally, a choice of the predeterminer does not pose problems. However, in
practice, one has to resolve a contradiction between the conditions imposed
on the matrix $W$: ''closeness'' to $A^{-1}$ to reduce the number of
iterations; a ''rapid'' calculation of a product of the type $W\psi $ \cite
{2.51}. In the above-mentioned publication, I. E. Kaporin analyzes different
approaches to the construction of predeterminers for systems of linear
algebraic equations of a general type. An analogous issue, in the
interpretation of J. Ortega \cite{2.32}, is oriented mainly towards sparse
matrices.

The complexity of problems of linear algebra that arise in the realization
of modern methods of investigations in the field of the mechanics of a
continuous medium are characterized as follows \cite{2.51}: ''The matrices
of corresponding systems are rather large (up to a hundred thousand nonzero
elements), rather densely filled (up to hundreds or even thousands of
nonzero elements in each line), have no diagonal predominance, are not $M$%
-matrices and are rather ill-defined. In general, one can expect only
symmetry and positive definiteness of the matrix of the system''.\footnote{%
The non-diagonal elements of an $M$-matrix are non-positive, and all the
elements of its inverse are non-negative.}

Note that, for example, in seismic tomography \cite{2.44}, one has to be
satisfied with a numerical realization of discrete analogs of integral
equations of the first kind, because their kernels cannot be represented
analytically and parameters of the considered models are determined with the
help of natural experiments.

In light of the above, the considerations of R. W. Hamming (\cite{2.52}, p.
360) may seem to be archaic: ''A system of linear equations is said to be
ill-defined, if, roughly speaking, the equations are almost linearly
dependent. Many efforts were made to investigate the problem of the solution
of ill-defined systems. However, one may pose the question: Is it necessary
to solve such systems in practical situations? In what physical situation
may the solutions prove to be useful, if they depend in such a substantial
manner on the coefficients of the systems? Usually, the following is true:
Instead of the solution, one is looking for a system of almost linearly
independent equations. In light of this information, the problem can be
better understood and is usually reformulated again in a more satisfactory
way. It is rather probable that ill-defined systems of equations, provided
that round-off and measurement errors are eliminated, are actually linearly
dependent and thus do not reflect the physical situation''.

Note that, in contrast to the above-mentioned constructors of methods of
computational mathematics, the renowned practitioner adheres to the position
of correctness according to Hadamard. Let us quote P. S. Guter's preface to 
\cite{2.52}: ''The name of R. W. Hamming, a renowned American scientist,
former President of the Computer Association, Head of the Mathematical
Service of Bell Telephone Laboratories, and his works in the field of
computational mathematics and the theory of information are rather
well-known and do not need special recommendations. ... The book 'Numerical
Methods for Scientists and Engineers' is without any doubt an outstanding
phenomenon in mathematical literature''.

Of special interest is R. W. Hamming's opinion about the priority of
computational procedures (\cite{2.52}, p. 90): ''It is often believed that
the main problems of numerical analysis are concentrated on interpolation,
but this is not the case. They are mostly related to such operations as
integration, differentiation, finding zeros, maximization, etc., in those
cases when all we have or can compute are some nodes of functions that are
usually known not exactly, but approximately, because they are spoiled by
the round-off error''.

Thus, the problem should be posed correctly under the conditions of an
inevitable error in the data, which is equivalent to the preference of
algorithmic efficiency to the quality of initial information. Interpolation,
mentioned in the above quotation, implies approximate representation of the
latter for the performance of computer operations by means of a
finite-dimensional approximation.

However, in computational mathematics, alternative concepts are rather
wide-spread, which is reflected in K. I. Babenko's remark \cite{2.25}: ''In
some spheres of numerical analysis, the theory of approximation serves as
the foundation for the building of the numerical algorithm'' (p. 138).
''Information, inputted into the algorithm, is characterized, in the first
place, by its volume... All other characteristics, such as, e.g., accuracy,
are its derivatives and do not present a true picture of the input'' (p.
281).

Here, information is understood in the sense of Kolmogorov's theory of $%
\epsilon $-entropy that identifies it with the length of a given table or an
alphabet, whose words are manipulated by the algorithm. Correspondingly, the
issue of numerical analysis is interpreted in terms of, figuratively, the
deficiency in the search for necessary words and of the deletion of tables
in the course of operations.

Nevertheless, R. W. Hamming's point of view on the relation between the
method of investigations and the employed information is actively developed
by a group of specialists with J. Traub and G. Wasilkovski at the head. The
authors of \cite{2.53} point out (pp. 9, 6): ''In this book, we construct a
general mathematical theory of optimal reduction of uncertainty. We
interested in the two main questions: 1) Is it possible to reduce
uncertainty to a given level? 2) What will it cost? The aim of the theory of
informational complexity is to provide a unified approach to investigations
of optimal algorithms and their complexity for the problems that involve
incomplete, imprecise or paid information and to employ the general theory
to concrete problems from different fields''.

Here, complexity implies the number of arithmetic operations, the time of
their realization, computer memory resources, etc. It is interesting to note
that the interpretation of the notion of probability \cite{2.52,2.53}
correlates with the expressive statement of R. Bellman and S. Dreyfus (\cite
{2.54}, p. 342): ''Fortunately, in some cases, there is a very simple way to
overcome this difficulty. Instead of trying to study information as the
''smile of Cheshire Cat'', we consider the actual physical process, where
information is used to work out solutions.\footnote{%
The smile of Cheshire Cat, according to L. Carrols "Alice in th Miracle
Land", existed separately from this cat (editor's note to [54]).} The value
of information can then be measured by the efficiency of the solutions.

Thus, the usefulness of information depends on its application, which is the
most reasonable concept!''

It should be noted that the procedure of finite-dimensional approximation of
problems of mathematical physics is, of course, also very important, which
is accentuated by K. I. Babenko. Indeed, the obtained discrete model can
turn out to be incorrect, and the employed algorithms of the numerical
realization may prove to be divergent even in the solution of rather
ordinary problems. An example of instability of a finite-difference scheme
is given by S. K. Godunov and V. S. Ryabenkii (\cite{2.55}, section 4.9).

K. I. Babenko also emphasized the absence of any general methods of the
construction of finite-dimensional analogs (\cite{2.25}, p. 622): ''...the
provision of an approximation alone is insufficient''... one has to ensure
that the discrete problem ''retains the type of the original continuous
problem''. In his opinion, to achieve the above goal, ''a detailed
investigation in each concrete case is required, which is the most
nontrivial part of work''.

\section{The crisis of the technology of numerical simulations}

Of considerable interest is, in fact, a program statement of O. M.
Belotserkovskii and V. V. Stchennikov in the preface to \cite{2.56}:

''A rapid development of computers, especially during the last 10-15 years,
with a special acuteness posed the problem of the construction of a
principally new technology of the solution of problems by computers. ...
Historically, the problems of numerical simulations (in this notion, we
include the actual mathematical simulations related to a numerical
experiment), being rather advanced already in the ''precomputer'' period and
rapidly developing during the next periods, turned out to be the most
conservative component of the modern technology of the solution of problems
on the computer. Using, probably, redundant from the point of view of a
mathematician expressiveness of the description, one can characterize the
existent situation by two stable tendencies:

- an increase of the complexity of mathematical models;

- construction of rather sophisticated mathematical methods.

Both the tendencies inevitably lead to a technological deadlock, because
they create complications in the solution of the problem of the construction
of software-hardware means of the support of the whole technological chain.
... Without any pretension to profoundness and importance of the analogy, we
dare say that the present situation in numerical simulations is similar to
that in mechanics before the appearance of main ideas and concepts of
quantum mechanics''.

In the introductory article \cite{2.56} the same authors emphasize the
phenomenon of the accumulation of the round-off error in the numerical
realization of algorithms that include up to $10^{12}$ operations and the
absence of real means to estimate the error of solutions to, in particular,
evolution problems. In their opinion, ''...the following conclusion is quite
justified: a priori, any evolution problem for large times is numerically
(or computationally) incorrect in the sense of the absence of a practically
important solution...

In the case, when a priori or a posteriori information about the error of an
approximate solution is absent, it is impossible to claim that the solution
exists. This conclusion fairly agrees with A. N. Tikhonov's theorem that
states that the problem with the data on the operator and the right-hand
side has no solution in the manifold of approximate numbers''.

O. M. Belotserkovskii and V. V. Stchennikov regard as constructive the idea
that discrete models of the considered problems should be assembled with the
aim of increasing the accuracy of information by means of special
superposition. They also suggest to search for the solution in the class of
function with a bounded variation, with would endow the difference operator
of the problem with smoothing properties.

As is well-known, N. N. Yanenko paid considerable attention to the
methodology of mathematical simulations (see \cite{2.2}). His concept of
overcoming the above-mentioned crisis is explained by O. M. Belotserkovskii
( \cite{2.57}, p. 106):

''An investigation of finite-difference schemes, approximating different
classes of equations of mathematical physics, led N. N. Yanenko to an
extension of the notion of the scheme. For the first time, he begins to
consider the finite-difference scheme as an independent object of the
investigation, as a mathematical model, adequate to this or that physical
model. This fundamental concept is based on profound understanding of the
fundamentals of differential and integral calculus.

Indeed, physical and mathematical models, described by differential,
integral or integrodifferential equations, are obtained from discrete models
by means of averaging and passing to the limit with respect to certain
parameters. This is the case, for example, in the model of a continuous
medium, where for a sufficiently large number of elements in the unit volume
one comes to the notion of the continuous medium by averaging and passing to
the limit with respect to the volume. In this regard, one can interpret a
finite-difference scheme as an independent mathematical model with certain
properties''.

Note the fundamental, as it seems, considerations of N. N. Yanenko \cite{2.2}%
: ''The objects of modern mathematics, whose theoretical ''nucleus''
comprises topology, geometry, algebra and functional analysis, are ideal
logical constructions forming a certain operational system. We will call
them ideal objects, which underlines, on the one hand, their practical
inaccessibility and, on the other hand, their excellent operational
properties that allow one to make operations without loss of information.
Ideal objects of mathematics are essentially infinite and require an
infinite number of operations'' (p. 12).

''The development of the experimental foundation and the tool of
investigations, the computer, increased interest in such objects as computer
numbers, programs, finite automata. In this regard, the definition of
mathematics as studies of the infinite, accepted in the 20th century, should
be replaced by another one, more correctly reflecting its essence, i.e., as
studies of the relationship between the finite and the infinite'' (p. 18).

We think that on the basis of the above one can come to a very important
conclusion: In their construction of the conceptual basis of mathematical
simulation, the leading specialists were guided by the concept of
inapplicability of Banach's theorem on the inverse operator. Note that N.
Dunford and J. Schwartz considered this theorem as one of the three
principles of linear functional analysis, characterized as being rather
fruitful (\cite{2.58}, p. 61).\footnote{%
The other two are the principle of linear boundedness and Hahn-Banach's
theorem.}

A quotation from K. Maurin's manual (\cite{2.59}, p. 51) reads: ''This
theorem [on the closed graph], in the last years, has gained itself a
reputation of being the most important theorem of functional analysis, if
this one is considered from the point of view of applications''.

An attempt to renew the above-mentioned fundamentals in the context of the
accentuation of peculiarities of computational mathematics was made by A. V.
Chechkin \cite{2.60}, who suggested a division of sections of mathematics
into classical and non-classical ones, respectively: ''arithmetics,
mathematical analysis, algebra, geometry, probability theory, etc.;
mathematical logic, the theory of information and statistics, the theory of
sets, the theory of algorithms and recursive functions, methods of
computational mathematics, the theory of finite-difference schemes, the
theory of cubic formulas, methods of the solution of incorrect problems,
etc.'' (p. 8). As a criterion, the authors choose the fact of availability
of absolutely complete or partial information about the considered objects
(points, functions etc.).

Let us quote the abstract of section (\cite{2.60}, p. 78): '' We define and
study a new type of mappings that generalize classical notions. Classical
mappings realize correspondence between the points of a set. This implies
that the points are known with absolute precision. The new mappings, termed
ultramappings, realize correspondence between pieces of information about
points of sets. The main construction of the ultramappings, termed
ultraoperators, allows one to obtain separate information about the image
point from separate information about the inverse image point.

Ultracontinuity of ultraoperators is defined, which is a broad
generalization of the notion of the stability of methods. It is found that,
for an arbitrary base operator, one can construct an ultracontinuous
operator over it. A class of ultracontinuous operators, termed Tikhonov's
operators, is singled out. For these operators, the base operators are not
continuous''. Furthermore, ''they are related to A. N. Tikhonov's ideas and
methods of the solution of incorrect mathematical problems''.

Returning to the question of adequate discretization, we quote the abstract
of the monograph by A. A. Dezin \cite{2.61}: ''It is devoted to the
description of the basic structures of multidimensional analysis and to the
consideration of internally defined discrete problems of analysis and
mathematical physics. It implies not merely an approximation of a given
continuous object, but the construction its analog, starting from the notion
allowing for discrete interpretation''.

Arguments for contradiction to physical sense of differential models of
certain classes of problems of the mechanics of a continuous medium are
given by M. A. Zak \cite{2.62}. In this regard, he developed a general
approach, wholly based on the concepts of theoretical mechanics with a
special interpretation of Gauss' principle of least action.

The position of C. Truesdell is alternative. He thinks that continuum
mechanics of a deformed body ''is, in essence, not only subtler, more
beautiful, majestic than a rather sparse particular case, called
''analytical mechanics'', but it is much more suitable for the simulation of
real bodies'' (\cite{2.63}, page 10).

\chapter{Short comments on the material of the above sections and some
general considerations}

\section{The correctness of the formulation of problems of mathematical
physics}

The conditions of correctness, formulated by J. Hadamard at the turn of the
20th century (see \cite{3.1}) and insistently advocated by him thereafter,
amaze us by their ever-increasing importance for practical applications. It
forms the conceptual basis for methods of the simulation of physically
meaningful problems, which, in fact, is disputed by nobody. At the same
time, nowadays, the prevailing opinion is that Hadamard's concepts are
principally invalid.

This concerns the basic statement that the properties of existence and
uniqueness, considered by Hadamard as inherent to mathematical models of
real processes, lead to the correctness of the formulation of adequate
boundary-value (initial-boundary-value) problems, which implies the
stability of the employed algorithms of a numerical realization. Although,
of course, not all generally quite valid algorithms can ensure continuous
dependence of the solution on the data of a concrete problem: Implied is a
possibility of their construction.

The above-said concerns, in the first place, the key notion of this
statement, namely, correct formulation of problems of mathematical physics.
Indeed, only the existence of such formulation is claimed unambiguously,
which implies, in particular, that the Fredholm integral equation of the
first kind is simply unsuitable as a reflection of objective reality.

Naturally, the course of investigations with the aim to confirm or disprove
the hypothesis, or, in our opinion a prophecy, of J Hadamard, seemingly had
to be conducted from the position of variability of formulations of the
considered problems, which was not the case. The main reason is, in our
opinion, an erroneous understanding of a special mission of computational
means of numerical simulations that lightheartedly neglected even one of the
main principles of functional analysis, i.e., Banach's theorem on inverse
operator (\cite{3.2,3.3},section 9, and \cite{3.4}).

One can hardly explain the absence in special literature even of a thesis of
the necessity to coordinate the formulation of problems of mathematical
physics with algorithms of their numerical realization. The roots of this
situation seem to be in systemic character of the giant computer-supply
complex oriented at commercial efficiency at the expense of high costs of
provided services.

As a result, the alternative school of A. N. Tikhonov builds up the
criticism of J. Hadamard ideas according to the following scheme:

- the solution of the Fredholm integral equation of the first kind 
\begin{equation}  \label{3.1}
\left( A\psi \right) \left( x\right) \equiv \int\limits_0^1k\left( x,\xi
\right) \psi \left( \xi \right) d\xi =f\left( x\right) ,\quad x\in \left[ 0,1%
\right]
\end{equation}
is, in general, an incorrect problem (which is undisputable);

- integral equations of this type are adequate to a variety of real
phenomena, which is actually supported by a rather transparent
interpretation of corresponding direct problems (calculations of $f$ from
given $k$ and $\psi $).

However, what are the grounds for the formulation of the problem, inverse to
the calculation of $f$, by means of trivial renaming the given and sought
functions in (\ref{3.1})? The fact that the procedure of the restoration of $%
\psi \left( x\right) $ for given $f\left( x\right) $ and $k\left( x,\xi
\right) $ is incorrect does not imply any consequences.

The Fredholm integral equation of the first kind is, without any doubt, a
rather graphic object for an illustration of the suggested arguments. As
regards problems of mathematical physics for partial differential equations,
their evaluation is not so unambiguous. Thus, Hadamard's example (\ref{1.3}%
), (\ref{1.4}), concerned with the restoration of the function $u\left(
x,t\right) $ from the initial data, is interpreted as a direct problem,
whereas the inverse one, in contrast, has to be related to the determination
of the right-hand sides from the corresponding information about the
solution. Hence, the formulation of the direct problem is incorrect.

It is usually believed that the problem (\ref{1.3}), (\ref{1.4}) has
physical applications (see, e.g., \cite{3.2}). In this case, following J.
Hadamard''s methodology, its formulation can be reduced to a correct one. As
will be shown below, the difference between integral equations of the first
kind and boundary-value (initial-boundary-value) problems of mathematical
physics are, in a sense, conditional. Indeed, the problem (\ref{1.3}), (\ref
{1.4}) can be reduced by elementary means to an equation of the
above-mentioned type with respect to $\partial _x^2$ or $\partial _t^2u$.

The reproaches to J. Hadamard, whose typical elements are reproduced in
section 2.1, can be summarized as follows: The great scientist slowed down
the progress of science by refusing to admit that incorrectly formulated
problems were adequate to a variety of real processes (see \cite
{3.3,3.4,3.05}). However, J. Hadamard had no doubts that such problems
existed. As a matter of fact, put forward the thesis of inadequacy of the
employed means of numerical simulations. Indeed, is it justified to raise
the issue of someone's monopoly of a mathematical model as regards its
objective correspondence to the realities that we observe?

Clearly, the answer can be only negative: Mathematical models are
constructed with different degrees of detail, they have the form of integral
equations, boundary-value problems, etc., inducing, by the way, rather
non-identical difficulties of computational character. At the same time, the
results of reliable versions of numerical simulations for the considered
processes must, in general, be identical.

One can raise an argument that a transition from an incorrect formulation of
the problem to a correct one is a cardinal step and, accordingly, should be
accompanied by a substantial deformation of computational relations, which
in turn should influence the solution. This conclusion is seemingly
supported by the well-known fact of functional analysis that the property of
incorrect solvability is stable with respect to small perturbations \cite
{3.6}. However, this property is connected with a pair of spaces and,
generally speaking, looses its validity when Eq. (\ref{3.1}) is considered
from the point of view of the mapping between the spaces $L_2\left(
0,1\right) $ and $l_2$, as it was mentioned above.

It seems that we have pointed out an important issue. As a development of
this issue, we state the absence of serious arguments against a possibility
of the correct formulation of the problem, e.,g., of the restoration of the
function $\psi \left( x\right) $ from the kernel $k\left( x,\xi \right) $
and the right-hand side $f\left( x\right) $, obtained by integration
according to Eq. (\ref{3.1}). Purely heuristically, if small variations of
the data can substantially influence the solution of incorrectly formulated
problem, why not suggest that this effect can be overcome in a satisfactory
way by a rather nonessential correction of the traditional mathematical
model?

Among supporters of studies of problems of mathematical physics exclusively
in the correct formulation are: A. Poincar\'e, D. Hilbert, V. A. Steklov, I.
G. Petrovsky, I. Prigogine \cite{3.7,3.8,3.9,3.10,3.11}. On the other hand,
the role of the three absolutely independent conditions of the correctness
(existence, uniqueness and continuous dependence on the data of the
problem), introduced by R. Courant and D. Hilbert \cite{3.12}, can hardly be
called positive. We think that the constructive potential of the fact that
the third condition is a corollary of the previous ones could facilitate the
activation of studies of different aspects of the correct formulation of
problems of mathematical physics in the classes of functions inherent to
them.

When considering the Fredholm integral equation of the first kind (\ref{3.1}%
), one had to be more careful with respect to operations involving the
function $f\left( x\right) $ from $l_2$ as opposed, figuratively, to a
surrogate of continuous inversion under the condition that it belongs to $%
L_2\left( 0,1\right) $. Accordingly, an analysis of the premises of the
correct solvability of this equation could lead to a conclusion that the
problem of the restoration on its basis of the function $\psi \left(
x\right) $ from the given $f\left( x\right) $ and $k\left( x,\xi \right) $
cannot be interpreted as the inverse of the calculation of $f\left( x\right) 
$ from Eq. (\ref{3.1}).

\section{A relationship to the theorem on the inverse operator}

The above-mentioned fact that the third condition of the correctness has the
character of a corollary results from Banach's theorem on the inverse
operator \cite{3.13} whose optimistic meaning consists in the following: If
the solution to Eq. (\ref{3.1}), with $D\left( A\right) =B_1$ and $f\in B_2$%
, where $B_1$, $B_2$ are Banach spaces, exists and is unique, the inverse
operator $A^{-1}$ from $B_2$ into $B_1$ is bounded. As a consequence, the
ordinary procedure of the solution of the Fredholm integral equation of the
first kind (\ref{3.1}) is by Picard's theorem stable under the condition
that $B_2$ coincides with $l_2$.

However, in the general case both the verification and the fulfillment of
the condition $f\in l_2$ are practically infeasible. Therefore, such spaces
are called ''inconvenient'' (see \cite{3.14,3.15}). Thus, we are obviously
in a principle dilemma as to the choice of the methodology of the
investigation:

- an orientation at overcoming the difficulties resulting from the use of
the space $l_2$ related to the boundedness of the operator $A^{-1}$;

- the loss of this property in exchange for a possibility of studying
mathematical models in ''convenient'' spaces.

With the beginning of large-scale applications of computational methods to
mathematical investigations, the second way became dominant.
Correspondingly, the canonical formulation of discussed S. Banach's theorem,
given, e.g., by A. N. Kolmogorov and S. V. Fomin \cite{3.16} (see section
2.3), has been accompanied by a specification that it is valid only in the
case when the operator $A$ realizes a mapping on the whole space $B_2$ \cite
{3.17,3.18}. It should be noted that for $f\in L_2\left( 0,1\right) $ this
condition is rather important, because the space $l_2$ inherent to Eq. (\ref
{3.1}) is a subspace or part of $L_2$.

The dynamics of the point of view of S. G. Mikhlin, reflected in his courses
of mathematical physics and the theory of errors of 1968, 1977 and 1988 \cite
{3.19,3.20,3.21}. At the beginning, the author considers Eq. (\ref{3.1})
under the traditional assumption that the operator $A$ is compact, that is,
according to the type of the mapping inside the space $L_2\left( 0,1\right) $%
. In this case, the inverse operator $A^{-1}$ is unbounded. As a result, the
standard algorithms of the numerical realization are inapplicable, and one
has to turn to the methodology of A. N. Tikhonov.

Later, S. G. Mikhlin drew attention to the fact that if the Fredholm
integral equation of the first kind (\ref{3.1}) is interpreted from the
point of view of a mapping between the spaces $L_2\left( 0,1\right) $ and $%
l_2$, the corresponding operator $A$ is no longer compact and, as a
consequence, the inverse $A^{-1}$ can be bounded, and the problem of the
restoration of the function $\psi \left( x\right) $ becomes correct. In this
way, the completeness of the conditions of correctness is restored, whereas
the third condition was initially singled out by the author.

Thus, the use of the pair of spaces $L_2\left( 0,1\right) -l_2$ in a sense
transfers the canonical incorrect problem to the mainstream of fundamentals
of functional analysis. Note the fact that, being a mathematician, S. G.
Mikhlin did not devalue the importance of the outlined step by banal
reasoning in terms of ''convenient/inconvenient'' or ''bad'' and ''good''
spaces.

Such a position apparently incurred criticism: In his concluding monograph,
S. G. Mikhlin somewhat irritably readdresses actual formulation of problems
of mathematical physics to specialists in applied sciences, including
sociologists, who are interested in their solution. Simultaneously, the
author has found it reasonable not to consider infinite dimensional models
with inherent aspects of incorrectness.

The statement that a mathematical problem, being posed (in terms of
premises), must be solved by rigorous methods belongs to V. A. Steklov. At
the same time, why not consider the procedure of the formulation of problems
of mathematical physics as an additional reserve of increasing the
efficiency of employed techniques of numerical realization? Moreover, maybe
rigidly predetermined formulations of problems themselves prose artificial
complications of computational character under the conditions when physical
considerations admit a small, in a sense, variation? In our opinion, the
formulation of problems of mathematical physics and the algorithm of its
numerical realization are essentially interrelated categories.

The material of section 2.4 develops the arguments for the advantages of
studies of the Fredholm integral equations of the first kind in the pair of
spaces $L_2\left( 0,1\right) -l_2$. Thus, an obvious mismatch, in the
general case, between the free term $f\left( x\right) $ and the space $l_2$
allowed us to outline a constructive algorithm of numerical realization by
means of an adequate representation of the arising error of closure.

\section{The methodology of the solution of incorrect problems}

Incorrectly formulated problems of mathematical physics are deceptively
transparent from the point of view of the interpretation of considered
processes. This results from their adequacy to spaces that in the
computational sense are practically infeasible. If the data of such problems
are specified in their natural classes of functions, the corresponding
formulations loose a mathematical sense because of insolubility. In this
situation, of crucial importance is a role of general methodological
concepts, that is, one has to be guided by a system of global principles.

From this point of view, J. Hadamard's insistence on the correct formulation
of problems describing physical phenomena \cite{3.1} still can be
interpreted as a kind of hypothesis, whereas in fact related Banach's
theorem on the inverse operator is a universally accepted element of the
foundation of modern mathematics \cite{3.22}. Nevertheless, A. N. Tikhonov
chose an alternative orientation by a revision of the actual notion of the
solution of the incorrect problem and the use with respect to it of special
algorithms of numerical realization \cite{3.2}. It seems that this choice
was in full harmony with the attitude of the scientific community of the
last decades towards the unprecedented revolutionary role of computational
mathematics in natural science (see \cite{3.3,3.23,3.24,3.25}).

A notion of correctness according to A. N. Tikhonov appeared that played up
a rather simple version of a search for the solution in a reduced class of
functions meeting the criterion of the correct formulation \cite{3.14}. By
the way, any a priori premises for finding such a class on the basis of
reasonable information are missing.

Shaky character of the conceptual basis led to the failure of the idea of a
limiting transition with respect to a small parameter in the solution of a
family of problems related to incorrectly formulated ones (the method of
regularization \cite{3.2}). The reason, in our opinion, lies in the same
inadequacy of the use of functional spaces. Given that $l_2$ is
characterized by an infinite number of features, whereas $L_2$ is
characterized by only one, is it possible, even on the heuristic basis, to
expect to overcome this cardinal disagreement with the help of the
regularization parameter $\alpha $?

The situation in the sphere of activity of numerous followers of A. N.
Tikhonov looks rather deplorable. The actual efforts are concentrated on a
surrogate with a small factor $\alpha $, formed on the basis of (\ref{3.1}): 
\begin{equation}  \label{3.2}
\alpha \psi \left( x\right) +\int\limits_0^1k\left( x,\xi \right) \psi
\left( \xi \right) d\xi =f\left( x\right) ,\quad x\in \left[ 0,1\right] .
\end{equation}
This is called the Fredholm integral equation of the second kind, without
any mentioning of its insufficiency in this respect. Despite a large number
of investigations devoted to the determination of the regularization
parameter $\alpha $, any more or less constructive algorithms are absent.
The main reason seems to be the inconsistency of the thesis that implies a
possibility of efficient matching between the solution and the data of
incorrectly formulated problems (see, e.g., \cite{3.2,3.26}).

As a matter of fact, one has to be satisfied only by a comparison of
solutions to (\ref{3.2}) obtained in the range of the decrease of $\alpha $.
Because of great labor input of numerical realization for small values of
the regularization parameter, a large-scale application of A. N. Tikhonov
methodology to the practice of scientific investigations incurred
considerable economic damage. As regards attempts to investigate the
Fredholm integral equation if functional spaces of its correct solvability,
they were isolated and were not accompanied by constructive implementation 
\cite{3.27}.

V. M. Fridman, whose papers \cite{3.28,3.29} are considered in section 3.3,
approached the solution of the Fredholm integral equation of the first kind
as a mathematically formulated problem regardless of its applicability to
modeling of concrete processes.From the point of view of our consideration,
the iterative algorithms of V. M. Fridman may be of interest, because they
allow one to achieve maximal possible efficiency in the framework of the
chosen object of investigation, which is indirectly confirmed by their
simplicity and brevity.

In other words, it is hardly possible to obtain more from the traditional
interpretation of Eq. (\ref{3.1}) (under the condition that formal
convergence exists). By approaching the solution, the employed corrections
become negligibly small compared to the values of the sought function: 
\[
\psi _{n+1}\left( x\right) =\psi _n\left( x\right) +\lambda \left[ f\left(
x\right) -\left( A\psi _n\right) \left( x\right) \right] . 
\]

As is well-known, in the absence of a timely halt of such a procedure,
computational ''noise'' from operations with incommensurate numbers can
radically distort the solution \cite{3.05,3.15}. It becomes obvious that the
Fredholm integral equation of the first kind, by virtue of its nature,
contains an inherent defect that principally disagrees with pithy
formulation of the problem of the restoration of the function $\psi \left(
x\right) $ from the right-hand side of (\ref{3.1}).

Therefore, it seems to be quite natural to complete the integral equation (%
\ref{3.1}) by a component that, by adaptively compensating for the negative
factor of computational ''noise'', would not substantially influence the
initial mathematical model by virtue of its smallness (see the constructive
premises in section 2.4).

In section 3.5, we have given the argument of K. I. Babenko \cite{3.23} for
the necessity to take into account the fact of the loss of information when
evaluating comparative efficiency of computational algorithms. This argument
seems to be even more important at the stage of the formulation of the
problem. Since calculations of $f\left( x\right) $ from (\ref{3.1})
objectively delete the information on the function $\psi \left( x\right) $,
its restoration in the framework of the traditional approach quite naturally
reduces to an incorrect problem.

If we hypothetically assume that the sought function $\psi \left( x\right) $
is explicitly present in Eq. (\ref{3.1}) [i.e., the parameter $\alpha $ of
Eq. (\ref{3.2}) is commensurate with the values of the kernel], all the
outlined problems disappear. Nevertheless, any actions of the type of a
coarse substitution are absolutely unacceptable. However, such an appearance
of $\psi \left( x\right) $ can be viewed in the context of the modeling of
computational ''noise'' with the participation of the integral term also.
Below, we will specify this exclusively important point.

\section{Concepts of numerical simulations at the modern stage}

The predetermined method of conjugate gradients is considered to be one of
the most efficient methods for the solution of ill-posed systems of linear
algebraic equations that appear as a result of discretization of different
problems of mathematical physics \cite{3.30}. The predeterminer, a
non-degenerate matrix, allows one to reduce the procedure of numerical
realization to a sequence of algebraic problems with desired favorable
properties. On the other hand, however, the number of necessary iterations
and their difficulty increase (section 3.7).

It should be noted that the structure of the predeterminer is deprived of
adaptive basis with respect to the orientation at specific characteristics
of a concrete computational procedure. In our opinion, an analogous
situation is, in general, typical of discrete models. From this point of
view, the potential of the techniques of continuous analysis is at a
qualitatively higher level.

One of the key problems of computational mathematics is the development of
the conceptual basis for a relationship between a representation of the data
and the efficiency of the employed algorithms. In this regard, the ideas of
K. I. Babenko \cite{3.23}, completely based on a qualitative interpretation
of the notion of information put forward by A. N. Kolmogorov, can be
estimated as rather pessimistic. Indeed, almost all computational operations
of this guide are accompanied by a ''colossal'' loss of information, whereas
rare exceptions correspond only to a special representation of initial
tables, which, as a rule, is not realized in practice.

The position of R. W. Hamming \cite{3.31}, who is a direct follower of the
ideas of J. Hadamard in the field of computational mathematics, is
alternative. In his opinion, methods of numerical realization must be
adapted to the available information. As regards principal difficulties,
such as the incorrectness of the formulation, the main attention should be
concentrated on a modification of mathematical models. The arguments of P.
Bellman and S. Dreyfus for the expediency of the evaluation of the quality
of information on the basis of its efficiency indices \cite{3.32} are also
rather attractive.

O. M. Belotserkovsky and V. V. Shennikov \cite{3.24} stated a crisis in the
sphere of numerical simulations resulting from the complexity of both the
formulations of practical problems and the techniques of their numerical
realization. (section 3.8). As a reason, they have pointed out an
inapplicability of methods of ''pre-computer'' mathematics to situations,
when owing to the accumulation of round-off errors actually any algorithm
becomes computationally incorrect. As a matter of fact, the authors proposed
to develop more intensively approaches in the style of A. N. Tikhonov,
without any mentioning of the alternative way, i.e., matching the
formulations of considered problems with Banach's theorem on the inverse
operator.

Note that generations of specialists in different fields of mathematical
physics were brought up under slogans of the type ''all real problems of the
mechanics of continuum medium are ill-posed'' that were repeatedly
reiterated without any explanations by ''greats'' at different conferences.
As a result, we have an implementation at a folklore level of the thesis
supported only by the practice of scientific research.

N. N. Yanenko, who, in contrast to some colleagues, was well aware of the
losses of numerical simulations from the breakup of ties of the techniques
of numerical realization with the basics of functional analysis, can be
called a flagship of this ideology. However, he considered to be of crucial
importance the principal difference between classical and computational
mathematics consisting in the fact that the former dealt with abstract
symbols without the loss of information, whereas the objects of the latter
were numerical arrays whose transformation was inevitably accompanied by
errors of different kinds (see \cite{3.3,3.33}).

A methodological orientation of the arguments of the works of N. N. Yanenko
allows us to suggest that a certain role in the formation of his ideas was
played by ambitious motivations of being a co-participant of the emergence
of ''new'' mathematics that, while partly employing the ''old'' one, was, in
general, substantially superior. A grotesque manifestation of this position
is represented by the monograph of A. V. Chechkin \cite{3.25} completely
based on A. N. Tikhonov concepts.

It seems that we facing a distortion of the essence of the problem, because
Banach's theorem on the inverse operator is an entity of a higher level than
numerical operations and, at the same time, is indispensable to them.
Indeed, the boundedness of the inverse operator yields practically a unique
possibility to prevent both inadequate dependence of the solution on the
data and the accumulation of computational errors.

The outstanding ideas of J. Hadamard and S. Banach who foresaw the
development of the technique of numerical simulations should be regarded as
its fundamental basis.

\section{Ideas of the development of a constructive theory}

Let us suppose that the kernel $k\left( x,\xi \right) $ of the Fredholm
integral equation of the first kind (\ref{3.1}) is closed in $L_2\left(
0,1\right) $ and its solution exists.\footnote{%
In the interests of continuity of our consideration, we repeat some of the
above-discussed points.} We also assume a possibility of exact evaluation of
the right-hand side by means of integration, or the function is assumed to
belong to the space $l_2$, which implies the condition 
\begin{equation}  \label{3.3}
\sum_{n=1}^\infty \alpha _n^2\lambda _n^2<\infty ,\quad \alpha
_n=\int\limits_0^1f\left( x\right) \bar \psi _n\left( x\right) dx,
\end{equation}
where $\lambda _n$, $\bar \psi _n\left( x\right) $ are the characteristic
numbers and the eigenfunctions of the kernel $k\left( x,\xi \right) $.

Because of the closure of $k\left( x,\xi \right) $, the solution to (\ref
{3.1}) is unique as well; the operator $A$ that maps from $L_2\left(
0,1\right) $ into $l_2$ is continuous: Hence all the conditions of Banach's
theorem on the inverse operator are fulfilled. This theorem states that the
inverse operator $A^{-1}$ that maps from $l_2$ into $L_2\left( 0,1\right) $
is continuous as well. In other words, the procedure of the evaluation of $%
\psi \left( x\right) $ is stable against small perturbations of the given $%
k\left( x,\xi \right) $ and $f\left( x\right) $. Therefore, it can be
realized without an accumulation of round-off errors of significant digits.

In this regard, the Inverse World of S. Banach is rather captivating and, at
the same time, it is based on the absence of any differentiation of the
employed spaces with respect to preference. The dominant ideas in the sphere
of computational mathematics are purely alternative. Therefore, both openly
and mainly implicitly, the arguments for non-constructive character of this
fundamental theorem are put forward.

The arguments are seemingly reasonable. Indeed, the function $f\left(
x\right) $ is determined mostly with an error of measurement. Thus, spaces
of the type $L_2$ that restrict certain superpositions on an interval are
natural for its estimate. As already mentioned, the space $l_2$ is illusive
because it deals with an infinite set of features of the data that, as a
result, cannot be identified in practice. Even a comparatively small
variation of $f\left( x\right) $ may violate the criterion (\ref{3.3}) and,
as a consequence, distort the solution to (\ref{3.1}) by Picard's theorem.

Thus, the actually given function $f\left( x\right) $, as a rule, does not
belong to the space $l_2$. A trivial case of the representation of $f\left(
x\right) $ as series in terms of the elements $\bar \psi _n\left( x\right) $
for exactly determined $\lambda _n$ is an exception. However, such premises
cannot serve as a basis for the neglect of the space $l_2$ in the studies of
Eq. (\ref{3.1}). It seems that constructiveness is possible here only in the
context of the agreement of, generally speaking, different orientations:

- the function $f\left( x\right) \in L_2\left( 0,1\right) $;

- the operator $A$ maps from $L_2\left( 0,1\right) $ into $l_2$.

The motivation is obvious: to preserve the potential of continuous inversion
of the operator $A$ for practical realization. At the same time, the
outlined contradiction is clear, and it cannot be overcome exclusively in
the framework of the Fredholm integral equation of the first kind (\ref{3.1}%
). In this situation, it is quite natural to turn, figuratively speaking, to
the origin of this equation, that is, to the issues related to the
formulation of the problem.

Consider a certain process described by the operator $A$. The direct problem
consists in the evaluation of the integral according to (\ref{3.1}) under
the substitution of the given function $\psi \left( x\right) $. If it can be
expanded in a series of $\sin \left( n\pi x\right) $, in particular,
corresponding terms acquire a factor of $n^{-1}$ and, as a result, the
convergence strengthens. This procedure has a lot of physical and other
interpretations and is mathematically correct.

A key element is the formulation of the inverse problem for the same
operator $A$, which is related to the restoration of the function $\psi
\left( x\right) $ from given realization of the procedure of integration,
that is, $f\left( x\right) $. What is actually implied is the determination
of the cause from its consequence. Whereas the scheme of the solution of the
direct problem is transparent, the status of the inverse problem is
diametrically opposed. A priority of its solution is the actual algorithmic
procedure on the basis of an adequate mathematical model that is not an
analog of the process occurring in the regime of real time.\footnote{%
Indeed, the cause as an outcome of the consequence has no physical sense.}

In general, the traditional formulation of inverse problems by means of
formal renaming of known and unknown components of mathematical models
describing objectively occurring processes has no grounds.

In light of the above, it is quite natural to recall the statement of J.
Hadamard that all problems having practical interpretation admit a
mathematically correct formulation. From this point of view, since the
solution of the problem, inverse of the evaluation of the integral (\ref{3.1}%
), objectively exists and is unique, it has to be only correctly posed. At
the same time, J. Hadamard did mot give corresponding recommendations of
constructive character, and at the present stage of the development of
mathematical physics his methodology turned out to be, in essence,
completely rejected.\footnote{%
The first publication of J. Hadamard on this problem dates back to 1902. One
may suggest that the idea of incorrectness initiated S. Bamach's
investigations in 1920s.}

Let us try to follow the formulation of the problem, inverse of the
evaluation of the integral (\ref{3.1}), that is carried out, in general,
with a certain error: 
\begin{equation}  \label{3.4}
A\psi =f+\delta f^{\prime },\quad x\in \left[ 0,1\right] .
\end{equation}

In the direct formulation, taking into account this error has no principal
importance. Nevertheless, solutions to the Fredholm integral equations of
the first kind (\ref{3.1}) and (\ref{3.4}) can be completely different. At
this point, any quantitative interpretation of $\delta f^{\prime }$ is
senseless. Moreover, even under the condition of the existence of an
analytical solution ($\delta f^{\prime }\equiv 0$), its restoration by means
of a discretization reduces to the solution of ill-posed systems of linear
algebraic equations.

By general considerations, the presence of $\delta f^{\prime }$ in (\ref{3.4}%
) increases the potential of the formulation of the inverse problem, and the
question of a model of the allowed error arises alongside. One must take
into account that the mechanism of its generation is governed by the factor
of smoothing of $\psi \left( x\right) $ by the integration procedure.
Therefore, the structure of $\delta f^{\prime }$ must reflect, in a sense,
informational incompatibility between this function and its mapping $A\psi $
(i.e., between explicit and implicit representations).

By these arguments, we use an operator model of the error in the form 
\begin{equation}  \label{3.5}
\delta _{\bullet }=I-\lambda B,
\end{equation}
where $B$ is an integral operator, $\lambda $ is a parameter that serve for
an adequate approximation of $\delta f^{\prime }$ in the range from zero to
finite distortions related to measurements.

In this way, instead of the Fredholm integral equation of the first kind (%
\ref{3.1}), we propose to solve the following problem: 
\begin{equation}  \label{3.6}
\mu A\psi =\mu f+\delta f;\quad \delta f=0,~x\in \left[ 0,1\right] .
\end{equation}
Here, the parameter $\mu $, like $\lambda $, serves to prevent the Fredholm
integral equation of the second kind, obtained as a result of
transformations, from positioning itself on the spectrum, which is
equivalent to existence and uniqueness of its solution (see, e.g., \cite
{3.34}); $\delta f=\mu \delta f^{\prime }$.

Note that the traditional formulation of the inverse problem in the form of
the Fredholm integral equation of the first kind (\ref{3.1}) is by no means
violated, because we have only added a function representing zero in $%
L_2\left( 0,1\right) $ to its free term. At the same time, the
transformation of the incorrect problem (\ref{3.1}) into the formulation (%
\ref{3.6}) creates conditions for a radical change of the situation. Indeed,
we can demand, generally speaking, that $\delta f$ adaptively compensate for
the errors of numerical operations that take $f\left( x\right) $ out of the
space $l_2$. As a result, prospects for a constructive realization of the
operator $A^{-1}$ emerge. For $f+\delta f\in R\left( A\right) $, the
negative factor of the incorrectness of Eq. (\ref{3.1}) is fully neutralized.

As will be shown below, it is reasonable to represent the operator $B$ from (%
\ref{3.5}), when $\delta f=0$ in $L_2\left( 0,1\right) $, as 
\[
B_{\bullet }=\int\limits_{-1}^1h\left( x,\xi \right) _{\bullet }d\xi , 
\]
imposing the condition of closure on the kernel $h\left( x,\xi \right) $,
transformed by means of a linear change of variables. As a result, the
problem (\ref{3.6}) takes the form 
\begin{equation}  \label{3.7}
\psi \left( x\right) =\mu \int\limits_0^1k\left( x,\xi \right) \psi \left(
\xi \right) d\xi +\lambda \int\limits_{-1}^1h\left( x,\xi \right) \psi
\left( \xi \right) d\xi -\mu f\left( x\right) ,
\end{equation}
\begin{equation}  \label{3.8}
\psi \left( x\right) =\lambda \int\limits_{-1}^1h\left( x,\xi \right) \psi
\left( \xi \right) d\xi ,\quad x\in \left[ 0,1\right] .
\end{equation}

It is easy to notice that the condition on $\delta f$, equivalent to Eq. (%
\ref{3.8}), is supposed to be satisfied by means of an extension of $\psi
\left( x\right) $ on $x\in \left[ -1,0\right) $, which is equivalent to the
use of a new unknown function.

There exists a well-known opinion that prospects of obtaining new
substantial results by simple transformation of mathematical relations are
not great. Indeed, by applying to Eqs. (\ref{3.7}), (\ref{3.8}) a
subtraction operation we again obtain the initial problem which is
incorrect. However, first, we are not going to do this, and, second, the
integral equation with the sought function in an explicit form will allow us
to feel a constructive potential.

From this point of view, a ''refusal'' of the well-known demonstration of
smoothing of peculiarities of the sought solution by means of integration of
(\ref{3.1}) seems to be very significant. Indeed, assuming that the function 
$\psi =\psi _{*}\left( x\right) $ satisfying the system of equations (\ref
{3.7}), (\ref{3.8}) is known, we give it a perturbation of the type $%
\epsilon \sin \left( n\pi x\right) $. A substitution into (\ref{3.6}) shows
that this perturbation influences the free term $f\left( x\right) $ both via
a reduction coefficient $n^{-1}$ and without it, at the expense of
corresponding integration and an explicit presence of $\psi \left( x\right) $%
.

What is said does not apply to $\psi \left( x\right) $, $x\in \left[
-1,0\right) $. However, the determination of this function is beyond the
scope of the considered problem. We want to emphasize that the above
arguments are exclusively heuristic, and, as will be shown, the fulfillment
of the condition $\delta f=0$ is related to some rather subtle points.

In conclusion of this section, we want to point out the inconsistency of the
wide-spread opinion that the formulation of problems of numerical simulation
should be left to specialists in applied sciences, whereas pure
mathematicians should be concerned exclusively with rigorous analytical
investigations, the development of computational methods and participation
in their realization.

It seems that specialists in applied sciences should be concerned with the
formulation of direct and, generally, correct problems. The factor of
incorrectness is directly related to the method of the solution. Therefore,
the main concern of pure mathematicians should be a reduction of
formulations of problems describing the considered processes to the
conditions of efficient implementation of Banach's theorem on the inverse
operator.

\chapter{A method of the reduction of problems, traditionally associated
with Fredholm integral equations of the first kind, to Fredholm integral
equations of the second kind}

\section{The structure of the representation of the error}

In light of the arguments of section 4.5, we proceed with the consideration
of the same Fredholm integral equation of the first kind 
\begin{equation}  \label{4.1}
\left( A\psi \right) \left( x\right) \equiv \int\limits_0^1k\left( x,\xi
\right) \psi \left( \xi \right) d\xi =f\left( x\right) ,\quad x\in \left[ 0,1%
\right] ,
\end{equation}
with a closed kernel $k\left( x,\xi \right) $. In the determination of the
right-hand sides by one of methods, including the integration of the
initially given function $\psi \left( x\right) $, an error was made $\delta
f/\mu $, where $\mu $ is a constant. In other words, the property of
correctness is inherent to the equation 
\begin{equation}  \label{4.2}
\mu \left( A\psi \right) \left( x\right) =\mu f\left( x\right) +\left(
\delta f\right) \left( x\right) ,\quad x\in \left[ 0,1\right]
\end{equation}
in its natural space $l_2$.

Following (\ref{3.5}), we present the error as a difference between the
sought function and the integral component 
\begin{equation}  \label{4.3}
\left( \delta f\right) \left( x\right) =\psi \left( x\right) -\lambda \left(
B\psi \right) \left( x\right) ,\quad x\in \left[ 0,1\right] ,
\end{equation}
where $\lambda $ is a constant, by setting\footnote{%
The use of an indefinite integral in Eq. (5.4) will be explained below.} 
\begin{equation}  \label{4.4}
B_{\bullet }=\int\limits_{-1}^1h\left( x,\xi \right) _{\bullet }d\xi .
\end{equation}

The substitution of $\delta f$ from (\ref{4.3}) into (\ref{4.2}) leads to
the equation 
\begin{equation}  \label{4.5}
\psi \left( x\right) =\mu \left( A\psi \right) \left( x\right) +\lambda
\left( B\psi \right) \left( x\right) -\mu f\left( x\right) ,\quad x\in \left[
0,1\right]
\end{equation}
with two unknown functions: $\psi \left( x\right) $ on $x\in \left[ 0,1%
\right] $ and on $x\in \left[ -1,0\right) $. As the second equation of the
system, equation (\ref{4.3}) could be used, however, the function $\left(
\delta f\right) \left( x\right) $ requires a specification.

If exclusively the loss of information is implied, resulting from the
evaluation of $f\left( x\right) $ by means of the procedure of integration
of (\ref{4.1}), it is logical to assume that 
\begin{equation}  \label{4.6}
\left( \delta f\right) \left( x\right) =0,\quad x\in \left[ 0,1\right] ,
\end{equation}
or 
\begin{equation}  \label{4.7}
\left\| \delta f\right\| _{L_2\left( 0,1\right) }=0.
\end{equation}

In this case, a modeling of the error occurs as a result of mutual
inadequacy of the components of the right-hand side of (\ref{4.3}) of a
qualitative character (in the first place, with respect to smoothness). It
should be also noted that, except for the trivial case $f\in l_{2}$, the
substitution (\ref{4.1}) has no mathematical sense, and the problem can be
considered only from the point of view of a search for an approximate
solution: 
\begin{equation}
\min_{\psi }\left\| A\psi -f\right\| _{L_{2}\left( 0,1\right) }.  \label{4.8}
\end{equation}
Taking into account (\ref{4.6}), equation (\ref{4.3}) can be given the form 
\begin{equation}
\psi \left( x\right) =\lambda \left( B\psi \right) \left( x\right) ,\quad
x\in \left[ 0,1\right] ,  \label{4.9}
\end{equation}
or 
\begin{equation}
\int\limits_{-1}^{0}h\left( x,\xi \right) \psi \left( \xi \right) d\xi
=g\left( x\right) ,\quad x\in \left[ 0,1\right] ,  \label{4.10}
\end{equation}
where 
\begin{equation}
g\left( x\right) =\frac{1}{\lambda }\psi \left( x\right)
-\int\limits_{0}^{1}h\left( x,\xi \right) \psi \left( \xi \right) d\xi .
\label{4.11}
\end{equation}

Making in (\ref{4.10}) a change of variables $\zeta =2\pi x-\pi $; $\theta
=2\pi \xi +\pi $, we get 
\begin{equation}  \label{4.12}
\int\limits_{-\pi }^\pi h\left( \zeta ,\theta \right) \psi \left( \theta
\right) d\theta =g\left( \zeta \right) ,\quad \zeta \in \left[ -\pi ,\pi %
\right] .
\end{equation}

As is obvious, the satisfaction of (\ref{4.9}) by $\psi \left( x\right) $ on 
$x\in \left[ -1,0\right) $ is equivalent to the solvability of Eq. (\ref
{4.12}). Let the kernel $h\left( \zeta ,\theta \right) $ be closed and $%
g\left( \zeta \right) \in L_2\left( -\pi ,\pi \right) $.\footnote{%
Under these conditions, by Eq. (5.11), $\psi \left( x\right) \in L_2\left(
0,1\right) $, which is, by the way, implied.} Then Eq. (\ref{4.12}) is a
Fredholm integral equation of the first kind, whose solution, if it exists,
is unique (see \cite{4.1}). To simplify the following transformations, it is
reasonable to assume that its kernel is symmetric, depends on the difference
of the arguments and satisfies the condition of periodicity on $\zeta \in 
\left[ -\pi ,\pi \right] $. For example, 
\begin{equation}  \label{4.13}
h\left( \zeta ,\theta \right) =\frac{1-r^2}{2\pi \left[ 1-2r\cos \left(
\zeta -\theta \right) +r^2\right] },\quad 0<\left| r\right| <1.
\end{equation}

In this case, equation (\ref{4.12}) takes the form of the Poisson integral 
\cite{4.2}. Correspondingly, in (\ref{4.4}), 
\begin{equation}  \label{4.14}
h\left( x,\xi \right) =\frac{1-r^2}{1-2r\cos \left[ 2\pi \left( x-\xi
\right) \right] +r^2}.
\end{equation}

By Picard's theorem, the necessary condition of convergence to the solution
of Eq. (\ref{4.12}) in $L_2\left( -\pi ,\pi \right) $ of the series 
\begin{equation}  \label{4.15}
\sum_{n=0}^\infty \alpha _n\lambda _n\bar \psi _n\left( \zeta \right)
\end{equation}
is 
\begin{equation}  \label{4.16}
\sum_{n=0}^\infty \alpha _n^2\lambda _n^2<\infty ,\quad \alpha
_n=\int\limits_{-\pi }^\pi g\left( \zeta \right) \bar \psi _n\left( \zeta
\right) d\zeta .
\end{equation}

Here, the characteristic numbers and the eigenfunctions of the kernel (\ref
{4.13}) have the form,respectively [\cite{4.1}]: 
\[
\lambda _n=r^{-n},\quad n=0,1,2,\ldots ; 
\]
\[
\bar \psi _0\left( \zeta \right) =1/\sqrt{2},\quad \bar \psi _n\left( \zeta
\right) =\left\{ \cos \left( n\zeta \right) ;\sin \left( n\zeta \right)
\right\} ,\quad n=1,2,\ldots . 
\]

However, because of (\ref{4.11}), the free term $g\left( \zeta \right) $
and, hence, the coefficients $\alpha _n$, depend on the sought function $%
\psi \left( x\right) $. Therefore, it is impossible to verify the
fulfillment of the condition (\ref{4.16}). From this point of view, E.
Goursat's remark on Picard's theorem (\cite{4.3}, pages 141-143) is rather
appropriate. It consists in the following.

The kernel $h\left( \zeta ,\theta \right) $ is assumed to be closed, whereas
the function $g\left( \zeta \right) $ is such that the condition (\ref{4.16}%
) is not fulfilled. Nevertheless, there exists a function $\psi \left( \zeta
\right) $, represented by the convergent series (\ref{4.15}), whose
substitution assures the fulfillment of the Fredholm integral equation of
the first kind (\ref{4.12}) in $L_2\left( -\pi ,\pi \right) $.

A proof is based on the fact that 
\begin{equation}  \label{4.17}
g_n\left( \zeta \right) =\int\limits_{-\pi }^\pi h\left( \zeta ,\theta
\right) \psi _n\left( \theta \right) d\theta ,
\end{equation}
where 
\begin{equation}  \label{4.18}
\psi _n\left( \theta \right) =\sum_{i=0}^n\alpha _i\lambda _i\bar \psi
_i\left( \theta \right) ,
\end{equation}
coincides with the sum of the first $n$ terms of the Fourier expansion of
the function $g\left( \zeta \right) $ in elements $\bar \psi _i\left( \zeta
\right) $.

It should be noted that by Mercer's theorem \cite{4.4} 
\begin{equation}  \label{4.19}
h\left( \zeta ,\theta \right) =\sum_{m=0}^\infty \frac{\bar \psi _m\left(
\zeta \right) \bar \psi _m\left( \theta \right) }{\lambda _m}.
\end{equation}

Because of this fact, $n$ can be chosen such that the integral 
\begin{equation}  \label{4.20}
\int\limits_{-\pi }^\pi \left[ g\left( \zeta \right) -g_n\left( \zeta
\right) \right] ^2d\zeta
\end{equation}
will be smaller than an infinitesimal $\epsilon >0$.

It seems that this result can be considered as a specific addition to the
Riesz-Fischer theorem with preceding Weyl's lemma \cite{4.4}. According to
the classification \cite{4.5}, the series (\ref{4.18}), by (\ref{4.17}),
converges to the solution of (\ref{4.12}) in the sense of a distribution.

If the condition (\ref{4.16}) is not fulfilled, the series (\ref{4.15})
diverges in $L_2$.\footnote{%
This is not equivalent to the conclusion $\psi \left( x\right) \notin
L_2\left( -1,0\right) $.} In this regard, it is interesting to establish
restrictions on the data of (\ref{4.1}) to satisfy (\ref{4.16}). First, one
has to make Eq. (\ref{4.1}) symmetric by multiplying by the conjugate
operator 
\[
A_{\bullet }^{*}=\int\limits_0^1k\left( \xi ,x\right) _{\bullet }d\xi . 
\]

We get: 
\begin{equation}  \label{4.21}
\int\limits_0^1k^{\prime }\left( x,\xi \right) \psi \left( \xi \right) d\xi
=f^{\prime }\left( x\right) ,\quad x\in \left[ 0,1\right] .
\end{equation}

Here, 
\[
k^{\prime }\left( x,\xi \right) =\int\limits_0^1k\left( \zeta ,x\right)
k\left( \zeta ,\xi \right) d\zeta ;\quad f^{\prime }\left( x\right)
=\int\limits_0^1k\left( \xi ,x\right) f\left( \xi \right) d\xi , 
\]
and, correspondingly, $k^{\prime }\left( x,\xi \right) =k^{\prime }\left(
\xi ,x\right) $.

Suppose that the function $f^{\prime }\left( x\right) $ is such that the
solution of the Fredholm integral equation of the first kind (\ref{4.21}) is
a correctly formulated problem in the space $l_2$. Then 
\begin{equation}  \label{4.22}
\sum_{n=0}^\infty \alpha _n^{\prime 2}\lambda _n^{\prime 2}<\infty ,\quad
\alpha _n^{\prime }=\int\limits_0^1f^{\prime }\left( x\right) \bar \psi
_n^{\prime }\left( x\right) dx,
\end{equation}
where $\lambda _n^{\prime }$ and $\bar \psi _n^{\prime }\left( x\right) $
are the characteristic numbers and the eigenfunctions of the kernel $%
k^{\prime }\left( x,\xi \right) $.

In this case, by Picard's theorem, the solution to (\ref{4.21}) is
determined by the series 
\begin{equation}  \label{4.23}
\psi \left( x\right) =\sum_{n=0}^\infty \alpha _n^{\prime }\lambda
_n^{\prime }\bar \psi _n^{\prime }\left( x\right)
\end{equation}
that converges in $L_2\left( 0,1\right) $. Correspondingly, in (\ref{4.12}), 
\[
g\left( \zeta \right) =\sum_{n=0}^\infty \alpha _n^{\prime }\lambda
_n^{\prime }\left[ \frac 1\lambda \bar \psi _n^{\prime }\left( \zeta \right)
-\int\limits_{-\pi }^\pi h\left( \zeta ,\theta \right) \bar \psi _n^{\prime
}\left( \theta \right) d\theta \right] , 
\]
and, taking account of (\ref{4.19}), 
\[
\alpha _0=\frac{1-2\pi }\lambda \sum_{m=0}^\infty \alpha _m^{\prime }\lambda
_m^{\prime }\int\limits_{-\pi }^\pi \bar \psi _m^{\prime }\left( \zeta
\right) d\zeta ; 
\]
\begin{equation}  \label{4.24}
\alpha _n=\frac{\lambda _n-\pi }{\lambda _n}\sum_{m=0}^\infty \alpha
_m^{\prime }\lambda _m^{\prime }\int\limits_{-\pi }^\pi \bar \psi _m^{\prime
}\left( \zeta \right) \psi _n\left( \zeta \right) d\zeta ,\quad n=1,2,\ldots
.
\end{equation}

Thus, the use of the additional condition (\ref{4.9}) implies that not only
the condition (\ref{4.22}) but also (\ref{4.16}) are fulfilled with the
coefficients (\ref{4.24}).

However, in the framework of the formulation (\ref{4.8}), it is more correct
to assume that the stated requirements are not fulfilled in principle.
Correspondingly, the series (\ref{4.18}) turn out to be divergent for $%
n\rightarrow \infty $.

At the same time, if the value $n$, although arbitrarily large, is still
finite, the situation changes diametrically. Indeed, the conditions of the
solvability of the Fredholm integral equations of the first kind (\ref{4.1})
and (\ref{4.21}) are fulfilled automatically; the series (\ref{4.18}) is not
merely convergent in the sense of (\ref{4.20}), but is a $L_2$-function.

We emphasize that these are only heuristic arguments that do not allow us to
claim that $\psi \left( x\right) \in L_2\left( -1,0\right) $. For what
follows, this condition is necessary. Therefore, we turn below to the
modeling of the error from a somewhat different position.

In the case of an experimental determination, the right-hand side of (\ref
{4.1}), as a rule, differs from the expression for $f_{*}\left( x\right) $,
obtained as a result of hypothetically exact integration of the given
function $\psi _{*}\left( x\right) $, by a certain quantity $\left( \Delta
f\right) \left( x\right) $. We only have to require that this quantity be
bounded, otherwise, on the one hand, the adequacy of the actual mathematical
model is violated, and, on the other hand, a following reduction to the
Fredholm integral equation of the first kind loses a practical sense.

In general, compared to $\delta f$, the error $\Delta f$ is not of principle
importance. Indeed, if 
\[
\left\| \Delta f\right\| \ll \left\| f\right\| ,~L_2\left( 0,1\right) , 
\]
under the condition of the construction of a stable algorithm of the
numerical realization, the error of the sought solution $\psi \left(
x\right) $ turns out to be comparatively small. If this condition is not
fulfilled, both the formulation of the concrete considered problem and its
solution will be merely distorted.

If we assume in (\ref{4.4}) 
\begin{equation}  \label{4.25}
B_{\bullet }=\int\limits_{-1}^xh\left( x,\xi \right) _{\bullet }d\xi ,
\end{equation}
this fact will influence the modeling of the error via a modification of the
expression (\ref{4.11}): 
\[
g\left( x\right) =\frac 1\lambda \psi \left( x\right)
-\int\limits_0^xh\left( x,\xi \right) \psi \left( \xi \right) d\xi . 
\]

The following method of the representation of the error can also be
considered: 
\begin{equation}  \label{4.26}
\left( \delta f\right) \left( x\right) =\psi \left( x\right) -\lambda \left(
B\psi \right) \left( x\right) -\sigma \left( x\right) ,\quad x\in \left[ 0,1%
\right] ,
\end{equation}
where $\sigma \left( x\right) $ is another unknown function; the operator is
given by 
\[
B_{\bullet }=\int\limits_0^xh\left( x,\xi \right) _{\bullet }d\xi . 
\]

It is obvious that $\sigma \left( x\right) \neq 0$, because otherwise the
condition (\ref{4.6}) for the $L_2$-kernel $h\left( x,\xi \right) $ reduces
to the homogeneous Volterra integral equation of the second kind 
\[
\psi \left( x\right) =\int\limits_0^xh\left( x,\xi \right) \psi \left( \xi
\right) d\xi ,\quad x\in \left[ 0,1\right] , 
\]
whose solution is trivial.

\section{A transformed formulation of the problem}

Let us turn to the model of the error given by (\ref{4.3}), (\ref{4.4}) and (%
\ref{4.6}). In order to use more completely the possibilities of the
techniques of integral equations, we introduce new unknown functions $%
\varphi \left( x\right) $ and $\chi \left( x\right) $ by 
\begin{equation}  \label{4.27}
\Psi \left( x\right) =\lambda \left( B\Psi \right) \left( x\right) +\left\{ 
\begin{array}{c}
\chi \left( x\right) ,\quad x\in \left[ 0,1\right] ;~ \\ 
0,\quad x\in \left[ -1,0\right) .
\end{array}
\right.
\end{equation}

Here, 
\[
\Psi \left( x\right) =\left\{ 
\begin{array}{c}
\psi \left( x\right) ,\quad x\in \left[ 0,1\right] ;~ \\ 
\varphi \left( x\right) ,\quad x\in \left[ -1,0\right) ,
\end{array}
\right. 
\]
where $\psi \left( x\right) $ is the solution of the problem in the sense of
(\ref{4.8}); $\varphi \left( x\right) $ is determined via $\psi \left(
x\right) $ from the Fredholm integral equation of the second kind 
\[
\varphi \left( x\right) =\lambda \int\limits_{-1}^0h\left( x,\xi \right)
\varphi \left( \xi \right) d\xi +p\left( x\right) ,\quad x\in \left[
-1,0\right) , 
\]
with the free term 
\[
p\left( x\right) =\lambda \int\limits_0^1h\left( x,\xi \right) \psi \left(
\xi \right) d\xi 
\]
and the characteristic numbers $\lambda _n=r^{-n}$, $n=0,1,2,\ldots $ (see a
transition to the Poisson integral in section 5.1).

In other words, $\chi \left( x\right) $ is a combination of the
hypothetically known solution to the considered problem and of the function $%
\varphi \left( x\right) $, by means of which $\Psi \left( x\right) $
satisfies an equation analogous to (\ref{4.9}), however not in the mean but
uniformly on $x\in \left[ -1,0\right) $. In this case, equation (\ref{4.27})
is a Fredholm integral equation of the second kind with respect to $\Psi
\left( x\right) $.

Taking into account (\ref{4.1}), on the basis of (\ref{4.27}), we can obtain
an equation 
\begin{equation}  \label{4.28}
\Psi \left( x\right) =\lambda \left( B\Psi \right) \left( x\right) +\left\{ 
\begin{array}{c}
\mu \left( A\Psi \right) \left( x\right) -\mu f\left( x\right) +\chi \left(
x\right) ,\quad x\in \left[ 0,1\right] ;~ \\ 
0,\quad x\in \left[ -1,0\right) ,
\end{array}
\right.
\end{equation}
which, by (\ref{4.8}), strictly speaking, is neither integral nor of the
Fredholm second kind type. At the same time, this objection can be easily
overcome, if, analogously to (\ref{4.2}), we subtract $\mu \left( \delta
f\right) \left( x\right) $ in the right-hand side of (\ref{4.28}).

By the well-known stability of the properties of Fredholm integral equations
of the second kind against small perturbations \cite{4.6}, the fulfillment
of the condition (\ref{4.7}) is fully sufficient. Thus, the actual absence
of $\delta f$ in (\ref{4.28}) will not substantially influence the results
of subsequent transformations with the use of this equation.

Using the solution of the Fredholm integral equation of the second kind 
\begin{equation}  \label{4.29}
\Psi \left( x\right) =\lambda \left( B\Psi \right) \left( x\right) +\left\{ 
\begin{array}{c}
\mu \left( A\Psi \right) \left( x\right) +\chi \left( x\right) ,\quad x\in 
\left[ 0,1\right] ;~ \\ 
0,\quad x\in \left[ -1,0\right)
\end{array}
\right.
\end{equation}
on $x\in \left[ -1,0\right) $ that will be denoted as $\varphi _0\left(
x\right) $, we can eliminate the function $\chi \left( x\right) $ in (\ref
{4.27}).

As a result, we get an equation of the type 
\begin{equation}  \label{4.30}
\Psi ^{\prime }\left( x\right) =\lambda \left( B\Psi ^{\prime }\right)
\left( x\right) +\left\{ 
\begin{array}{c}
0,\quad x\in \left[ 0,1\right] ;~ \\ 
\kappa \left( x\right) ,\quad x\in \left[ -1,0\right) .
\end{array}
\right.
\end{equation}

Here, 
\begin{equation}  \label{4.31}
\Psi ^{\prime }\left( x\right) =\left\{ 
\begin{array}{c}
\psi \left( x\right) ,\quad x\in \left[ 0,1\right] ;~ \\ 
\varphi ^{\prime }\left( x\right) ,\quad x\in \left[ -1,0\right) ;
\end{array}
\right.
\end{equation}
with 
\[
\varphi ^{\prime }\left( x\right) =\varphi \left( x\right) -\varphi _0\left(
x\right) , 
\]
and $\kappa \left( x\right) $ being a new unknown $L_2$-function.\footnote{%
This is determined, in particular, by the type of Eqs. (5.27)-(5.29).}

In this case, $\kappa \neq 0$: otherwise, the solution to Eq. (\ref{4.30})
would be trivial. One can note that a relationship between Eqs. (\ref{4.30})
and (\ref{4.27}), (\ref{4.28}) is realized by means of a procedure of a
specific ''leakage'' of the function $\chi \left( x\right) $ into $\kappa
\left( x\right) $ via their solutions on $x\in \left[ -1,0\right) $.

Thus, starting from the traditional formulation (\ref{4.1}), we have
proceeded to the Fredholm integral equations of the second kind (\ref{4.27}%
), (\ref{4.28}) and (\ref{4.30}). In light of the general methodological
considerations of sections 2-4, these equations reveal powerful potential of
a constructive orientation.

Indeed, the structure of the free terms is such that any solution on a part
of the domain [$x\in \left[ 0,1\right] $ or $x\in \left[ -1,0\right) $]
contains explicitly new unknown variables $\chi \left( x\right) $ and $%
\kappa \left( x\right) $. Simultaneously, in the expressions for the
solutions on the remaining part, only integral terms are present.\footnote{%
We must take into account the form of the solution of the Fredholm integral
equation of the second kind by means of the resolvent (see, e.g., Ref. [7]).}
This makes a premise for an analogous representation of $\delta f$, which
is, in fact, a strategic objective of the proposed method.

Returning to the questions of section 5.1, note that, on the basis of (\ref
{4.30}), we can make an immediate extension of Eqs. (\ref{4.5}), (\ref{4.9}%
): 
\[
\psi \left( x\right) =\lambda \left( B\psi \right) \left( x\right) +\left\{ 
\begin{array}{c}
\mu \left( A\psi \right) \left( x\right) -\mu q\left( x\right) ,\quad x\in 
\left[ 0,1\right] ;~ \\ 
\kappa \left( x\right) ,\quad x\in \left[ -1,0\right) ;
\end{array}
\right. 
\]
\begin{equation}  \label{4.32}
\psi \left( x\right) =\lambda \left( B\psi \right) \left( x\right) +\left\{ 
\begin{array}{c}
0,\quad x\in \left[ 0,1\right] ;~ \\ 
\kappa \left( x\right) ,\quad x\in \left[ -1,0\right) .
\end{array}
\right.
\end{equation}

In this case, in fact, 
\[
\varphi ^{\prime }\left( x\right) \equiv \psi \left( x\right) \in L_2\left(
-1,0\right) . 
\]

In what follows, we will need the inverse of the operator $I-\lambda B$, or,
taking into account the change of the variables $\zeta =\pi x$, $\theta =\pi
x$, 
\begin{equation}  \label{4.33}
I-\lambda \int\limits_{-\pi }^\pi h\left( \zeta ,\theta \right) _{\bullet
}d\theta ,\quad \zeta \in \left[ -\pi ,\pi \right] ,
\end{equation}
where 
\[
h\left( \zeta ,\theta \right) =\frac{1-r^2}{\pi \left\{ 1-2r\cos \left[
2\left( \zeta -\theta \right) \right] +r^2\right\} }. 
\]

This kernel depends on the difference of the arguments, the function $%
h\left( \zeta \right) $ is periodic on $x\in \left[ -\pi ,\pi \right] $ .
Correspondingly, the eigenfunctions and the characteristic numbers of the
operator (\ref{4.33}) are \cite{4.1}: 
\[
\bar \psi _0\left( \zeta \right) =1/\sqrt{2},\quad \bar \psi _n\left( \zeta
\right) =\left\{ \cos \left( n\zeta \right) ;\sin \left( n\zeta \right)
\right\} ,\quad n=1,2,\ldots ; 
\]
\[
\lambda _n^{-1}\left( r\right) =\int\limits_{-\pi }^\pi h\left( \zeta
\right) \cos \left( n\zeta \right) d\zeta =\frac{1-r^2}\pi \int\limits_{-\pi
}^\pi \frac{\cos \left( n\zeta \right) d\zeta }{1-2r\cos \left( 2\zeta
\right) +r^2} 
\]
\[
=\frac{1-r^2}{2\pi }\int\limits_{-2\pi }^{2\pi }\frac{\cos \left( \frac
12n\zeta \right) d\zeta }{1-2r\cos \left( \zeta \right) +r^2}=\frac{1-r^2}%
\pi \int\limits_0^{2\pi }\frac{\cos \left( \frac 12n\zeta \right) d\zeta }{%
1-2r\cos \left( \zeta \right) +r^2}, 
\]
\[
n=1,2,\ldots . 
\]

It is known (\cite{4.2}, p. 161) that 
\[
\frac{1-r^2}{2\left[ 1-2r\cos \left( \zeta \right) +r^2\right] }=\frac
12+\sum_{m=1}^\infty r^m\cos \left( m\zeta \right) , 
\]
hence 
\[
\lambda _n^{-1}\left( r\right) =\frac 2\pi \int\limits_0^{2\pi }\left[ \frac
12+\sum_{m=1}^\infty r^m\cos \left( m\zeta \right) \right] \cos \left( \frac
12n\zeta \right) d\zeta ,\quad n=1,2,\ldots . 
\]

As a result, $\lambda _n=0.5r^{-n}$, $n=0,1,2,\ldots $, and the resolvent of
the considered operator is \cite{4.4} 
\begin{equation}  \label{4.34}
H\left( x,\xi ,\lambda \right) =\sum_{n=0}^\infty \frac{\bar \psi _n\left(
x\right) \bar \psi _n\left( \xi \right) }{\lambda _n-\lambda }%
=2\sum_{n=0}^\infty \frac{r^n}{1-2\lambda r^n}\bar \psi _n\left( x\right)
\bar \psi _n\left( \xi \right) ,
\end{equation}
where 
\[
\bar \psi _0\left( x\right) =1/\sqrt{2},\quad \bar \psi _n\left( x\right)
=\left\{ \cos \left( 2n\pi x\right) ;\sin \left( 2n\pi x\right) \right\}
,\quad n=1,2,\ldots . 
\]

\section{A constructive algorithm of practical realization}

By inverting the operator $I-\lambda B$ in (\ref{4.27}), (\ref{4.29}), we
get, respectively, 
\begin{equation}  \label{4.35}
\varphi \left( x\right) =\lambda \int\limits_0^1H\left( x,\xi ,\lambda
\right) \chi \left( \xi \right) d\xi ;
\end{equation}
\begin{equation}  \label{4.36}
\psi \left( x\right) =\chi \left( x\right) +\lambda \int\limits_0^1H\left(
x,\xi ,\lambda \right) \chi \left( \xi \right) d\xi ,\quad x\in \left[ 0,1%
\right]
\end{equation}
and 
\begin{equation}  \label{4.37}
\varphi _0\left( x\right) =\mu \lambda \int\limits_0^1l\left( x,\xi \right)
\varphi _0\left( \xi \right) d\xi +\lambda \int\limits_0^1H\left( x,\xi
,\lambda \right) \chi \left( \xi \right) d\xi ,\quad x\in \left[ -1,0\right)
,
\end{equation}
where 
\begin{equation}  \label{4.38}
l\left( x,\xi \right) =\int\limits_0^1H\left( x,\zeta ,\lambda \right)
k\left( \zeta ,\xi \right) d\zeta .
\end{equation}

From Eq. (\ref{4.30}), 
\[
\kappa \left( x\right) =\psi ^{\prime }\left( x\right) -\lambda
\int\limits_{-1}^0h\left( x,\xi \right) \psi ^{\prime }\left( \xi \right)
d\xi -\lambda \int\limits_0^1h\left( x,\xi \right) \psi \left( \xi \right)
d\xi . 
\]

Upon substitution of the expressions (\ref{4.31}) and (\ref{4.36}), taking
into account (\ref{4.35}) and the functional relation \cite{4.7} 
\begin{equation}  \label{4.39}
\lambda \int\limits_{-1}^1h\left( x,\zeta \right) H\left( \zeta ,\xi
,\lambda \right) d\zeta =H\left( x,\xi ,\lambda \right) -h\left( x,\xi
\right) ,
\end{equation}
we get 
\begin{equation}  \label{4.40}
\kappa \left( x\right) =\lambda \int\limits_{-1}^0h\left( x,\xi \right)
\varphi _0\left( \xi \right) d\xi -\varphi _0\left( x\right) .
\end{equation}

As the basis for the construction of the system, i.e., of one more Fredholm
integral equation of the second kind in addition to (\ref{4.37}), we will
use (\ref{4.28}): 
\[
\chi \left( x\right) =\psi \left( x\right) -\lambda
\int\limits_{-1}^0h\left( x,\xi \right) \varphi \left( \xi \right) d\xi
-\lambda \int\limits_0^1h\left( x,\xi \right) \psi \left( \xi \right) d\xi 
\]
\begin{equation}  \label{4.41}
-\mu \int\limits_0^1k\left( x,\xi \right) \psi \left( \xi \right) d\xi +\mu
f\left( x\right) ,\quad x\in \left[ 0,1\right] .
\end{equation}

It is obvious, that the explicitly appearing function $\psi $$\left(
x\right) $ cannot be represented with the help of (\ref{4.36}), because $%
\chi \left( x\right) $ in the right-hand side cancels out.\footnote{%
In this way, we would trivially return to the initial object (5.1).}
Therefore, we employ the solution of Eq. (\ref{4.30}) 
\begin{equation}  \label{4.42}
\psi \left( x\right) =\lambda \int\limits_{-1}^0H\left( x,\xi ,\lambda
\right) \kappa \left( \xi \right) d\xi ,\quad x\in \left[ 0,1\right] .
\end{equation}

Taking into account (\ref{4.40}), (\ref{4.39}), the property of
orthogonality 
\[
\int\limits_{-1}^0\bar \psi _n\left( x\right) \bar \psi _m\left( x\right)
dx=\int\limits_0^1\bar \psi _n\left( x\right) \bar \psi _m\left( x\right)
dx=\frac 12\int\limits_{-1}^1\bar \psi _n\left( x\right) \bar \psi _m\left(
x\right) dx 
\]
\begin{equation}  \label{4.43}
=\left\{ 
\begin{array}{c}
0,\quad n\neq m; \\ 
\frac 12,\quad n=m,
\end{array}
\right.
\end{equation}
that follows from the fact that the kernel (\ref{4.14}) is periodic not only
on $x\in \left[ -1,1\right] $ but also on $x\in \left[ -1,0\right) $, $x\in 
\left[ 0,1\right] $, and, besides, using 
\[
\int\limits_0^1h\left( x,\zeta \right) H\left( \zeta ,\xi ,\lambda \right)
d\zeta =\int\limits_0^1H\left( x,\xi ,\lambda \right) h\left( \zeta ,\xi
\right) d\zeta 
\]
that follows from (\ref{4.19}), (\ref{4.34}), we find 
\begin{equation}  \label{4.44}
\psi \left( x\right) =-\frac 12\lambda \int\limits_{-1}^0\left[ h\left(
x,\xi \right) +H\left( x,\xi ,\lambda \right) \right] \varphi _0\left( \xi
\right) d\xi ,\quad x\in \left[ 0,1\right] .
\end{equation}

To determine the other components of the right-hand side of (\ref{4.41}), we
can employ the expression (\ref{4.36}). Substitution of (\ref{4.35}), (\ref
{4.36}) and (\ref{4.44}) into (\ref{4.41}), taking into account (\ref{4.39}%
), leads to the equation 
\[
\chi \left( x\right) =-\int\limits_0^1\left\{ \lambda H\left( x,\xi ,\lambda
\right) +\mu \left[ k\left( x,\xi \right) +\int\limits_0^1k\left( x,\zeta
\right) H\left( \zeta ,\xi ,\lambda \right) d\zeta \right] \right\} \chi
\left( \xi \right) d\xi 
\]
\begin{equation}  \label{4.45}
-\frac 12\lambda \int\limits_{-1}^0\left[ h\left( x,\xi \right) +H\left(
x,\xi ,\lambda \right) \right] \varphi _0\left( \xi \right) d\xi +\mu
f\left( x\right) ,\quad x\in \left[ 0,1\right] .
\end{equation}

Note that on the basis of (\ref{4.34}), in (\ref{4.38}) and (\ref{4.45}), 
\[
\int\limits_0^1H\left( x,\zeta ,\lambda \right) k\left( \zeta ,\xi \right)
d\zeta =2\sum_{n=0}^\infty \frac{r^nc_n^{\prime }\left( \xi \right) }{%
1-2\lambda r^n}\bar \psi _n\left( x\right) ; 
\]
\[
\int\limits_0^1k\left( x,\zeta \right) H\left( \zeta ,\xi ,\lambda \right)
d\zeta =2\sum_{n=0}^\infty \frac{r^nc_n^{\prime \prime }\left( x\right) }{%
1-2\lambda r^n}\bar \psi _n\left( \xi \right) , 
\]
where 
\[
c_n^{\prime }\left( \xi \right) =\int\limits_0^1k\left( \zeta ,\xi \right)
\bar \psi _n\left( \zeta \right) d\zeta ;\quad c_n^{\prime \prime }\left(
x\right) =\int\limits_0^1k\left( x,\zeta \right) \bar \psi _n\left( \zeta
\right) d\zeta . 
\]

As a result of symmetry and periodicity $H\left( x,\xi ,\lambda \right) $, $%
x\in \left[ 0,1\right] $, the operator 
\[
I+\lambda \int\limits_0^1H\left( x,\xi ,\lambda \right) _{\bullet }d\zeta 
\]
can be inverted analytically (by analogy with the above-described procedure,
one can construct the corresponding resolvent). At the same time, the
realization of this procedure would not yield a considerable simplification
of Eq. (\ref{4.45}).

The inversion in (\ref{4.37}) of the operator 
\[
I-\mu \lambda \int\limits_0^1I\left( x,\xi \right) _{\bullet }d\zeta 
\]
for $\mu \neq \mu _n$, where $\mu _n$ are its characteristic numbers, leads
to the equation 
\begin{equation}  \label{4.46}
\varphi _0\left( x\right) =\lambda \int\limits_0^1\left[ H\left( x,\xi
,\lambda \right) +\mu \int\limits_0^1L\left( x,\zeta ,\mu \right) H\left(
\zeta ,\xi ,\lambda \right) d\zeta \right] \chi \left( \xi \right) d\xi .
\end{equation}

Here, $L\left( x,\xi ,\mu \right) $ is the resolvent of the kernel (\ref
{4.38}).

In practice, for its construction, the solution can be represented as a
series 
\[
\varphi _0\left( x\right) =\sum_{n=0}^\infty \alpha _n\bar \psi _n\left(
x\right) , 
\]
where $\alpha _n$ are constants to be determined from a system of linear
algebraic equations, obtained by a reduction of factors of the elements $%
\bar \psi _n\left( x\right) $. In this regard, we mention also a method of
numerical realization of the resolvent developed by S. G. Mikhlin (\cite{4.8}%
, section 12).

Substitution of the expression (\ref{4.46}) into (\ref{4.45}) allows us to
reduce the considered problem to the solution of the Fredholm integral
equation of the second kind with respect to $\chi \left( x\right) $.

In particular, in order to give the system of the Fredholm integral
equations of the second kind (\ref{4.37}) and (\ref{4.45}) a canonical form,
we set 
\begin{equation}  \label{4.47}
\lambda \equiv \mu \neq \left\{ 0.5r^{-n};r^{-n}\right\} ,\quad 0<\left|
r\right| <1;\quad n=0,1,2,\ldots .
\end{equation}

We also introduce the notation 
\begin{equation}  \label{4.48}
\chi \left( x\right) \equiv \chi _1\left( x\right) ,\quad \varphi _0\left(
x\right) \equiv \chi _1\left( x\right) ;
\end{equation}
\[
K_{11}\left( x,\xi \right) =-\left[ H\left( x,\xi ,\mu \right) +k\left(
x,\xi \right) +\mu \int\limits_0^1k\left( x,\zeta \right) H\left( \zeta ,\xi
,\mu \right) d\zeta \right] ; 
\]
\begin{equation}  \label{4.49}
K_{12}\left( x,\xi \right) =-\frac 12\lambda \left[ h\left( x,\xi \right)
+H\left( x,\xi ,\mu \right) \right] ;\quad F_1\left( x\right) =\mu f\left(
x\right) ;
\end{equation}
\[
K_{21}\left( x,\xi \right) =H\left( x,\xi ,\mu \right) ;\quad K_{22}\left(
x,\xi \right) =\mu \int\limits_0^1H\left( x,\zeta ,\lambda \right) k\left(
\zeta ,\xi \right) d\zeta . 
\]

By virtue of (\ref{4.43}) and the obvious property 
\[
\bar \psi _n\left( x\right) \equiv \bar \psi _n\left( x+1\right) ,\quad
n=0,1,2,\ldots , 
\]
equations (\ref{4.45}) and (\ref{4.37}) take the form, respectively, 
\[
\chi _1\left( x\right) =\mu \int\limits_0^1\left[ K_{11}\left( x,\xi \right)
\chi _1\left( \xi \right) +K_{12}\left( x,\xi \right) \chi _2\left( \xi
\right) \right] d\xi +F_1\left( x\right) ; 
\]
\begin{equation}  \label{4.50}
\chi _2\left( x\right) =\mu \int\limits_0^1\left[ K_{21}\left( x,\xi \right)
\chi _1\left( \xi \right) +K_{22}\left( x,\xi \right) \chi _2\left( \xi
\right) \right] d\xi ;\quad x\in \left[ 0,1\right] .
\end{equation}

As shown in (\cite{4.4}, pages 195-196), this system is equivalent to the
Fredholm integral equation of the second kind 
\begin{equation}  \label{4.51}
\chi \left( x\right) =\mu \int\limits_0^2K\left( x,\xi \right) \chi \left(
\xi \right) d\xi +F\left( x\right) ,\quad x\in \left[ 0,2\right] ,
\end{equation}
where 
\begin{equation}  \label{4.52}
\chi \left( x\right) =\left\{ 
\begin{array}{c}
\chi _1\left( x\right) ,\quad x\in \left[ 0,1\right] ; \\ 
\chi _2\left( x-1\right) ,\quad x\in \left( 1,2\right] ;
\end{array}
\right.
\end{equation}
\begin{equation}  \label{4.53}
F\left( x\right) =\left\{ 
\begin{array}{c}
F_1\left( x\right) ,\quad x\in \left[ 0,1\right] ; \\ 
0,\quad x\in \left( 1,2\right] ;
\end{array}
\right.
\end{equation}
\begin{equation}  \label{4.54}
K\left( x,\xi \right) =\left\{ 
\begin{array}{c}
K_{11}\left( x,\xi \right) ,\quad x,\xi \in \left[ 0,1\right] ; \\ 
K_{12}\left( x,\xi -1\right) ,\quad x\in \left( 0,1\right) ,\quad \xi \in
\left( 1,2\right] ; \\ 
K_{21}\left( x-1,\xi \right) ,\quad x\in \left( 1,2\right) ,\quad \xi \in 
\left[ 0,1\right) ; \\ 
K_{22}\left( x-1,\xi -1\right) ,\quad x,\xi \in \left[ 1,2\right] .
\end{array}
\right.
\end{equation}

In the case 
\begin{equation}  \label{4.55}
\left[ \mu \right] <\frac 1M,\quad
M^2=\sum_{i,j=1}^2\int\limits_0^1\int\limits_0^1\left| K_{ij}\left( x,\xi
\right) \right| ^2dxd\xi ,
\end{equation}
the solution of Eq. (\ref{4.51}) can be obtained by means of successive
approximations \cite{4.9} 
\begin{equation}  \label{4.56}
\chi _{n+1}\left( x\right) =\mu \int\limits_0^2K\left( x,\xi \right) \chi
_n\left( \xi \right) d\xi +F\left( x\right) ,\quad x\in \left[ 0,2\right] ,
\end{equation}
where $\chi _0\left( x\right) \in L_2\left( 0,2\right) $ is arbitrary.%
\footnote{%
The so-called method of simple iterations.}

If, along with (\ref{4.55}), 
\begin{equation}  \label{4.57}
\int\limits_0^1\left| K_{ij}\left( x,\xi \right) \right| ^2d\xi \leq \alpha
_{ij}=const,
\end{equation}
this convergence is regular. Of course, the parameter $\mu $ cannot coincide
with characteristic numbers of the kernel (\ref{4.54}).

It should be noted that the procedure (\ref{4.56}) is, in fact, equivalent
to the construction of the resolvent \cite{4.7} 
\[
\Gamma \left( x,\xi ,\mu \right) =\sum_{n=1}^\infty \mu ^{n-1}K_n\left(
x,\xi \right) , 
\]
where 
\[
K_1\left( x,\xi \right) =K\left( x,\xi \right) ;\quad K_{m+1}\left( x,\xi
\right) =\int\limits_0^2K_m\left( x,\zeta \right) K_1\left( \zeta ,\xi
\right) d\zeta 
\]
are iterated kernels $K\left( x,\xi \right) $.

Iterations of the type (\ref{4.56}), under the conditions (\ref{4.55}), (\ref
{4.57}), can be directly applied to the system of the Fredholm integral
equations of the second kind (\ref{4.50}). For the determination of the
appropriate value of $\mu $, the means of numerical simulations can be used.
The identification (\ref{4.47}) somewhat restricts its variability.

Note that there are a number of efficient algorithms for the solution of
Fredholm integral equations of the second kind, including the determination
of spectral characteristics and the evaluation of quadratures. (The
literature on this issue is immense \cite{4.10,4.11,4.12,4.13}.) The
fundamentals of general theory of approximate methods of numerical
realization of this type of equations are comprehensively presented in \cite
{4.14}.

It should be emphasized that the structure of the kernel $K\left( x,\xi
\right) $ does not enforce on it any additional features or oscillations
beyond the framework of the initial formulation (\ref{4.1}).\footnote{%
In other words, those that result from internal difficulties of the employed
method.}

Given $\chi _1\left( x\right) $, from Eq. (\ref{4.36}) we get 
\begin{equation}  \label{4.58}
\psi _1\left( x\right) =\chi _1\left( x\right) +\lambda
\int\limits_0^1H\left( x,\xi ,\lambda \right) \chi _1\left( \xi \right) d\xi
.
\end{equation}

The function $\psi _1\left( x\right) $ is the solution of the considered
problem (\ref{4.1}) in the sense of (\ref{4.8}), that is, exactly for $\psi
\left( x\right) \equiv \psi _1\left( x\right) $, $\min \left\| f-f_1\right\|
_{L_2\left( 0,1\right) }$ is achieved, where 
\[
f_1\left( x\right) =\int\limits_0^1k\left( x,\xi \right) \psi _1\left( \xi
\right) d\xi . 
\]

As a matter of fact, we determine in this way a maximal possible
approximation in $L_2$ of the result of integration of (\ref{4.1}) to the
free term $f(x)$, containing some errors (of measurement, approximation,
etc.), that objectively belongs to the space $l_2$. Correspondingly, as an
extension of (\ref{4.8}), the solution of the problem (\ref{4.1}) acquires
the following meaning: 
\begin{equation}  \label{4.59}
\left\| A\psi -f-\delta f\right\| _{L_2\left( 0,1\right) }=0.
\end{equation}

This issue seems to be of principal importance.

\section{The issue of justification and the mechanism of achieved efficiency}

As follows from (\ref{4.50}), by virtue of (\ref{4.1}) and (\ref{4.49}), the
function $\chi \left( x\right) $ satisfying (\ref{4.51}) is also a solution
of the system of homogeneous Fredholm integral equations of the second kind 
\[
\chi _1\left( x\right) =\mu \int\limits_0^1\left[ K_{11}^{\prime }\left(
x,\xi \right) \chi _1\left( \xi \right) +K_{12}\left( x,\xi \right) \chi
_2\left( \xi \right) \right] d\xi ; 
\]
\begin{equation}  \label{4.60}
\chi _2\left( x\right) =\mu \int\limits_0^1\left[ K_{21}\left( x,\xi \right)
\chi _1\left( \xi \right) +K_{22}\left( x,\xi \right) \chi _2\left( \xi
\right) \right] d\xi ,\quad x\in \left[ 0,1\right] ,
\end{equation}
where 
\[
K_{11}^{\prime }\left( x,\xi \right) =K_{11}\left( x,\xi \right) +k\left(
x,\xi \right) . 
\]

The solution (\ref{4.60}) is trivial, if we do not take into account all the
subtleties related to the character of the dependence of the kernels $%
K_{11}^{\prime }$ and $K_{22}$ on $\mu $ whose value does not coincide with
the characteristic numbers of the homogeneous equation 
\[
\chi \left( x\right) =\mu \int\limits_0^2K^{\prime }\left( x,\xi \right)
\chi \left( \xi \right) d\xi ,\quad x\in \left[ 0,2\right] , 
\]
where the kernel is determined by (\ref{4.54}), (\ref{4.49}), with $%
K_{11}\left( x,\xi \right) $ replaced by $K_{11}^{\prime }\left( x,\xi
\right) $. This implies a contradiction.

Let us turn to Eq. (\ref{4.27}) 
\begin{equation}  \label{4.61}
\chi \left( x\right) =\psi \left( x\right) -\lambda
\int\limits_{-1}^0h\left( x,\xi \right) \varphi \left( \xi \right) d\xi
-\lambda \int\limits_0^1h\left( x,\xi \right) \psi \left( \xi \right) d\xi
,\quad x\in \left[ 0,1\right] ,
\end{equation}
assuming that the function $\psi \left( x\right) $ is explicitly determined
via (\ref{4.36}), while the rest of the components are determined as in (\ref
{4.41}). Taking into account (\ref{4.36}) and (\ref{4.44}), as a result of
subtraction from Eq. (\ref{4.61}), we get 
\[
\chi \left( x\right) =-\mu \int\limits_0^1H\left( x,\xi ,\mu \right) \chi
\left( \xi \right) d\xi -\frac 12\mu \int\limits_{-1}^0\left[ h\left( x,\xi
\right) \right. 
\]
\begin{equation}  \label{4.62}
\left. +H\left( x,\xi ,\mu \right) \right] \varphi _0\left( \xi \right) d\xi
+\mu q\left( x\right) ,\quad x\in \left[ 0,1\right] .
\end{equation}

Here, 
\begin{equation}  \label{4.63}
q\left( x\right) =\mu \left[ f\left( x\right) -\mu \int\limits_0^1k\left(
x,\xi \right) \psi _1\left( \xi \right) d\xi \right] ,
\end{equation}
with $\psi _1\left( x\right) $ being the solution of the problem (\ref{4.1}%
), calculated as described in section 5.3.

Substitution of (\ref{4.46}) into (\ref{4.62}), taking into account (\ref
{4.43}) and (\ref{4.47}), leads to the Fredholm integral equation of the
second kind 
\begin{equation}  \label{4.64}
\chi \left( x\right) =\mu \int\limits_0^1Q\left( x,\xi \right) \chi \left(
\xi \right) d\xi +q\left( x\right) ,\quad x\in \left[ 0,1\right] .
\end{equation}

Here, 
\[
Q\left( x,\xi \right) =-H\left( x,\xi ,\mu \right) -\frac 12\mu
\int\limits_0^1\left[ h\left( x,\zeta \right) +H\left( x,\zeta ,\mu \right) %
\right] \left[ H\left( \zeta ,\xi ,\mu \right) \right. 
\]
\[
\left. +\int\limits_0^1L\left( \zeta ,\theta ,\mu \right) H\left( \theta
,\xi ,\mu \right) d\theta \right] d\zeta , 
\]
with $L\left( x,\xi ,\mu \right) $ being the resolvent of the kernel (\ref
{4.38}).

Suppose that the free term of (\ref{4.1}) is determined without the error.
In other words, we are dealing with the situation when $A\psi _{*}=f_{*}$.
Then the algorithm of section 5.3 must lead to $\psi _1=\psi _{*}$, and as a
result of (\ref{4.63}), equation (\ref{4.64}) becomes homogeneous.

Contextually, this situation is rather natural, because it reflects the
equivalence of different representations, namely, of Eqs. (\ref{4.36}) and (%
\ref{4.44}), and the function $\psi \left( x\right) $ on $x\in \left[ 0,1%
\right] $. However, from a formal point of view, the solution of the
homogeneous equation 
\begin{equation}  \label{4.65}
\chi \left( x\right) =\mu \int\limits_0^1Q\left( x,\xi \right) \chi \left(
\xi \right) d\xi ,\quad x\in \left[ 0,1\right]
\end{equation}
is either trivial or, for $\mu =\mu _n$, with $\mu _n$ being the
characteristic numbers of the kernel $Q\left( x,\xi \right) $, coincides
with its eigenfunctions.

It should be pointed out that, using the assumption of exact fulfillment of (%
\ref{4.1}), we cannot consider Eq. (\ref{4.65}) in an abstract way: Its
solution should be sought in a narrower class of functions that are adequate
to the convergent series (\ref{4.23}). Substitution of the latter into (\ref
{4.65}) under the condition of hypothetically exact realization of
corresponding transformations should lead to an identity. It is logical to
interpret the cancellation of the parameter $\mu $ in the context of the
relation (\ref{4.39}).

As mentioned above, in the general case of approximate determination of $%
f\left( x\right) $, equations (\ref{4.28}) and (\ref{4.29}) should be
complemented by the error $\left( \delta f\right) \left( x\right) $.
Therefore, in light of the discussion of (\ref{4.2}), we add to the free
term of Eq. (\ref{4.28}) and subtract from the same term $\mu \delta f$. By
virtue of this fact, correction components appear in (\ref{4.37}), (\ref
{4.40}) and (\ref{4.42}). Using the technique of transformations of section
5.3 and the relation \cite{4.7} 
\[
\int\limits_{-1}^1H\left( x,\zeta ,\mu \right) H\left( \zeta ,\xi ,\mu
\right) d\zeta =\partial _\lambda H\left( x,\xi ,\mu \right) , 
\]
we get, respectively, 
\[
\left( \delta \varphi _0\right) \left( x\right) =\mu \int\limits_0^1H\left(
x,\xi ,\mu \right) \left( \delta f\right) \left( \xi \right) d\xi ; 
\]
\[
\left( \delta \kappa \right) \left( x\right) =-\frac 12\mu ^2\int\limits_0^1 
\left[ h\left( x,\xi \right) +H\left( x,\xi ,\mu \right) \right] \left(
\delta f\right) \left( \xi \right) d\xi ; 
\]
\[
\left( \delta \psi \right) \left( x\right) =\mu \int\limits_0^1t\left( x,\xi
\right) \left( \delta f\right) \left( \xi \right) d\xi , 
\]
where 
\[
t\left( x,\xi \right) =\frac 14\mu ^2\left[ h\left( x,\xi \right) -H\left(
x,\xi ,\mu \right) -\partial _\lambda H\left( x,\xi ,\mu \right) \right] . 
\]

Simultaneously, the term $\mu \delta f$ appears in (\ref{4.63}), and, taking
into account the definition (\ref{4.2}), instead of (\ref{4.65}), we get 
\begin{equation}  \label{4.66}
\chi \left( x\right) =\mu \int\limits_0^1Q\left( x,\xi \right) \chi \left(
\xi \right) d\xi +\left( \delta \psi \right) \left( x\right) ,\quad x\in 
\left[ 0,1\right] .
\end{equation}

By the use of the notation 
\[
\left( \delta f\right) \left( x\right) =\omega \left( x\right) \left( \delta
\chi \right) \left( x\right) 
\]
and of the representation 
\[
\omega \left( x\right) \left( \delta \chi \right) \left( x\right) =\left(
\delta \omega \right) \left( x\right) \chi \left( x\right) 
\]
(i.e., the arising correction is achieved as a result of the variation of
the connecting function $\omega $), equation (\ref{4.66}) takes the form 
\[
\chi \left( x\right) =\mu \int\limits_0^1\left[ Q\left( x,\xi \right)
+t\left( x,\xi \right) \left( \delta \omega \right) \left( \xi \right) %
\right] \chi \left( \xi \right) d\xi ,\quad x\in \left[ 0,1\right] . 
\]

In this regard, the algorithm of section 5.3 can be interpreted as follows.
There exists a function $\chi _1\left( x\right) $ that by means of (\ref
{4.58}) satisfies Eq. (\ref{4.1}) either exactly or with a certain error.
The existence of a homogeneous Fredholm integral equation of the second kind
with the eigenfunction $\chi _1\left( x\right) $ is also quite obvious.
These considerations allow us to draw a conclusion that (\ref{4.65}) plays
the role of such an equation.

Quite naturally, this is true within accuracy depending on the error of the
determination of the free term of (\ref{4.1}).\footnote{%
That is, of its misfit to the space $l_2$.} In this case, an adaptive
connection to the value of the parameter $\mu $ is realized by means of
variations of $\delta \psi $, which is equivalent to a shift of spectral
characteristics. In light of the above, the outlined contradiction can be
considered resolved.

It is clear that, as a rule, the error $\delta f$ and, respectively, the
correcting component $\delta \psi $ in (\ref{4.66}), are unknown, and we are
actually dealing with the Fredholm integral equation of the second kind 
\begin{equation}  \label{4.67}
\chi \left( x\right) =\mu \int\limits_0^1\left[ Q\left( x,\xi \right)
+k\left( x,\xi \right) \right] \chi \left( \xi \right) d\xi -\mu f\left(
x\right) ,\quad x\in \left[ 0,1\right] .
\end{equation}

The function $\chi \left( x\right) $, obtained from this equation, allows us
by (\ref{4.36}) to calculate the solution of the problem under consideration
in the sense of (\ref{4.59}).

Note that the structure of Eq. (\ref{4.65}) is determined by the separation
of the representation of $\psi \left( x\right) $ on $x\in \left[ 0,1\right] $%
: with the function $\chi \left( x\right) $ in the explicit form, Eq. (\ref
{4.36}), and by means of its integration, Eq. (\ref{4.44}). As a matter of
fact, this is a key element of the algorithm. The possibility of its
realization has been ensured, in the first place, by the transformation of
the formulation (\ref{4.1}), namely, the fact that the free term of (\ref
{4.27}) is equal to zero on a part of the domain. The latter follows from
the model of the error (\ref{4.3}), which is the main concept of our
investigation.

Thus, the problem (\ref{4.1}) is reduced to the solution of the Fredholm
integral equation of the second kind (\ref{4.51}) [or the system of
analogous equations (\ref{4.50})], whose components are determined according
to (\ref{4.52})-(\ref{4.54}) and (\ref{4.49}), as well as (\ref{4.14}), (\ref
{4.19}), (\ref{4.34}), under the conditions that the parameter $\mu $
satisfies (\ref{4.47}) and does not coincide with characteristic numbers of
the kernel (\ref{4.54}). To simplify the procedure of the numerical
realization, it is reasonable to assume the restrictions (\ref{4.55}), or
even (\ref{4.57}). In Eq. (\ref{4.67}), the value of $\mu $ should not
coincide with characteristic numbers of the kernel (\ref{4.38}).

Summarizing, our arguments for the representation of the error in the form (%
\ref{4.3}), using the transformed formulation (\ref{4.27}), (\ref{4.28}) and
(\ref{4.30}) instead of (\ref{4.1}), have been justified. It should be
emphasized that before the actual reduction of the considered problem to the
solution of Fredholm integral equations of the second kind the arguments for
the usefulness of our approach were merely heuristic.

From a practical point of view, of main importance is the modeling of the
error of the calculation of the function $f\left( x\right) $ according to (%
\ref{4.1}) by a difference between the sought function $\psi \left( x\right) 
$ in the explicit form and the integral of this function, Eq. (\ref{4.4}),
that is, a qualitatively adequate interpretation of the mechanism of
smoothing of information.

The representation of the operator $B$ by a Poisson integral is, of course,
not the only possible one. However, this point is not crucial. Therefore,
the expression for the kernel (\ref{4.14}) can be considered universal.
Should the importance of the above-mentioned structure of the representation
of (\ref{4.3}) be characterized in an ideological and methodological sense,
the operator $B$ with the kernel $h\left( x,\xi \right) $ facilitates the
implementation of constructive and technical function, which is also
necessary for the realization of the algorithm.

In the construction of the Fredholm integral equation of the second kind (%
\ref{4.51}), any additional information on the data of the problem (\ref{4.1}%
), except for belonging to the space $L_2\left( 0,1\right) $, was not used.
Therefore, it can be logically considered as being objectively equivalent to
the formulation that is erroneously supposed to be unambiguously incorrect.
As far as errors of the determination of $f\left( x\right) $ and $k\left(
x,\xi \right) $ are concerned, their influence on the solution of both (\ref
{4.51}) and related to the formulation (\ref{4.1}) problems is transformed
into the sphere of stable dependencies of the theory of operator equations
of the second kind \cite{4.14}.

\section{On other versions of the transformation}

Applying to Eqs. (\ref{4.27}), (\ref{4.30}) the subtraction procedure, we
get the Fredholm integral equation of the second kind 
\[
\varphi \left( x\right) -\varphi ^{\prime }\left( x\right) =\lambda
\int\limits_{-1}^0h\left( x,\xi \right) \left[ \varphi \left( \xi \right)
-\varphi ^{\prime }\left( \xi \right) \right] d\xi +\kappa \left( x\right)
,\quad x\in \left[ -1,0\right) , 
\]
whose solution has the form 
\begin{equation}  \label{4.68}
\varphi \left( x\right) -\varphi ^{\prime }\left( x\right) =\kappa \left(
x\right) +\lambda \int\limits_{-1}^0H^{\prime }\left( x,\xi ,\lambda \right)
\kappa \left( \xi \right) d\xi ,\quad x\in \left[ -1,0\right) .
\end{equation}

Here, the resolvent 
\begin{equation}  \label{4.69}
H^{\prime }\left( x,\xi ,\lambda \right) =\sum_{n=0}^\infty \frac{r^n}{%
1-2\lambda r^n}\bar \psi _n\left( x\right) \bar \psi _n\left( \xi \right)
\end{equation}
is obtained by analogy with section 5.1.

From Eq. (\ref{4.30}), we get 
\begin{equation}  \label{4.70}
\varphi ^{\prime }\left( x\right) =\kappa \left( x\right) +\lambda
\int\limits_{-1}^0H^{\prime }\left( x,\xi ,\lambda \right) \kappa \left( \xi
\right) d\xi ,\quad x\in \left[ -1,0\right) .
\end{equation}

The substitution of Eqs. (\ref{4.35}), (\ref{4.70}) into (\ref{4.68}) leads
to the Fredholm integral equation of the first kind 
\[
\int\limits_{-1}^0\left[ H\left( x,\xi ,\lambda \right) -H^{\prime }\left(
x,\xi ,\lambda \right) \right] \kappa \left( \xi \right) d\xi 
\]
\begin{equation}  \label{4.71}
=\int\limits_0^1H\left( x,\xi ,\lambda \right) \chi \left( \xi \right) d\xi
,\quad x\in \left[ -1,0\right) ,
\end{equation}
that, nevertheless, allows us to establish a relation between the Fourier
coefficients of the entering functions: 
\[
\kappa _n=\int\limits_{-1}^0\kappa \left( x\right) \bar \psi _n\left(
x\right) dx;\quad \chi _n=\int\limits_{-1}^0\chi \left( x\right) \bar \psi
_n\left( x\right) dx;\quad n=0,1,2,\ldots , 
\]
which may seem rather captivating.

However, the substitution of (\ref{4.34}), (\ref{4.69}) into (\ref{4.71})
yields 
\[
\kappa _n=\frac{1-\lambda r^n}{\lambda r^n}\chi _n,\quad n=0,1,2,\ldots , 
\]
and the expression (\ref{4.42}) reduces to (\ref{4.36}). Correspondingly,
the function $\chi \left( x\right) $ in the left-hand side of (\ref{4.41})
cancels out and the uselessness of the transformations is obvious.

The reason is that any attempts to represent $\delta f$ by use of an
explicit function for which the Fredholm integral equation of the second
kind is constructed, without any relation to the data of the problem (\ref
{4.1}), as it is done above, are futile.\footnote{%
The use of Eq. (5.29) is implied.}

In (\ref{4.3}), we replace the operator (\ref{4.4}) by (\ref{4.25}) and turn
to the system of equations (\ref{4.32}), where the kernel $h\left( x,\xi
\right) $ can be different from (\ref{4.14}), with the parameter $\lambda =1$%
.

It is obvious that 
\[
\psi \left( x\right) =\kappa \left( x\right) +\int\limits_{-1}^xH\left(
x,\xi \right) \kappa \left( \xi \right) d\xi ,\quad x\in \left[ -1,0\right)
, 
\]
where $H\left( x,\xi \right) $ is the resolvent of the kernel $h\left( x,\xi
\right) $, and the problem under consideration is reduced to the formulation
(\ref{4.26}) with a new unknown function 
\[
\sigma \left( x\right) =\int\limits_{-1}^0H\left( x,\xi \right) \psi \left(
\xi \right) d\xi . 
\]

The corresponding system of integral equations takes the form 
\[
\psi \left( x\right) =\mu \int\limits_0^1k\left( x,\xi \right) \psi \left(
\xi \right) d\xi +\int\limits_0^xh\left( x,\xi \right) \psi \left( \xi
\right) d\xi +\sigma \left( x\right) -\mu f\left( x\right) ; 
\]
\begin{equation}  \label{4.72}
\psi \left( x\right) =\int\limits_0^xh\left( x,\xi \right) \psi \left( \xi
\right) d\xi +\sigma \left( x\right) ,\quad x\in \left[ 0,1\right] .
\end{equation}

Now we use the procedure of successive approximations 
\[
\psi _{n+1}\left( x\right) =\frac 12\left[ \psi _{n+1}^{\prime }\left(
x\right) +\psi _{n+1}^{\prime \prime }\left( x\right) \right] , 
\]
with 
\[
\psi _{n+1}^{\prime }\left( x\right) =\mu \int\limits_0^1k\left( x,\xi
\right) \psi _n\left( \xi \right) d\xi +\int\limits_0^xh\left( x,\xi \right)
\psi _n\left( \xi \right) d\xi +\sigma \left( x\right) -\mu f\left( x\right)
; 
\]
\[
\psi _{n+1}^{\prime \prime }\left( x\right) =\int\limits_0^xh\left( x,\xi
\right) \psi _n\left( \xi \right) d\xi +\sigma \left( x\right) ,\quad
n=0,1,2,\ldots , 
\]
\[
\psi _0\left( x\right) =\sigma \left( x\right) -\frac 12\mu f\left( x\right)
, 
\]
to get 
\begin{equation}  \label{4.73}
\psi \left( x\right) =\sum_{n=0}^\infty \left[ \int\limits_0^xh\left( x,\xi
\right) +\frac 12\mu \int\limits_0^1k\left( x,\xi \right) \right] ^n\left[
\sigma \left( \xi \right) -\frac 12\mu f\left( \xi \right) \right] d\xi ,
\end{equation}
where the value of $\left| \mu \right| $ is supposed to be small enough to
ensure the convergence of this series.

The substitution of (\ref{4.73}) into (\ref{4.72}) leads to the cancellation
of both the function $\sigma \left( x\right) $ and the indefinite integrals
of this function. If we retain only $n=N$ terms of the series, a term of
this type, namely, 
\[
\int\limits_0^xh_{N+1}\left( x,\xi \right) \sigma \left( \xi \right) d\xi , 
\]
will survive. Here, the iterated kernel has the form 
\[
h_{N+1}\left( x,\xi \right) =\int\limits_0^xh_m\left( x,\zeta \right)
h_{N-m}\left( \zeta ,\xi \right) d\zeta ,\quad m=1,2,\ldots ,N. 
\]

Taking $h\left( x,\xi \right) $ such that 
\[
h_{N+1}\left( x,x\right) \neq 0,\quad x\in \left[ 0,1\right] , 
\]
by differentiating (\ref{4.73}), under the condition that $\partial
_xk\left( x,\xi \right) $ and $\partial _xf\left( x\right) $ exist, we can
get a Fredholm integral equation of the second kind for the function $\sigma
\left( x\right) $. Its structure is determined by evaluating the power of
the operator sum by the Newton binomial: 
\begin{equation}  \label{4.74}
\left( a+b\right) ^n=C_n^0a^n+C_n^1a^{n-1}b+\ldots +C_n^ma^{n-m}b^m+\ldots
+C_n^{n-1}ab^{n-1}+C_n^nb^n,
\end{equation}
where 
\[
C_n^m=\frac{n!}{m!\left( n-m\right) !}. 
\]

To increase the efficiency of the procedure of the numerical realization, it
is reasonable to maximize explicitly the multiplier of $\sigma \left(
x\right) $, i.e., $k\left( x,x\right) $, which is achieved at the expense of
the strengthening of the oscillations of the kernel $h\left( x,\xi \right) $%
. This situation manifests itself also in the growth of the integral terms,
which less important for the terms with prevailing coefficients $C_n^m$. On
the other hand, the convergence of the series (\ref{4.73}) objectively
worsens.

In general, for sufficiently large $N$, the obtained integral equation can
be represented in the form that does not contain a small multiplier of the
sought function $\sigma \left( x\right) $ that enters explicitly. At the
same time, both its actual structure and corresponding transformations are
rather cumbersome. From a constructive point of view, this approach is
disadvantageous compared to the algorithm of section 5.3.

Note that, as follows from (\ref{4.74}), the integral terms of the
above-mentioned equation have the form 
\begin{equation}  \label{4.75}
\int\limits_0^1t_{nm}\left( x,\xi \right) \sigma \left( \xi \right) d\xi .
\end{equation}

Here, 
\begin{equation}  \label{4.76}
t_{nm}\left( x,\xi \right) =\int\limits_0^xh_n\left( x,\zeta \right)
k_m\left( \zeta ,\xi \right) d\zeta ,
\end{equation}
with $h_n\left( x,\xi \right) $, $k_m\left( x,\xi \right) $ being the
iterated kernels of the corresponding order. However, it is impossible to
extract any advantages by means of a priori adaptation of $h_n\left( x,\xi
\right) $.

The reason lies not only in the difficulties of the restoration of $%
h_n\left( x,\xi \right) $ from the integral equation (\ref{4.76}): In
practice, it is impossible to find a non-singular kernel $t_{nm}\left( x,\xi
\right) $ for which the integral (\ref{4.75}) could be effectively inverted
(see \cite{4.15}).

Indeed, a specific exception is provided by Schl\"omilch's integral 
\[
\frac 2\pi \int\limits_0^{\frac \pi 2}\sigma \left( x\sin \xi \right) d\xi , 
\]
which, however, does not fit into the framework of the suggested scheme.

We briefly note one more version resulting from the representation 
\[
\sigma \left( x\right) =\int\limits_0^xh\left( x,\xi \right) \psi \left( \xi
\right) d\xi 
\]
in (\ref{4.26}), where the kernel $h\left( x,\xi \right) $ is to be
determined.

The problem reduces to the solution of the system of equations 
\[
\psi \left( x\right) =\mu \int\limits_0^1k\left( x,\xi \right) \psi \left(
\xi \right) d\xi +\int\limits_0^xh\left( x,\xi \right) \psi \left( \xi
\right) d\xi -\mu f\left( x\right) ; 
\]
\begin{equation}  \label{4.77}
\psi \left( x\right) =\int\limits_0^xh\left( x,\xi \right) \psi \left( \xi
\right) d\xi ,\quad x\in \left[ 0,1\right] ,
\end{equation}
where, aside from $\psi \left( x\right) $, the kernel $h\left( x,\xi \right) 
$ is also unknown.

The procedure of successive approximations, analogous to that outlined
above, for $\sigma \left( x\right) \equiv 0$, yields a solution in the form
of the series (\ref{4.73}). Substitution of this series into (\ref{4.77})
leads to an equation for the sought function $h\left( x,\xi \right) $
outside the integration sign, i.e., of the second kind, however, essentially
non-linear. Its higher dimension is yet another factor further complicating
its solution.

The model of the error (\ref{4.26}), analogous to the use of the operator (%
\ref{4.25}) in (\ref{4.3}), is more general in comparison with (\ref{4.4}).
This implies intuitive attractiveness of the interpretation of $\delta f$ in
a uniform metric. However, the above transformations witness that less rigid
conditions discarding (\ref{4.3}), (\ref{4.4}) and (\ref{4.6}), and also (%
\ref{4.7}) turn out to be fully sufficient for their implementation.
Therefore, the representation (\ref{4.26}) does not reflect the qualitative
peculiarity of the mechanism of smoothing of information by integration.

Nevertheless, the prospects of the reduction of the considered problem to
the Fredholm integral equation of the second kind on the basis of (\ref{4.26}%
), in principle, can be clearly seen. From the point of view of general
considerations, this situation is rather remarkable.

\chapter{A reduction of linear boundary-value and initial-boundary-value
problems to Fredholm integral equations of the first kind}

\section{Problems for ordinary differential equations}

Consider, for example, 
\begin{equation}  \label{5.1}
u^{\prime \prime }-a\left( x\right) u=f\left( x\right) ,\quad x\in \left[ 0,1%
\right] ;
\end{equation}
\begin{equation}  \label{5.2}
u^{\prime }\left( 0\right) =u\left( 1\right) =0,
\end{equation}
where $a\left( x\right) $ and $f\left( x\right) $ are given $L_2$-functions.

From the notation 
\begin{equation}  \label{5.3}
u^{\prime \prime }\left( x\right) =\psi \left( x\right) ,
\end{equation}
we get 
\begin{equation}  \label{5.4}
u^{\prime }\left( x\right) =\int\limits_0^x\psi \left( \xi \right) d\xi +c_1;
\end{equation}
\begin{equation}  \label{5.5}
u\left( x\right) =\int\limits_0^x\left( x-\xi \right) \psi \left( \xi
\right) d\xi +c_1x+c_0,
\end{equation}
where $c_0$, $c_1$ are the constants of integration.

The substitution of expressions (\ref{5.3}) and (\ref{5.5}) into (\ref{5.1})
leads to a Volterra integral equations of the second kind for a new unknown
function: 
\begin{equation}  \label{5.6}
\psi \left( x\right) =a\left( x\right) \int\limits_0^x\left( x-\xi \right)
\psi \left( \xi \right) d\xi +\left( c_1x+c_0\right) a\left( x\right)
+f\left( x\right) ,\quad x\in \left[ 0,1\right] ,
\end{equation}
whose solution is 
\begin{equation}  \label{5.7}
\psi \left( x\right) =\left( c_1x+c_0\right) a\left( x\right) +f\left(
x\right) +\int\limits_0^xQ\left( x,\xi \right) \left[ \left( c_1\xi
+c_0\right) a\left( \xi \right) +f\left( \xi \right) \right] d\xi ,
\end{equation}
where $Q\left( x,\xi \right) $ is the resolvent of the kernel $a\left(
x\right) \left( x-\xi \right) $.

Taking into account (\ref{5.4}), (\ref{5.5}) and (\ref{5.7}), we find from
the boundary conditions (\ref{5.2}): $c_1=0;$%
\begin{equation}  \label{5.8}
c_0=-\frac{\int\limits_0^1\left( 1-\xi \right) \left[ f\left( \xi \right)
+\int\limits_0^\xi Q\left( \xi ,\zeta \right) f\left( \zeta \right) d\zeta %
\right] d\xi }{1+\int\limits_0^1\left( 1-\xi \right) \left[ a\left( \xi
\right) +\int\limits_0^\xi Q\left( \xi ,\zeta \right) a\left( \zeta \right)
d\zeta \right] d\xi }.
\end{equation}

One can act in a different way: Namely, upon the substitution of expressions
(\ref{5.4}), (\ref{5.5}) into (\ref{5.2}), we get $c_1=0;$%
\[
c_0=-\int\limits_0^1\left( 1-\xi \right) \psi \left( \xi \right) d\xi , 
\]
and, as a result, 
\begin{equation}  \label{5.9}
u\left( x\right) =\left[ \int\limits_0^x\left( x-\xi \right)
-\int\limits_0^1\left( 1-\xi \right) \right] \psi \left( \xi \right) d\xi .
\end{equation}
Thus, the problem under consideration reduces to the Fredholm integral
equation of the second kind 
\begin{equation}  \label{5.10}
\psi \left( x\right) =a\left( x\right) \left[ \int\limits_0^x\left( x-\xi
\right) -\int\limits_0^1\left( 1-\xi \right) \right] \psi \left( \xi \right)
d\xi +f\left( x\right) ,\quad x\in \left[ 0,1\right] .
\end{equation}

Its solution is 
\begin{equation}  \label{5.11}
\psi \left( x\right) =f\left( x\right) +\int\limits_0^1Q\left( x,\xi \right)
f\left( \xi \right) d\xi ,
\end{equation}
where $Q\left( x,\xi \right) $ is the resolvent of the kernel 
\[
-a\left( x\right) \left\{ 
\begin{array}{c}
1-\xi ,\quad x<\xi \leq 1; \\ 
1-x,\quad 0\leq \xi \leq x
\end{array}
\right. 
\]
for the parameter $\lambda =1$ in the right-hand side of (\ref{5.10}).%
\footnote{%
It is clear that this resolvent is not identical to the resolvent of Eq.
(6.6).}

The substitution of (\ref{5.7}) into (\ref{5.5}), taking into account (\ref
{5.8}), or the substitution of (\ref{5.11}) into (\ref{5.9}), allows us to
calculate the solution of the problem (\ref{5.1}), (\ref{5.2}).

Note that the outlined approach is substantially indifferent to the order of
the differential equations, in view of initial or boundary conditions, and
to the data of the problem under consideration, such as the functions $%
a\left( x\right) $ and $f\left( x\right) $. The reduction of the problems to
integral equations of the Fredholm type, in general, requires an analysis of
their solvability. In other words, it is necessary to verify whether $%
\lambda =1$ belongs to the set of characteristic numbers of the
corresponding homogeneous equation, which is implemented in the framework of
the general theory.

Analogously transformations are traditionally discussed in courses of the
theory of integral equations (see, e.g., \cite{5.1,5.2}). At the same time,
as far as the solution of applied problems is concerned, the methodology of
the reduction to integral equations of the second kind did not gain
sufficient popularity, which can be characterized as a kind of a paradox. It
is rather surprising in light of rather active attempts of its
popularization: see, e.g., publications of S. E. Mikeladze, I. A. Birger,
and A. N. Golubentsev \cite{5.3,5.4,5.5}.

In our opinion, the reasons for this are, on the one hand, purely technical
difficulties of numerical realization of integral equations before
wide-spread implementation of computers, and, on the other hand, a
systematic orientation on the use of the techniques of fundamental
solutions, which was formed under the influence of general methodological
concepts of studies of problems of mathematical physics. Indeed, integral
equations, as a rule, were constructed on the basis of the theory of the
potential, when the role of the kernel $k\left( x,\xi \right) $ was played
by the solution for a localized source and the sought function was the
distribution of the intensity of the corresponding action.

In this context, it is rather interesting to point out the development of
the ideas of mutual relationships between the notions of the derivative, the
indefinite and definite integrals, as it was followed in detail by F. A.
Medvedev \cite{5.6}. As a matter of fact, these notions, although initially
invertible, diverged under the influence of increasing sophistication of the
structure of employed kernels and became practically independent. It should
be noted that the representation of the solution by means of different kinds
of local actions was in agreement with practical methods in a number of
scientific disciplines and was widely employed before the appearance of
technical methods of efficient realization of computational operations.

In general, the idea of a possibility of a comparatively simple reduction,
with respect to the highest-order derivative, of boundary-value problems for
ordinary differential equations to Volterra integral equations or Fredholm
integral equations of the second kind did not receive adequate understanding
in the field of applications.

\section{Illustrations of the procedure of the reduction}

Let us turn to a model of the bending of a membrane stretched along a
contour by a uniform load or, alternatively, of a bar under torsion: 
\begin{equation}  \label{5.12}
\partial _x^2u+\partial _y^2u=-1,
\end{equation}
\begin{equation}  \label{5.13}
u\left( 0,y\right) =u\left( 1,y\right) =0,
\end{equation}
\begin{equation}  \label{5.14}
u\left( x,0\right) =u\left( x,1\right) =0.
\end{equation}

Introducing the notation 
\begin{equation}  \label{5.15}
\partial _x^2u\left( x,y\right) =\psi \left( x,y\right) ,
\end{equation}
we get 
\begin{equation}  \label{5.16}
u\left( x,y\right) =\int\limits_0^x\left( x-\xi \right) \psi \left( \xi
,y\right) d\xi +xg_{11}\left( y\right) +g_{12}\left( y\right) ,
\end{equation}
where $g_{1j}\left( y\right) $ are functions of integration.

In view of (\ref{5.15}), equation (\ref{5.12}) takes the form 
\[
\partial _y^2u\left( x,y\right) =-1-\psi \left( x,y\right) , 
\]
and, respectively, 
\begin{equation}  \label{5.17}
u\left( x,y\right) =-\frac 12y^2-\int\limits_0^y\left( y-\eta \right) \psi
\left( x,\eta \right) d\eta +yg_{21}\left( x\right) +g_{22}\left( x\right) ,
\end{equation}
where $g_{2j}\left( y\right) $ are also functions of integration.

The substitution of expressions (\ref{5.16}), (\ref{5.17}) into the boundary
conditions (\ref{5.13}) and (\ref{5.14}), respectively, allows us to
determine $g_{12}=g_{22}=0;$ 
\[
g_{11}\left( y\right) =-\int\limits_0^1\left( 1-\xi \right) \psi \left( \xi
,y\right) d\xi ;\quad g_{21}\left( x\right) =\frac 12+\int\limits_0^1\left(
1-\eta \right) \psi \left( x,\eta \right) d\eta . 
\]

As a result, 
\begin{equation}  \label{5.18}
u\left( x,y\right) =\left[ \int\limits_0^x\left( x-\xi \right)
-x\int\limits_0^1\left( 1-\xi \right) \right] \psi \left( \xi ,y\right) d\xi
;
\end{equation}
\begin{equation}  \label{5.19}
u\left( x,y\right) =\frac 12y\left( 1-y\right) -\left[ \int\limits_0^y\left(
y-\eta \right) -y\int\limits_0^1\left( 1-\eta \right) \right] \psi \left(
x,\eta \right) d\eta .
\end{equation}

Eliminating $u\left( x,y\right) $ from these expressions, we get a Fredholm
integral equation of the first kind with respect anew unknown function: 
\[
\left[ \int\limits_0^x\left( x-\xi \right) -x\int\limits_0^1\left( 1-\xi
\right) \right] \psi \left( \xi ,y\right) d\xi 
\]
\begin{equation}  \label{5.20}
+\left[ \int\limits_0^y\left( y-\eta \right) -y\int\limits_0^1\left( 1-\eta
\right) \right] \psi \left( x,\eta \right) d\eta =\frac 12y\left( 1-y\right)
.
\end{equation}

Thus, a principal difference from the one-dimensional case consists in the
reduction of the problem (\ref{5.12})-(\ref{5.14}) to an incorrectly
formulated one. However, here we will be interested not in a numerical
realization of Eq. (\ref{5.20}) (note that the algorithm of the previous
section applies to it as well) but in the universality of the procedure of
transformation.

Indeed, let the domain of the problem be different from the canonical one,
and let, for example the second condition (\ref{5.13}) have the form $%
u\left( \gamma ,y\right) =0$, where $x=\gamma \left( y\right) $ a certain
single-valued function. Instead of (\ref{5.18}), we have 
\[
u\left( x,y\right) =\left[ \int\limits_0^x\left( x-\xi \right)
-x\int\limits_0^{\gamma \left( y\right) }\left[ \gamma \left( y\right) -\xi %
\right] \right] \psi \left( \xi ,y\right) d\xi , 
\]
and any specific changes are absent, because a transition to an ordinary
procedure of the evaluation of the integral requires only a non-orthogonal
mapping of the type $x=\gamma \bar x$, $y=\bar y$.

It is not difficult to notice that each of expressions (\ref{5.18}) and (\ref
{5.19}) satisfy identically the pair of boundary conditions (\ref{5.13}) and
(\ref{5.14}), respectively. The rest of the conditions are fulfilled
approximately, depending on the accuracy of the determination of $\psi
\left( x,y\right) $. At the same time, the solution can be represented in
the form that satisfies identically both the conditions (\ref{5.13}) and (%
\ref{5.14}): 
\[
U_1\left( x,y\right) =u_1\left( x,y\right) -\left( 1-y\right) u_1\left(
x,0\right) -yu_1\left( x,1\right) ; 
\]
\[
U_2\left( x,y\right) =u_2\left( x,y\right) -\left( 1-x\right) u_2\left(
0,y\right) -xu_2\left( 1,y\right) . 
\]

Here, the functions $u_1\left( x,y\right) $, $u_2\left( x,y\right) $ are
determined by (\ref{5.18}) and (\ref{5.19}), respectively.

The norm of the error of closure of the values of $u_1\left( x,y\right) $ or 
$U_1\left( x,y\right) $ allows us to estimate the error of the approximate
solution: 
\[
\delta =\frac{2\left\| U_1\left( x,y\right) -U_2\left( x,y\right) \right\| }{%
\left\| U_1\left( x,y\right) +U_2\left( x,y\right) \right\| }. 
\]

However, if instead of (\ref{5.13}) the conditions 
\[
\partial _xu\left( 0,y\right) =\partial _xu\left( 1,y\right) =0 
\]
would be imposed, they could not be satisfied by the expression for the
derivative 
\[
\partial _xu\left( x,y\right) =\int\limits_0^x\psi \left( \xi ,y\right) d\xi
+g_{11}\left( y\right) 
\]
that follows from (\ref{5.15}).

Nevertheless, this complication can be easily overcome by the use, in
particular, of the relation 
\[
\partial _x^2u+u=\psi 
\]
that allows as to retain both the functions of integration $g_{1j}\left(
y\right) $.

In the general case, it is reasonable to turn to an equivalent formulation
of the problem (\ref{5.12})-(\ref{5.14}): 
\begin{equation}  \label{5.21}
\partial _x^2u_1+\partial _y^2u_2=-1;\quad u_1\left( x,y\right) =u_2\left(
x,y\right) ,
\end{equation}
\begin{equation}  \label{5.22}
u_1\left( 0,y\right) =u_1\left( 1,y\right) =u_2\left( x,0\right) =u_2\left(
x,1\right) =0,
\end{equation}
using a representation of the solution of the type 
\[
u_1\left( x,y\right) =\int\limits_0^xk_1\left( x,y,\xi \right) \psi _1\left(
\xi ,y\right) d\xi +\sum_{j=1}^2\mu _{1j}\left( x\right) g_{1j}\left(
y\right) ; 
\]
\[
u_2\left( x,y\right) =\int\limits_0^yk_2\left( x,y,\eta \right) \psi
_2\left( x,\eta \right) d\eta +\sum_{j=1}^2\mu _{2j}\left( y\right)
g_{2j}\left( x\right) . 
\]

We assume that the kernels are given and satisfy the conditions 
\[
k_1\left( x,y,x\right) =k_1\left( x,y,y\right) =0; 
\]
\begin{equation}  \label{5.23}
\partial _xk_1\left( x,y,x\right) \neq 0;\quad \partial _yk_2\left(
x,y,y\right) \neq 0,\quad x,y\in \left[ 0,1\right] ;
\end{equation}
$\mu _{1j}\left( x\right) $, $\mu _{2j}\left( y\right) $ are also given; $%
g_{1j}\left( y\right) $, $g_{2j}\left( x\right) $ are to be determined from
the boundary conditions as discussed above.

Thus, in agreement with the previous scheme, we can set $\mu _{11}=x$, $\mu
_{21}=y$, $\mu _{12}=\mu _{22}=1$. The conditions (\ref{5.22}) lead to the
expressions 
\begin{equation}  \label{5.24}
u_1\left( x,y\right) =\left[ \int\limits_0^xk_1\left( x,y,\xi \right)
-x\int\limits_0^1k_1\left( 1,y,\xi \right) \right] \psi _1\left( \xi
,y\right) d\xi ;
\end{equation}
\begin{equation}  \label{5.25}
u_2\left( x,y\right) =\left[ \int\limits_0^yk_2\left( x,y,\eta \right)
-y\int\limits_0^1k_2\left( x,1,\eta \right) \right] \psi _2\left( x,\eta
\right) d\eta ,
\end{equation}
and, respectively, 
\begin{equation}  \label{5.26}
\partial _xk_1\left( x,y,x\right) \psi _1\left( x,y\right)
+\int\limits_0^x\partial _x^2k_1\left( x,y,\xi \right) \psi _1\left( \xi
,y\right) d\xi =\partial _x^2u_1\left( x,y\right) ;
\end{equation}
\begin{equation}  \label{5.27}
\partial _yk_2\left( x,y,y\right) \psi _2\left( x,y\right)
+\int\limits_0^y\partial _y^2k_2\left( x,y,\eta \right) \psi _2\left( x,\eta
\right) d\eta =\partial _y^2u_2\left( x,y\right) .
\end{equation}

Let, in addition to the conditions (\ref{5.23}), $\partial _x^2k_1\left(
x,y,\xi \right) $ and $\partial _y^2k_2\left( x,y,\eta \right) $ are $L_2$%
-kernels. Taking into account the a priori information about the solution of
the considered problem, (\ref{5.26}), (\ref{5.27}) are Volterra integral
equations of the second kind with respect to the functions $\psi _1\left(
x,y\right) $, $\psi _2\left( x,y\right) $, whose solutions, according to the
general theory, exist and are unique. Therefore, expressions (\ref{5.24})
and (\ref{5.25}) are adequate to the physical content of the problem (\ref
{5.21}), (\ref{5.22}).

Note that, for a different choice of the elements $\mu _{1j}\left( x\right) $%
, $\mu _{2j}\left( y\right) $, integral equations constructed in this way
could belong to the Fredholm type of the second kind. In this case, $%
k_1\left( x,y,\xi \right) $, $k_2\left( x,y,\eta \right) $ must also satisfy
a solubility condition with the parameter $\lambda =-1$, which, in fact,
does not pose any problem.

Taking into account (\ref{5.23}), we can set 
\[
k_1\left( x,y,\xi \right) =\left( x-\xi \right) k_1^{\prime }\left( x,y,\xi
\right) ;\quad k_2\left( x,y,\eta \right) =\left( y-\eta \right) k_2^{\prime
}\left( x,y,\eta \right) , 
\]
where 
\[
k_1^{\prime }\left( x,y,x\right) \neq 0;\quad k_2^{\prime }\left(
x,y,y\right) \neq 0,\quad x,y\in \left[ 0,1\right] , 
\]
using expressions $k_1^{\prime }\left( x,y,x\right) $, $k_2^{\prime }\left(
x,y,y\right) $ to refract the a priori information about the solution in
order to smooth the sought functions $\psi _i\left( x,y\right) $ and, in
general, to simplify the procedure of calculations. It is clear that this
point is important for more complicated problems with different
singularities of the solution, and we mention it here only for the sake of
completeness.

The substitution of expressions (\ref{5.24}), (\ref{5.25}) into (\ref{5.21})
produces a system of integral equations 
\begin{equation}  \label{5.28}
\psi _2\left( x,y\right) =-\frac 1{\partial _yk_2\left( x,y,y\right)
}\int\limits_0^y\partial _y^2k_2\left( x,y,\eta \right) \psi _2\left( x,\eta
\right) d\eta +F\left( x,y,\psi _1\right) ,
\end{equation}
where 
\[
F\left( x,y,\psi _1\right) =-\frac 1{\partial _yk_2\left( x,y,y\right) } 
\left[ 1+\partial _xk_1\left( x,y,x\right) \psi _1\left( x,y\right) \right. 
\]
\[
\left. +\int\limits_0^x\partial _x^2k_1\left( x,y,\xi \right) \psi _1\left(
\xi ,y\right) d\xi \right] ; 
\]
\[
\left[ \int\limits_0^xk_1\left( x,y,\xi \right) -x\int\limits_0^1k_1\left(
1,y,\xi \right) \right] \psi _1\left( \xi ,y\right) d\xi 
\]
\begin{equation}  \label{5.29}
-\left[ \int\limits_0^yk_2\left( x,y,\eta \right) -y\int\limits_0^1k_2\left(
x,1,\eta \right) \right] \psi _2\left( x,\eta \right) d\eta =0,\quad x,y\in %
\left[ 0,1\right] .
\end{equation}

From Eq. (\ref{5.28}), we find 
\begin{equation}  \label{5.30}
\psi _2\left( x,y\right) =F\left( x,y,\psi _1\right) +\int\limits_0^yQ\left(
x,y,\eta \right) F\left( x,\eta ,\psi _1\right) d\eta ,
\end{equation}
where $Q\left( x,y,\eta \right) $ is the resolvent of the kernel $-\partial
_y^2k_2\left( x,y,\eta \right) /\partial _yk_2\left( x,y,y\right) $.

Substitution of the expression (\ref{5.30}) into (\ref{5.29}) leads to a
Fredholm integral equation of the first kind with respect to the function $%
\psi _1\left( x,y\right) $.

The described procedure can be directly extended to differential equations
of other types. As an illustration we consider the simplest problem of
thermal conductivity: 
\begin{equation}  \label{5.31}
\partial _tu-\partial _x^2u_2=0,
\end{equation}
\begin{equation}  \label{5.32}
u\left( x,0\right) =u_0\left( x\right) ;\quad u\left( 0,t\right) =u\left(
1,t\right) =0.
\end{equation}

From $\psi =\partial _x^2u$, equation (\ref{5.31}) and conditions (\ref{5.32}%
), we get 
\[
u\left( x,t\right) =\left[ \int\limits_0^x\left( x-\xi \right)
-x\int\limits_0^1\left( 1-\xi \right) \right] \psi \left( \xi ,t\right) d\xi
; 
\]
\[
u\left( x,t\right) =\int\limits_0^t\psi \left( x,\eta \right) d\eta
+u_0\left( x\right) . 
\]

As a consequence, 
\[
\left[ \int\limits_0^x\left( x-\xi \right) -x\int\limits_0^1\left( 1-\xi
\right) \right] \psi \left( \xi ,t\right) d\xi -\int\limits_0^t\psi \left(
x,\eta \right) d\eta =u_0\left( x\right) ;\quad x,y\in \left[ 0,1\right] . 
\]

In order to make an analogous reduction of the problem of the bending of a
rectangular plate of variable stiffness $D$, fixed along a contour, we write 
\cite{5.7}: 
\[
D\Delta \Delta u+2\partial _xD\partial _x\Delta u+2\partial _yD\partial
_y\Delta u+\Delta D\Delta u 
\]
\begin{equation}  \label{5.33}
-\left( 1-\nu \right) \left( \partial _x^2D\partial _y^2u-2\partial
_{xy}D\partial _{xy}u+\partial _y^2D\partial _x^2u\right) =q,
\end{equation}
\begin{equation}  \label{5.34}
\partial _x^nu\left( 0,y\right) =\partial _x^nu\left( a,y\right) =\partial
_y^nu\left( x,0\right) =\partial _y^nu\left( x,b\right) =0,\quad n=0,1,
\end{equation}
where $\Delta =\partial _x^2+\partial _y^2$; $\nu $ is the Poisson
coefficient; $q\left( x,y\right) $ is the intensity of the transverse load.
To calculate the derivatives with respect to $x$, we can set 
\begin{equation}  \label{5.35}
u\left( x,y\right) =\int\limits_0^xk\left( x,y,\xi \right) \psi \left( \xi
,y\right) d\xi +\sum_{j=1}^4x^{j-1}g_{1j}\left( y\right) .
\end{equation}

Here, 
\[
\partial _x^nk_1\left( x,y,x\right) =0,\quad n=0,1;\quad \partial
_x^3k_2\left( x,y,y\right) \neq 0,\quad x\in \left[ 0,a\right] ;\quad y\in %
\left[ 0,b\right] , 
\]
and the functions $g_{1j}\left( y\right) $ are intended to satisfy the
conditions (\ref{5.34}) for $x=0$, $x=a$.

The second representation of the solution via $\psi \left( x,y\right) $ is
determined by means of the substitution of (\ref{5.35}) into Eq. (\ref{5.33}%
) and four-fold integration over the variable $y$. The appearing functions $%
g_{2j}\left( x\right) $ allow us to satisfy the conditions (\ref{5.34}) for $%
y=0$, $y=b$. After that, $u\left( x,y\right) $ is eliminated from the
representation of the solution..

Note that with the help of a special structure of the kernel $k\left(
x,y,\eta \right) $ one can easily satisfy conditions of the type $u\left(
x_i,y_i\right) =0$ at isolated points inside the considered domain. The
procedure of the reduction also applies to mixed boundary conditions ( a
change of the type along a side) and to the case of a connection of plates.
Analogously, three-dimensional problems of mathematical physics can also be
reduced to Fredholm integral equations of the first kind.

\section{Universality and analogous approaches}

Thus, a rather elementary method of the reduction of linear boundary-value
and initial-boundary-value problems to Fredholm integral equations of the
first kind is practically universal from the point of view of its
realizations as far as the following aspects are concerned:

- the order and structure of differential equations;

- the form of boundary conditions;

- the availability of variable coefficients;

- the form of the domain;

- the dimensionality of the problem.

In this situation, all the information about a concrete problem is
transferred into a functional equation, whose solution does not require any
conditions on the contour of the domain, which poses a substantial
advantage. Thus, its solution can be sought in the form of a series of a
system of coordinate elements intended exclusively to ensure the efficiency
of the procedure of the numerical realization.

However, the problem obtained as a result of transformations is incorrect,
hence its numerical realization requires adequate methods. At the same time,
in applications, the solution of such a problem can be acceptably
approximated by a series with the number of terms that does not affect the
stability of the numerical algorithms. Therefore, one can hardly explain the
absence of interest to a systematic use of this procedure, especially in the
period before the general orientation at the discretization of problems of
mathematical modeling.

One may state that special literature did not point out the existence of a
formalized method of the reduction of practically arbitrary
initial-boundary-value problems to Fredholm integral equations of the first
kind. At the same time, there a number of examples of applications of
analogous transformations in rather particular situations. As a rule, they
were a given physical interpretation that considerably disguised the
generality of this approach.

Thus, Yu. V. Repman used as new unknown variables boundary forces of a plate
of a canonical configuration that allowed one to satisfy conditions on an
internal contour of complex configuration \cite{5.8}. L. A. Rozin has
developed a method of separation that admits a reduction of the problems of
calculations of membranes under the forces of interaction of isolated bars
to systems of Fredholm integral equations of the first kind (\cite{5.9},
section 9). Some publications point out the advantages of the approximation
of higher-order derivative of differential equations with respect to one of
the variables that, compared to numerical differentiation, are much more
accurate (see \cite{5.10}). It should be noted, however, that an actual
transition to an incorrect formulation, as a rule, passed unnoticed.

In general, we think that there occurred a kind of assimilation of the
discussed procedure of the reduction by the methods of the theory of the
potential, based on the use of integral relations along the boundaries of
the domains and by the techniques of fundamental solutions \cite{5.11}. The
reduction of the dimensionality of the sought functions, achieved in this
way, seems to have overweighed by its importance the above-mentioned
universality. Moreover, the construction of integral equations with strong
singularities in the kernels that partly smooth over the factor of
incorrectness attracted certain attention \cite{5.12}.

Some problems for differential equations, and, in particular, the following
one: 
\[
\partial _{xy}u=a\partial _xu+b\partial _yu+cu+f, 
\]
where $a$, $b$, $c$ and $f$ are given functions of the variables $x$ and $y$%
, can be reduced directly to Volterra and Fredholm integral equations of the
second kind with respect to the higher-order derivative ($\psi =\partial
_{xy}u$). These issues are studied in detail by G. M\"untz \cite{5.13}. Of
considerable interest is the fact, established by this author, that
analogous transformations cannot be extended to the case of the simplest
equation of the elliptic type.

\section{A connection to the algorithm of the previous section}

Fredholm integral equations of the first kind, appearing in the
above-discussed case of the reduction of two-dimensional boundary-value
problems, can be represented in the form 
\begin{equation}  \label{5.36}
\left( A\psi \right) \left( x,y\right) \equiv \int\limits_0^1\tau _1\left(
x,y,\xi \right) \psi \left( \xi ,y\right) d\xi +\int\limits_0^1\tau _2\left(
x,y,\eta \right) \psi \left( x,\eta \right) d\eta =f\left( x,y\right) ,
\end{equation}
\[
x,y\in \left[ 0,1\right] . 
\]

A typical discontinuity of the kernels $\tau _1\left( x,y,\xi \right) $, $%
\tau _2\left( x,y,\eta \right) $ on the diagonal $\xi =x$, $\eta =y$, as
well as other analogous complications, are of no principal importance.

The algorithm of section 5 does not require in this case any substantial
changes. The differences are related only to the structure of the operator $%
A $. The analog of the system of Fredholm integral equations of the second
kind (\ref{4.37}) and (\ref{4.45}) takes the form 
\begin{equation}  \label{5.37}
\varphi _0\left( x,y\right) =\mu \lambda \int\limits_0^1H\left( x,\xi
,\lambda \right) \left( A\varphi _0\right) \left( \xi ,y\right) d\xi
+\lambda \int\limits_0^1H\left( x,\xi ,\lambda \right) \chi \left( \xi
,y\right) d\xi ;
\end{equation}
\[
\chi \left( x,y\right) =-\lambda \int\limits_0^1H\left( x,\xi ,\lambda
\right) \chi \left( \xi ,y\right) d\xi -\mu \left( A\psi \right) \left(
x,y\right) 
\]
\[
-\frac 12\lambda \int\limits_{-1}^0\left[ h\left( x,\xi \right) +H\left(
x,\xi ,\lambda \right) \right] \varphi _0\left( \xi ,y\right) d\xi +\mu
f\left( x,y\right) ,\quad x,y\in \left[ 0,1\right] , 
\]
where 
\begin{equation}  \label{5.38}
\psi \left( x,y\right) =\chi \left( x,y\right) +\lambda
\int\limits_0^1H\left( x,\xi ,\lambda \right) \chi \left( \xi ,y\right) d\xi
.
\end{equation}

Under the condition (\ref{4.47}), taking into account (\ref{4.43}) and the
notation (\ref{4.48}), we can reduce the problem to the solution of a system
of two dimensional Fredholm integral equations of the second kind: 
\[
\chi _1\left( x,y\right) =\mu \left[ \int\limits_0^1K_{11}\left( x,y,\xi
\right) \chi _1\left( \xi ,y\right) d\xi +\int\limits_0^1K_{12}\left(
x,y,\eta \right) \chi _1\left( x,\eta \right) d\eta \right. 
\]
\[
\left. +\int\limits_0^1\int\limits_0^1K_{13}\left( x,y,\xi ,\eta \right)
\chi _1\left( \xi ,\eta \right) d\xi d\eta +\int\limits_0^1K_{14}\left(
x,\xi \right) \chi _2\left( \xi ,y\right) d\xi \right] +F_1\left( x,y\right)
; 
\]
\[
\chi _2\left( x,y\right) =\mu \left[ \int\limits_0^1K_{21}\left( x,\xi
\right) \chi _1\left( \xi \right) d\xi +\int\limits_0^1K_{22}\left( x,y,\xi
\right) \chi _2\left( \xi ,y\right) d\xi \right. 
\]
\[
\left. +\int\limits_0^1\int\limits_0^1K_{23}\left( x,y,\xi ,\eta \right)
\chi _2\left( \xi ,\eta \right) d\xi d\eta \right] ,\quad x,y\in \left[ 0,1%
\right] . 
\]

Here, 
\[
K_{11}\left( x,y,\xi \right) =-\left[ H\left( x,\xi ,\mu \right) +\tau
_1\left( x,y,\xi \right) +\int\limits_0^1\tau _1\left( x,y,\zeta \right)
H\left( \zeta ,\xi ,\mu \right) d\zeta \right] ; 
\]
\[
K_{12}\left( x,y,\eta \right) =-\tau _2\left( x,y,\eta \right) ;\quad
K_{13}\left( x,y,\xi ,\eta \right) =-\tau _2\left( x,y,\eta \right) H\left(
x,\xi ,\mu \right) ; 
\]
\[
K_{14}\left( x,\xi \right) =-\frac 12\left[ h\left( x,\xi \right) +H\left(
x,\xi ,\mu \right) \right] ;\quad F_1\left( x,y\right) =\mu f\left(
x,y\right) ; 
\]
\[
K_{21}\left( x,\xi \right) =H\left( x,\xi ,\mu \right) ;\quad K_{22}\left(
x,y,\xi \right) =\mu \int\limits_0^1H\left( x,\zeta ,\mu \right) \tau
_1\left( \zeta ,y,\xi \right) d\zeta ; 
\]
\[
K_{23}\left( x,y,\xi ,\eta \right) =\mu H\left( x,\xi ,\mu \right) \tau
_2\left( \xi ,y,\eta \right) . 
\]

In addition to (\ref{4.47}), the parameter $\mu $ should not coincide with
the characteristic numbers of the above system of the equations. To simplify
the procedure of the numerical realization, it is reasonable to choose this
parameter in agreement with conditions of the type (\ref{4.55}), (\ref{4.57}%
).

The function $\psi \left( x,y\right) $ satisfying Eq. (\ref{5.36}) in the
sense of (\ref{4.59}) is determined from (\ref{5.38}) with $\chi \left(
x,y\right) \equiv \chi _1\left( x,y\right) $. It should be noted that the
techniques of the numerical realization of Fredholm integral equations of
the second kind, pointed out in section 5.3, can be directly applied to this
case. One can also use special methods of the evaluation of two-dimensional
integrals (\cite{5.14}, section 3).

\chapter{Other classes of problems}

\section{The initial-boundary-value problem for the Korteweg-de Vries
equation}

Consider the formulation 
\begin{equation}  \label{6.1}
\partial _tu-6u\partial _xu+\partial _x^3u=0,
\end{equation}
\begin{equation}  \label{6.2}
u\left( x,0\right) =u_0\left( x\right) ;\quad u\left( 0,t\right) =u_1\left(
t\right) ;\quad \partial _xu\left( 0,t\right) =u_2\left( t\right) ;\quad
u\left( 1,t\right) =u_3\left( t\right) ,
\end{equation}
where the given functions $u_0\left( x\right) $; $u_i\left( t\right) $, $%
i=1,2,3$ are supposed to impose no additional restrictions on the model of
the process under consideration.

It should be noted that up to now there is no any general theory that allows
one to investigate a priori the solvability of the problems of the type (\ref
{6.1}), (\ref{6.2}) in adequate classes of functions. In this situation, the
main means of the specification of the mathematical model are results of
numerical simulations as well as solutions of specially simplified equations
near the boundary (see \cite{6.1}, section 10).

Using the procedure of the previous section, it is not difficult to reduce
the problem (\ref{6.1}), (\ref{6.2}) to an integral equation of the first
kind with respect to 
\[
\psi \left( x,t\right) =\partial _x^3u\left( x,t\right) , 
\]
which yields 
\begin{equation}  \label{6.3}
u\left( x,t\right) =\frac 12\int\limits_0^x\left( x-\xi \right) ^2\psi
\left( \xi ,t\right) d\xi +\frac 12x^2g_3\left( t\right) +xg_2\left(
t\right) +g_1\left( t\right) ,
\end{equation}
with the functions of integration determined from the boundary conditions: 
\[
g_1\left( t\right) =u_1\left( t\right) ;\quad g_2\left( t\right) =u_2\left(
t\right) ; 
\]
\[
g_3\left( t\right) =2\left[ u_3\left( t\right) -u_2\left( t\right)
-u_1\left( t\right) \right] -\int\limits_0^1\left( 1-\xi \right) ^2\psi
\left( \xi ,t\right) d\xi . 
\]

Substitution into (\ref{6.3}) leads to the expression 
\[
u\left( x,t\right) =\frac 12\left[ \int\limits_0^x\left( x-\xi \right)
^2-x^2\int\limits_0^1\left( 1-\xi \right) ^2\right] \psi \left( \xi
,t\right) d\xi 
\]
\begin{equation}  \label{6.4}
+x^2u_3\left( t\right) +x\left( 1-x\right) u_2\left( t\right) +\left(
1-x^2\right) u_1\left( t\right) .
\end{equation}

Now we rewrite Eq. (\ref{6.1}) in the form 
\begin{equation}  \label{6.5}
\partial _tu=6u\partial _xu-\partial _x^3u.
\end{equation}

The substitution of (\ref{6.4}) into the right-hand side of (\ref{6.5}) and
integration from $0$ to $t\,$ under the initial condition (\ref{6.2}) allows
us to determine 
\[
u\left( x,t\right) =6\int\limits_0^t\left[ \int\limits_0^x\left( x-\xi
\right) ^2+x^2\int\limits_0^1\left( 1-\xi \right) ^2\right] \psi \left( \xi
,\eta \right) d\xi \left\{ \left[ \int\limits_0^x\left( x-\xi \right)
\right. \right. 
\]
\[
\left. \left. -x\int\limits_0^1\left( 1-\xi \right) \right] \psi \left( \xi
,\eta \right) d\xi +2xu_3\left( \eta \right) +\left( 1-2x\right) u_2\left(
\eta \right) \right\} d\eta +\left[ 2xu_3\left( \eta \right) \right. 
\]
\[
\left. \left. +\left( 1-2x\right) u_2\left( \eta \right) +2xu_1\left( \eta
\right) \right] \left[ \int\limits_0^x\left( x-\xi \right)
-x\int\limits_0^1\left( 1-\xi \right) \right] \psi \left( \xi ,\eta \right)
d\xi \right\} d\eta 
\]
\[
-\int\limits_0^t\psi \left( x,\eta \right) d\eta +6\int\limits_0^t\left[
x^2u_3\left( \eta \right) +x\left( 1-x\right) u_2\left( \eta \right) +\left(
1-x^2\right) u_1\left( \eta \right) \right] 
\]
\begin{equation}  \label{6.6}
\times \left[ 2xu_3\left( \eta \right) +\left( 1-2x\right) u_2\left( \eta
\right) -2xu_1\left( \eta \right) \right] d\eta +u_0\left( x\right) .
\end{equation}

Eliminating $u\left( x,t\right) $ from (\ref{6.4}), (\ref{6.6}), we get the
equation 
\begin{equation}  \label{6.7}
\left( A\psi \right) \left( x,t\right) =f\left( x,t\right) ,\quad x,t\in
\Omega :0\leq x,t\leq 1,
\end{equation}
where the expressions for the operator $A$ and the free term $f$ are rather
obvious.

Let the data $u_0\left( x\right) $, $u_i\left( t\right) $ be such that $%
f\left( x,t\right) \in L_2\left( \Omega \right) $ and, besides, the solution
of the problem (\ref{6.1}), (\ref{6.2}) be a $L_2$-function. In this case,
following section 5.1, we can use the relation 
\[
\psi \left( x,t\right) =\lambda \left( B\psi \right) \left( x,t\right)
,\quad x,t\in \Omega , 
\]
where the operator $B$ and the parameter $\lambda $ are the same.

The algorithm of section 5.1 allows us to reduce the considered problem to
the solution of a system of two nonlinear integral equations of the second
kind (\ref{5.37}), where the variable $y$ should be replaced by $t$.

By virtue of continuous dependence of the integrand of (\ref{6.6}) on $\psi
\left( x,t\right) $ and the principle of contracted mappings for
sufficiently small values of the parameter $\mu $ satisfying the condition (%
\ref{4.47}), the solution of this system of equations is achieved by means
of simple iterations with an arbitrary initial approximation from $L_2$. An
estimate of the influence of the error of calculations, the issue of the
acceleration of the convergence, as well as other aspects of the numerical
realization are discussed in \cite{6.2}.

It should be emphasized that the realization of the conditions of
convergence with the help of the parameter $\mu $, which, in fact, allows
one to determine particular solutions by comparatively simple means, is of
special importance in the nonlinear case. It is implied that other solutions
can be determined in the framework of numerical simulations for large $\mu $%
, related to studies of implicit functions, branching and bifurcations.

Using expressions (\ref{6.4}), (\ref{6.6}), the solution can be given a form
that satisfies identically both the initial and boundary conditions (\ref
{6.2}). We can also estimate the error by analogy with the discussion in
section 6.2.

\section{A boundary-value problem for a substantially nonlinear differential
equation}

Here, we discuss nonlinearity related to higher-order derivatives. As an
example, consider Monge-Amp\`ere's equation that plays an important role in
a number of problems of geometry: 
\begin{equation}  \label{6.8}
\partial _x^2u\partial _y^2u-\left( \partial _{xy}u\right) ^2=s_1\partial
_x^2u+s_2\partial _y^2u+s_3\partial _{xy}u+q,
\end{equation}
where $s_i$, $i=1,2,3\,$ and $q$ depend on the variables $x$, $y$, the
sought function $u\left( x,y\right) $ and its first derivatives $\partial
_xu $, $\partial _yu$ \cite{6.3}.

For definitiveness, we set $s_i=s_i\left( x,y\right) $, $q=q\left(
x,y\right) $ and 
\begin{equation}  \label{6.9}
u\left( 0,y\right) =u\left( 1,y\right) =u\left( x,0\right) =u\left(
x,1\right) =0.
\end{equation}

Without going into analysis of the conditions imposed on the data of this
problem to ensure the existence and uniqueness of the solution, we outline a
scheme of a numerical simulation. It should be noted that corresponding
estimates are rather non-trivial and, as a rule, cover only certain
particular cases \cite{6.4}.

We use the notation 
\[
\partial _x^2u\left( x,y\right) =\psi _1\left( x,y\right) ;\quad \partial
_y^2u\left( x,y\right) =\psi _2\left( x,y\right) ; 
\]
\[
\partial _{xy}u\left( x,y\right) =\psi \left( x,y\right) ,\quad x,y\in
\Omega :0\leq x,y\leq 1. 
\]
As a result, taking into account (\ref{6.9}), we get 
\[
u\left( x,y\right) =\left[ \int\limits_0^x\left( x-\xi \right)
-x\int\limits_0^1\left( 1-\xi \right) \right] \psi _1\left( \xi ,y\right)
d\xi ; 
\]
\[
u\left( x,y\right) =\left[ \int\limits_0^y\left( y-\eta \right)
-y\int\limits_0^1\left( 1-\eta \right) \right] \psi _2\left( x,\eta \right)
d\eta ; 
\]
\[
u\left( x,y\right) =\int\limits_0^xd\xi \int\limits_0^y\psi \left( \xi ,\eta
\right) d\eta . 
\]

Upon the substitution of these expressions into (\ref{6.8}) and the
elimination of the function $u\left( x,y\right) $, we reduce the problem to
the following system of equations: 
\[
\psi _1\left( x,y\right) \psi _2\left( x,y\right) -\psi ^2\left( x,y\right)
=s_1\left( x,y\right) \psi _1\left( x,y\right) 
\]
\begin{equation}  \label{6.10}
+s_2\left( x,y\right) \psi _2\left( x,y\right) +s_3\left( x,y\right) \psi
\left( x,y\right) +q\left( x,y\right) ;
\end{equation}
\begin{equation}  \label{6.11}
\left[ \int\limits_0^x\left( x-\xi \right) -x\int\limits_0^1\left( 1-\xi
\right) \right] \psi _1\left( \xi ,y\right) d\xi -\int\limits_0^xd\xi
\int\limits_0^y\psi \left( \xi ,\eta \right) d\eta =0;
\end{equation}
\begin{equation}  \label{6.12}
\left[ \int\limits_0^y\left( y-\eta \right) -y\int\limits_0^1\left( 1-\eta
\right) \right] \psi _2\left( x,\eta \right) d\eta -\int\limits_0^xd\xi
\int\limits_0^y\psi \left( \xi ,\eta \right) d\eta =0,\quad x,y\in \Omega .
\end{equation}

The existence of $\partial _x^2u$, $\partial _y^2u$ implies a possibility of
two-fold differentiation of Eqs. (\ref{6.11}), (\ref{6.12}) with respect to $%
x$ and $y$, respectively, which yields 
\[
\psi _1\left( x,y\right) =\int\limits_0^y\partial _x\psi \left( x,\eta
\right) d\eta ;\quad \psi _2\left( x,y\right) =\int\limits_0^x\partial
_y\psi \left( \xi ,y\right) d\xi . 
\]

Equation (\ref{6.10}) takes the form 
\[
\left[ \int\limits_0^y\partial _x\psi \left( x,\eta \right) d\eta \right] %
\left[ \int\limits_0^x\partial _y\psi \left( \xi ,y\right) d\xi \right]
-\psi ^2\left( x,y\right) =s_1\left( x,y\right) \int\limits_0^y\partial
_x\psi \left( x,\eta \right) d\eta 
\]
\[
+s_2\left( x,y\right) \int\limits_0^x\partial _y\psi \left( \xi ,y\right)
d\xi +s_3\left( x,y\right) \psi \left( x,y\right) +q\left( x,y\right) ,\quad
x,y\in \Omega , 
\]
and after integration in the limits $0,x$ and $0,y$: 
\begin{equation}  \label{6.13}
\left( A\psi \right) \left( x,y\right) =f\left( x,y\right) ,\quad x,y\in
\Omega ,
\end{equation}
where 
\[
\left( A\psi \right) \left( x,y\right) =\int\limits_0^x\psi \left( \xi
,y\right) d\xi \int\limits_0^y\psi \left( x,\eta \right) d\eta
-\int\limits_0^x\psi \left( \xi ,\eta \right) d\xi \int\limits_0^y\psi
\left( x,\eta \right) d\eta 
\]
\[
-\int\limits_0^x\psi \left( \xi ,y\right) d\xi \int\limits_0^y\psi \left(
\xi ,\eta \right) d\eta -\int\limits_0^yd\eta \int\limits_0^\eta \left[
s_1\left( x,\eta \right) \psi \left( \xi ,\eta _1\right) -s_1\left( 0,\eta
\right) \psi \left( 0,\eta _1\right) \right. 
\]
\[
\left. -\int\limits_0^x\partial _\xi s_1\left( \xi ,\eta \right) \psi \left(
\xi ,\eta _1\right) d\xi \right] d\eta _1-\int\limits_0^xd\xi
\int\limits_0^\xi \left[ s_2\left( \xi ,y\right) \psi \left( \xi _1,\eta
\right) -s_2\left( \xi ,0\right) \psi \left( \xi _1,0\right) \right. 
\]
\[
\left. -\int\limits_0^y\partial _\eta s_2\left( \xi ,\eta \right) \psi
\left( \xi _1,\eta \right) d\eta \right] d\xi _1-\int\limits_0^xs_3\left(
\xi ,\eta \right) d\xi \int\limits_0^y\psi \left( \xi ,\eta \right) d\eta ; 
\]
\[
f\left( x,y\right) =\int\limits_0^xd\xi \int\limits_0^yq\left( \xi ,\eta
\right) d\eta . 
\]

The above implies the existence of the derivatives $\partial _xs_1$, $%
\partial _ys_2$.

As far as a numerical realization of Eq. (\ref{6.13}) is concerned, the
considerations of the previous section can be applied.

\section{Nonlinearity of the boundary condition}

Consider a typical problem of the irradiation of an infinite plate with a
thermally insulated surface into a medium whose absolute temperature is
equal to zero \cite{6.5}:

\begin{equation}  \label{6.14}
a_T\partial _x^2u-\partial _tu=0,
\end{equation}
\begin{equation}  \label{6.15}
u\left( x,0\right) =u_0\left( x\right) ;\quad p\partial _xu\left( 0,t\right)
+u^m\left( 0,t\right) =0;\quad \partial _xu\left( 1,t\right) =0.
\end{equation}

Here, $u\left( x,t\right) $ is the temperature gradient; $u_0\left( x\right) 
$ is a given function; $a_T$ is the temperature conductivity; $p=\lambda
/\alpha $, with $\lambda $, $\alpha $ being the thermal-conductivity and the
heat-transfer coefficients, respectively.

Introduce the notation 
\begin{equation}  \label{6.16}
\partial _x^2u\left( x,y\right) =\psi \left( x,t\right) ,
\end{equation}
which leads to 
\[
u\left( x,t\right) =\int\limits_0^x\left( x-\xi \right) \psi \left( \xi
,t\right) d\xi +xg_1\left( t\right) +g_2\left( t\right) , 
\]
where $g_i\left( t\right) $ are functions of integration.

The boundary conditions (\ref{6.15}) yield 
\[
g_1\left( t\right) =-\int\limits_0^1\psi \left( \xi ,t\right) d\xi ;\quad
g_1\left( t\right) =pg_2^m\left( t\right) , 
\]
and, respectively, 
\[
u\left( x,t\right) =\left[ \int\limits_0^x\left( x-\xi \right)
-x\int\limits_0^1\right] \psi \left( \xi ,t\right) d\xi +\left[
p\int\limits_0^1\psi \left( \xi ,t\right) d\xi \right] ^{\frac 1m}. 
\]

Using (\ref{6.14}), (\ref{6.15}) and taking into account the initial
condition (\ref{6.15}), we get 
\[
u\left( x,t\right) =a_T\int\limits_0^t\psi \left( x,\eta \right) d\eta
+u_0\left( x\right) , 
\]
and the problem under consideration is reduced to the solution of the
nonlinear integral equation of the first kind (\ref{6.7}), where 
\[
A_{\bullet }=\left[ \int\limits_0^x\left( x-\xi \right) -x\int\limits_0^1%
\right] _{\bullet }d\xi +\left( p\int\limits_0^1\left. {}\right. _{\bullet
}d\xi \right) ^{\frac 1m}-a_T\int\limits_0^t\left. {}\right. _{\bullet
}d\eta ; 
\]
\[
f\left( x,t\right) =u_0\left( x\right) . 
\]

\section{A small parameter by the highest-order derivative of the
differential equation of the problem}

As an illustration of general considerations, we consider the problem of
heat transport induced by the processes of thermal conduction and convection
(the first and the second terms, respectively, of the differential equation) 
\cite{6.6}:

\begin{equation}  \label{6.17}
\partial _tu=\epsilon \partial _x^2u+\beta \partial _xu.
\end{equation}

Here, $\beta >0$ is a constant; $\epsilon \,$ is a small parameter, 
\begin{equation}  \label{6.18}
u\left( 0,x\right) =0;\quad u\left( t,0\right) =0;\quad u\left( t,1\right)
=u_1\left( t\right) ,
\end{equation}
with $u_1\left( t\right) $ being a given function.

The notation (\ref{6.16}) under the boundary conditions (\ref{6.18}) leads
to 
\begin{equation}  \label{6.19}
u\left( x,t\right) =\left[ \int\limits_0^x\left( x-\xi \right)
-\int\limits_0^1\left( 1-\xi \right) \right] \psi \left( \xi ,t\right) d\xi
+u_1\left( t\right) .
\end{equation}

The integration of (\ref{6.17}) in the limits $0,t$ with the use of (\ref
{6.19}) and of the initial condition (\ref{6.18}) yields 
\[
u\left( x,t\right) =\int\limits_0^t\left[ \epsilon \psi \left( x,\eta
\right) +\beta \int\limits_0^x\psi \left( \xi ,\eta \right) d\xi \right]
d\eta . 
\]

The problem is reduced to the linear integral equation of the first kind (%
\ref{6.7}), where 
\[
A=A_1+A_2, 
\]
\[
A_{1\bullet }=\epsilon \int\limits_0^t\left. {}\right. _{\bullet }d\eta
;\quad A_{2\bullet }=\beta \int\limits_0^xd\xi \int\limits_0^t\left.
{}\right. _{\bullet }d\eta -\left[ \int\limits_0^x\left( x-\xi \right)
-\int\limits_0^1\left( 1-\xi \right) \right] _{\bullet }d\xi ; 
\]
\[
f\left( x,t\right) =u_1\left( t\right) . 
\]

The algorithm of section 5 allows us to go over to a Fredholm integral
equation of the second kind 
\begin{equation}  \label{6.20}
\chi \left( x,t\right) =\mu \left[ \left( \epsilon R_1+R_2\right) \chi %
\right] \left( x,t\right) +F\left( x,t\right) ,
\end{equation}
where $R_1$ and $R_2$ are corresponding operators.

As a result of the expansion \cite{6.7} 
\[
\chi \left( x,t\right) =\sum_{m=0}^\infty \epsilon ^m\chi _m\left(
x,t\right) , 
\]
we get a sequence of recursion relations 
\[
\chi _0\left( x,t\right) =\mu \left( R_2\chi _0\right) \left( x,t\right)
+F\left( x,t\right) ; 
\]
\[
\chi _1\left( x,t\right) =\mu \left( R_2\chi _1\right) \left( x,t\right)
+\mu \left( R_2\chi _0\right) \left( x,t\right) ; 
\]
\[
\ldots 
\]
\[
\chi _{m+1}\left( x,t\right) =\mu \left( R_2\chi _{m+1}\right) \left(
x,t\right) +\mu \left( R_2\chi _m\right) \left( x,t\right) 
\]
that are integral equation of the same type of the canonical structure.

The above allows us to state that the proposed approach is rather efficient
in the problems of mathematical physics with a singular perturbation, whose
numerical realization, as a rule, meets with considerable difficulties (see,
in particular, \cite{6.8}). Indeed, we managed to transform the singular
perturbation (\ref{6.17}) into the regular one (\ref{6.20}), which
facilitated a radical simplification of the problem.\footnote{%
According to a standard classification, singular and regular perturbations
affect,respectively, main and dependent terms of the operators.}

\section{Equations of a mixed type}

Boundary-value problems for equations of this type are characterized by
non-triviality of the investigation of the issues of existence and
uniqueness (see \cite{6.9}). As a consequence, one has to consider such
equations on rather special domains, which restricts the field of practical
applications.

As an illustration, consider well-known Tricomi's equation 
\begin{equation}  \label{6.21}
y\partial _x^2u+\partial _y^2u=0.
\end{equation}
this equation belongs both to the hyperbolic and elliptical types for $y<0$
and $y>0$, respectively. We restrict ourselves to the framework of a
numerical simulation, that is, we do not consider a priori the solvability
of the problem and employ, for example, the following boundary conditions: 
\begin{equation}  \label{6.22}
u\left( 0,y\right) =u\left( 1,y\right) =u\left( x,-1\right) =0;\quad u\left(
x,1\right) =\nu \left( x\right) ,
\end{equation}
where the function $\nu \left( x\right) $ and satisfies the conditions $\nu
\left( 0\right) =\nu \left( 1\right) =0$.

Introduce the notation 
\begin{equation}  \label{6.23}
\partial _x^2u\left( x,y\right) =\psi \left( x,y\right) ,
\end{equation}
which, taking into account (\ref{6.22}), leads to 
\[
u\left( x,y\right) =\left[ \int\limits_0^x\left( x-\xi \right)
-x\int\limits_0^1\left( 1-\xi \right) \right] \psi \left( \xi ,y\right) d\xi
. 
\]

Two-fold integration of Eq. (\ref{6.21}) in the limits $-1$, $y$ under the
conditions (\ref{6.23}) and (\ref{6.33}) yields the expression 
\[
u\left( x,y\right) =-\left[ \int\limits_{-1}^y\left( y-\eta \right) -\frac{%
1+y}2\int\limits_{-1}^1\left( 1-\eta \right) \right] \eta \psi \left( x,\eta
\right) d\eta +\frac 12\left( 1+y\right) \nu \left( x\right) . 
\]

The problem is reduced to the solution of the Fredholm integral equation of
the second kind (\ref{6.13}) on the domain $\Omega :0\leq x\leq 1,-1\leq
y\leq 1$ with the operator 
\[
A_{\bullet }=\left[ \int\limits_0^x\left( x-\xi \right)
-x\int\limits_0^1\left( 1-\xi \right) \right] _{\bullet }d\xi +\left[
\int\limits_{-1}^y\left( y-\eta \right) -\frac{1+y}2\int\limits_{-1}^1\left(
1-\eta \right) \right] \eta _{\bullet }d\eta 
\]
and the free term 
\[
f\left( x,y\right) =\frac 12\left( 1+y\right) \nu \left( x\right) . 
\]

Note that the so-called condition of matching on the line of parabolic
degeneracy $y=0$, imposed on the solution of Eq. (\ref{6.21}) (\cite{6.9},
p. 27), is fulfilled in a natural way: 
\[
\lim _{y\rightarrow +0}u\left( x,y\right) =\lim _{y\rightarrow -0}u\left(
x,y\right) ,\quad x\in \left[ 0,1\right] ; 
\]
\[
\lim _{y\rightarrow +0}\partial _yu\left( x,y\right) =\lim _{y\rightarrow
-0}\partial _yu\left( x,y\right) ,\quad x\in \left[ 0,1\right] . 
\]

As in the previous subsection, this situation results from the fact that the
singularity of the problem is transferred from the main term of its operator
to the dependent one.

\section{The inverse problem of the restoration of the coefficient of the
differential equation}

Small oscillations in the transverse direction of a stretched string of
variable density are described by the equation 
\begin{equation}  \label{6.24}
\partial _t^2u=a\left( x\right) \partial _x^2u.
\end{equation}

Here, $x$, $t$ are dimensionless coordinates; 
\[
a\left( x\right) =NT^2/\rho \left( x\right) l^2, 
\]
with $N$ being the tension, $\rho \left( x\right) $ the density of the
material, $2l$ the length of the string, $T$ the time interval.

We assume that the ends of the string are fixed, whereas its density and the
oscillations are symmetric with respect to the coordinate $x=0$. The
corresponding boundary conditions have the form 
\begin{equation}  \label{6.25}
\partial _xu\left( 0,t\right) =u\left( 1,t\right) =0.
\end{equation}

We also employ the following initial conditions: 
\begin{equation}  \label{6.26}
u\left( x,0\right) =u_0\left( x\right) ;\quad \partial _tu\left( x,0\right)
=0.
\end{equation}

The coefficient $a\left( x\right) $ is to be determined from (\ref{6.24})-(%
\ref{6.28}) for given $u_0\left( x\right) $, $N$, $l$, $T$ and additional
information on the oscillations of the middle cross-section of the string:

\begin{equation}  \label{6.27}
u\left( 0,t\right) =\nu \left( t\right) .
\end{equation}

It is known \cite{6.10} that the solution of this problem can exist and be
unique in the classes of functions that are fully adequate to the method our
consideration. We assume that the necessary, in this sense, requirements are
fulfilled.

By analogy with what was done many time before, using the notation (\ref
{6.16}) and (\ref{6.24})-(\ref{6.26}), we find the representations 
\begin{equation}  \label{6.28}
u\left( x,y\right) =\left[ \int\limits_0^x\left( x-\xi \right)
-x\int\limits_0^1\left( 1-\xi \right) \right] \psi \left( \xi ,y\right) d\xi
.
\end{equation}
\[
u\left( x,y\right) =a\left( x\right) \int\limits_0^t\left( t-\eta \right)
\psi \left( x,\eta \right) d\eta +u_0\left( x\right) . 
\]

By eliminating $u\left( x,t\right) $, we obtain an equation of the type (\ref
{6.7}). The substitution of (\ref{6.28}) into (\ref{6.27}) leads to the
integral equation 
\[
\left( A_1\psi \right) \left( t\right) =f_1\left( t\right) ,\quad t\in \left[
0,1\right] , 
\]
where 
\[
A_{1\bullet }=\int\limits_0^1\left( 1-\xi \right) _{\bullet }d\xi ;\quad
f_1\left( t\right) =-\nu \left( t\right) . 
\]

One can estimate the convergence of successive approximations to the
solution of the thus formulated problem by the smoothing of the dependence
of the coefficient $a$ on the values of $t$ in the process of calculations.

\section{The problem of the Stefan type}

Consider the classical model \cite{6.11}: 
\begin{equation}  \label{6.29}
\partial _tu=\partial _x^2u,\quad 0<x<\gamma \left( t\right) ;\quad 0<t\leq
1,
\end{equation}
\begin{equation}  \label{6.30}
u\left( x,0\right) =u_0\left( x\right) ;\quad u\left( 0,t\right) =u\left(
\gamma \left( t\right) ,t\right) =0,\quad u_0\left( 0\right) =0.
\end{equation}

On the moving boundary that separates the phases an additional condition is
imposed: 
\begin{equation}  \label{6.31}
\alpha \partial _xu\left( \gamma \left( t\right) ,t\right) =\gamma ^{\prime
}\left( t\right) ,\quad \gamma \left( 0\right) =\gamma _0,
\end{equation}
where $\gamma _0>0$; the constant $\alpha $ can be both positive and
negative; $\gamma ^{\prime }\left( t\right) =d\gamma \left( t\right) /dt$.

Thus, the data of the problem are $u_0\left( x\right) $, $\alpha $ and $%
\gamma _0$; the functions $u\left( x,t\right) $ and $\gamma \left( t\right) $
are to be determined.

In Eqs. (\ref{6.29})-(\ref{6.31}), it is reasonable to make a non-orthogonal
mapping 
\begin{equation}  \label{6.32}
\bar x=x/\gamma \left( t\right) ,\quad \bar t=t
\end{equation}
on the canonical domain $\Omega :0\leq \bar x,\bar t\leq 1$.

We get: 
\begin{equation}  \label{6.33}
\partial _{\bar t}u-\left[ \bar x\gamma ^{\prime }\left( \bar t\right)
/\gamma \left( \bar t\right) \right] \partial _{\bar x}u=\partial _{\bar
x}^2u,
\end{equation}
\begin{equation}  \label{6.34}
u\left( \bar x,0\right) =0;\quad u\left( 0,\bar t\right) =u\left( 1,\bar
t\right) =0;
\end{equation}
\begin{equation}  \label{6.35}
\alpha \partial _{\bar x}u\left( 1,\bar t\right) =\gamma ^{\prime }\left(
\bar t\right) ,\quad \gamma \left( 0\right) =\gamma _0.
\end{equation}

By analogy with the above, the notation 
\[
\partial _{\bar x}^2u\left( \bar x,\bar y\right) =\psi \left( \bar x,\bar
y\right) , 
\]
conditions (\ref{6.34}) and equation (\ref{6.34}) lead to 
\begin{equation}  \label{6.36}
u\left( \bar x,\bar y\right) =\left[ \int\limits_0^{\bar x}\left( \bar x-\xi
\right) -\bar x\int\limits_0^1\left( 1-\xi \right) \right] \psi \left( \xi
,\bar t\right) d\xi ;
\end{equation}
\[
u\left( \bar x,\bar y\right) =\int\limits_0^t\left\{ \psi \left( x,\eta
\right) +\left[ \bar x\gamma ^{\prime }\left( \bar t\right) /\gamma \left(
\bar t\right) \right] \left[ \int\limits_0^x-\int\limits_0^1\left( 1-\xi
\right) \right] \psi \left( \xi ,\eta \right) d\xi \right\} d\eta 
\]
\begin{equation}  \label{6.37}
+u_0\left( x\right) .
\end{equation}

The substitution of (\ref{6.36}) into (\ref{6.35}) yields 
\[
\gamma ^{\prime }\left( \bar t\right) =\alpha \int\limits_0^1\xi \psi \left(
\xi ,\bar t\right) d\xi , 
\]
from which we get 
\begin{equation}  \label{6.38}
\gamma \left( \bar t\right) =\alpha \int\limits_0^1\xi d\xi
\int\limits_0^{\bar t}\psi \left( \xi ,\eta \right) d\eta +\gamma _0,
\end{equation}
and, respectively, in the expression (\ref{6.37}), we have 
\[
\frac{\bar x\gamma ^{\prime }\left( \bar t\right) }{\gamma \left( \bar
t\right) }=\alpha \bar x\int\limits_0^1\xi \psi \left( \xi ,\bar t\right)
d\xi /\left[ \alpha \int\limits_0^1\xi d\xi \int\limits_0^T\psi \left( \xi
,\eta \right) d\eta +\gamma _0\right] . 
\]

The elimination of $u\left( \bar x,\bar t\right) $ from (\ref{6.36}), (\ref
{6.37}) leads to the integral equation of the first kind (\ref{6.7}), whose
solution was discussed above. The function $\bar \psi $ determined from this
equation should be approximated by an analytical dependence on $\bar x$ in
order to make an inverse change of variables. The sought boundary is
determined from the nonlinear integral equation (\ref{6.38}). Finally, using
Eqs. (\ref{6.36}), (\ref{6.37}) and (\ref{6.32}), we calculate the function $%
u\left( x,t\right) $.

\chapter{Conclusions}

Let us briefly summarize the above consideration. Thus, the solution of the
Fredholm integral equation of the first kind 
\begin{equation}  \label{1}
\left( A\psi \right) \left( x\right) \equiv \int\limits_0^1k\left( x,\xi
\right) \psi \left( \xi \right) d\xi =f\left( x\right) ,\quad x\in \left[ 0,1%
\right]
\end{equation}
in the ''convenient'' for the numerical realization space $L_2$ is, without
any doubt, an incorrectly formulated problem.In the case of the space $l_2$,
the situation is diametrically different: the data of equation. (\ref{1})
may satisfy the conditions of its correctness, but, nevertheless, the
solution determined by Picard's theorem will constitute a series that
diverges as a result of the accumulation of round-off errors.

It should be noted that even an actual verification of whether the data of (%
\ref{1}) belongs to the space $l_2$ is, as a rule, infeasible. Moreover, as
an objective factor of incorrectness, there appear the error of experimental
determination of $f\left( x\right) $ as well as different errors of the
identification of the system under consideration, i.e., $k\left( x,\xi
\right) $.

As ma matter of fact, the basis of our consideration is formed by the
suggestion to connect adaptively Eq. (\ref{1}) to the space $l_2$ by means
of a functional model of the allowed error, including the smoothing of
information by the procedure of integration characterized by the condition 
\begin{equation}  \label{2}
\left\| \delta f\right\| _{L_2\left( 0,1\right) }=0.
\end{equation}

Starting by heuristic considerations that were later supported by more firm
arguments, we showed the usefulness of the representation of the error as a
difference between the sought function and of the integral component: 
\[
\left( \delta f\right) \left( x\right) =\psi \left( x\right) -\lambda \left(
B\psi \right) \left( x\right) ,\quad B_{\bullet }=\int\limits_{-1}^1h\left(
x,\xi \right) _{\bullet }d\xi , 
\]
in order to satisfy (\ref{2}) by $\psi \left( x\right) $, $x\in \left[
1,0\right) $.

We developed the proposed concept by transforming the formulation of the
problem under consideration: 
\begin{equation}  \label{3}
\Psi \left( x\right) =\lambda \left( B\Psi \right) \left( x\right) +\left\{ 
\begin{array}{c}
\chi \left( x\right) ,\quad x\in \left[ 0,1\right] ;~ \\ 
0,\quad x\in \left[ -1,0\right) ;
\end{array}
\right.
\end{equation}
\begin{equation}  \label{4}
\Psi \left( x\right) =\lambda \left( B\Psi \right) \left( x\right) +\left\{ 
\begin{array}{c}
\mu \left( A\Psi \right) \left( x\right) -\mu f\left( x\right) +\chi \left(
x\right) ,\quad x\in \left[ 0,1\right] ;~ \\ 
0,\quad x\in \left[ -1,0\right] ;
\end{array}
\right.
\end{equation}
\begin{equation}  \label{5}
\Psi ^{\prime }\left( x\right) =\lambda \left( B\Psi ^{\prime }\right)
\left( x\right) +\left\{ 
\begin{array}{c}
0,\quad x\in \left[ 0,1\right] ;~ \\ 
\kappa \left( x\right) ,\quad x\in \left[ -1,0\right) .
\end{array}
\right.
\end{equation}

Here, 
\[
\Psi \left( x\right) =\left\{ 
\begin{array}{c}
\psi \left( x\right) ,\quad x\in \left[ 0,1\right] ;~ \\ 
\varphi \left( x\right) ,\quad x\in \left[ -1,0\right) ;
\end{array}
\right. \quad \Psi ^{\prime }\left( x\right) =\left\{ 
\begin{array}{c}
\psi \left( x\right) ,\quad x\in \left[ 0,1\right] ;~ \\ 
\varphi ^{\prime }\left( x\right) ,\quad x\in \left[ -1,0\right) .
\end{array}
\right. 
\]

Note that the solution of the Fredholm integral equation of the second kind (%
\ref{3}), expressed via the resolvent of the kernel $h\left( x,\xi \right) $%
, ensures, in fact, the achievement of the set goal, because it contains on $%
x\in \left[ 0,1\right] $ and $\left[ -1,0\right) $ the function $\psi \left(
x\right) $, respectively, explicitly and under the integral sign: 
\begin{equation}  \label{6}
\psi \left( x\right) =\chi \left( x\right) +\lambda \int\limits_0^1H\left(
x,\xi ,\lambda \right) \chi \left( \xi \right) d\xi ,\quad x\in \left[ 0,1%
\right] ;
\end{equation}
\begin{equation}  \label{7}
\varphi \left( x\right) =\lambda \int\limits_0^1H\left( x,\xi ,\lambda
\right) \chi \left( \xi \right) d\xi ,\quad x\in \left[ -1,0\right) .
\end{equation}

Later, using the data of the problem (\ref{1}), which is important, we
expressed by means of integration of $\chi \left( x\right) $ the function $%
\varphi _0\left( x\right) $ that constitutes the solution of the Fredholm
integral equation of the second kind (\ref{4}) for $f\left( x\right) \equiv
0 $, as well as 
\[
\kappa \left( x\right) =\lambda \int\limits_{-1}^0h\left( x,\xi \right)
\varphi _0\left( \xi \right) d\xi -\varphi _0\left( x\right) . 
\]

From the Fredholm integral equation of the second kind (\ref{5}), a new
representation of the sought solution via $\chi \left( x\right) $ emerged: 
\begin{equation}  \label{8}
\psi \left( x\right) =\lambda \int\limits_{-1}^0H\left( x,\xi ,\lambda
\right) \kappa \left( \xi \right) d\xi ,\quad x\in \left[ 0,1\right] .
\end{equation}

The measure of non-identity of this representation to (\ref{6}) is reflected
exactly by the error $\left( \delta f\right) \left( x\right) $. Therefore,
the solution of the problem (\ref{1}), understood in the sense 
\begin{equation}  \label{9}
\left\| A\psi -f-\delta f\right\| _{L_2\left( 0,1\right) }=0,
\end{equation}
could be represented by expressions of principally different structure: with
the function $\chi \left( x\right) $ that enters explicitly, equation (\ref
{6}), and without this function, equation (\ref{8}).

This factor predetermined the reduction of the considered problem to the
solution of the Fredholm integral equation of the second kind 
\begin{equation}  \label{10}
\chi \left( x\right) =\mu \int\limits_0^1\left[ Q\left( x,\xi \right)
-k^{\prime }\left( x,\xi \right) \right] \chi \left( \xi \right) d\xi +\mu
f\left( x\right) ,\quad x\in \left[ 0,1\right] ,
\end{equation}
where 
\begin{equation}  \label{11}
k^{\prime }\left( x,\xi \right) =k\left( x,\xi \right) +\lambda
\int\limits_0^1k\left( x,\xi \right) H\left( \zeta ,\xi ,\lambda \right)
d\zeta ,
\end{equation}
by the substitution of expressions (\ref{6})-(\ref{8}) into (\ref{4}).

The structure of Eq. (\ref{10}) allows us to give the following
interpretation. As a result of (\ref{6}), (\ref{11}) and (\ref{9}), the
obtained solution must also satisfy the homogeneous equation 
\begin{equation}  \label{12}
\chi \left( x\right) =\mu \int\limits_0^1Q\left( x,\xi \right) \chi \left(
\xi \right) d\xi ,\quad x\in \left[ 0,1\right] ,
\end{equation}
with accuracy that depends on $\delta f$, which is quite natural. This means
that in the process of transformation we build up, in a way, the kernel $%
Q\left( x,\xi \right) $ for which the solution of the problem is an
eigenfunction, with equation (\ref{12}) positioned on the spectrum.

Thus, the traditional formulation (\ref{1}) is transformed into equation (%
\ref{10}), whose advantage lies in the fact of continuous dependence of the
solution on the data of the problem. an additional advantage consists in a
possibility to choose the parameter $\mu $ from the condition of convergence
to the solution by simple iteration 
\[
\chi _{n+1}\left( x\right) =\mu \int\limits_0^1\left[ Q\left( x,\xi \right)
-k^{\prime }\left( x,\xi \right) \right] \chi _n\left( \xi \right) d\xi +\mu
f\left( x\right) ,\quad x\in \left[ 0,1\right] . 
\]

Note that it proved to be sufficient to allot the kernel $h\left( x,\xi
\right) $ the property of completeness, whereas its symmetry, dependence on
the difference of arguments and periodicity were also very important and
facilitated analytical transformations.

We have shown that wide classes of problems of numerical simulations can be
rather elementarily reduced to Fredholm integral equations of the first
kind. After that, the discussed procedure of the correct formulation and
numerical realization can be applied to them without any substantial
changes. By the way, a traditional difference between direct and inverse
formulations of problems of mathematical physics practically vanishes. In
the situation of nonlinearity, the realization of iteration algorithms of
the solution to the obtained integral equation by means of the parameter $%
\mu \,$ is of special importance.

In light of the above, we can draw a conclusion that, if the phenomenon
(process) admits an adequate description by methods of numerical
simulations, the restoration of its underlying cause or of different
parameters from an objectively sufficient volume of additional information
does not pose principal difficulties, because the corresponding problems can
be correctly formulated. From this point of view, an analysis of actually
observed events, including multi-factor social-economic and ecological
processes, can be done with much larger efficiency.

Maybe, it would be reasonable to suggest that, in general, the process of
the understanding of the World is much simpler than a wide audience usually
supposes it to be under the influence of the sphere of applied science that,
at present, armed with means of electronic processing of information,
constitutes, in fact, a natural monopoly with an almost dominant role of
commercial component and, correspondingly, a systematic drive for investment?

Thus, colossal means are invested in problems of the restoration of
dependencies from empirical data and, in particular, in remote probing of
the surface of the Earth by spacecraft. What is actually realized is a
search for minimally and maximally acceptable values of the parameter $%
\alpha $ in the integral equation of the type 
\[
\alpha \psi \left( x\right) +\int\limits_0^1k\left( x,\xi \right) \psi
\left( \xi \right) d\xi =f\left( x\right) ,\quad x\in \left[ 0,1\right] . 
\]

The essence lies in the necessity to establish a balance between
computational and, respectively, financial abilities of the solution of an
ill-posed system of linear algebraic equations (which implies a realization
of one of discretization methods) and an approximation of the formulation of
the problem under consideration to an exact one that is erroneously
associated with the factor of incorrectness for $\alpha =0$.

In this regard, we note that, of course, it would be incorrect to suppose
that problems in science are altogether absent or that one can develop,
irrespective of the circumstances, efficient means to overcome these
problems. However, in our opinion, complications of principal character are
inherent, in the first place, to direct formulations of some problems, that
is, to the construction of mathematical models of insufficiently studied
processes and phenomena.

It is clear that the solution of some classes of inverse problems may also
pose substantial difficulties, but, nevertheless, the wide-spread thesis
that the procedure of the restoration of the cause from the consequence is
incorrect, in general, seems to be manifestly erroneous.

J. Hadamard's statement that the problems that adequately describe real
processes are correct is an ingenious idea, whose constructive development
allows one to attain a qualitatively higher level of the potential of
methods of numerical simulations.

March 21, 2001.

\appendix 

\chapter{A method of successive approximations for Fredholm integral
equations of the first kind}

\begin{center}
\smallskip\ {\bf V. M. Fridman\footnote{{\bf Uspekhi, Mat. Nauk 11, No 1,
233-234 (1956).}}\smallskip\ }
\end{center}

{\bf Theorem}. Let $K\left( x,s\right) $ be a symmetric square integrable
positive definite kernel, and let the equation 
\begin{equation}  \label{a1}
\int\limits_a^bK\left( x,s\right) \varphi \left( s\right) ds=f\left(
x\right) ,\quad f\left( x\right) \in L_2\left( a,b\right)
\end{equation}
be solvable. Then the sequence $\left\{ \varphi _n\left( x\right) \right\} $%
, determined by the recursion relation 
\begin{equation}  \label{a2}
\varphi _n\left( x\right) =\varphi _{n-1}\left( x\right) +\lambda \left[
f\left( x\right) -f_{n-1}\left( x\right) \right] ,
\end{equation}
where 
\[
\varphi _0\left( x\right) \in L_2\left( a,b\right) , 
\]
\begin{equation}  \label{a3}
f_{n-1}\left( x\right) =\int\limits_a^bK\left( x,s\right) \varphi
_{n-1}\left( s\right) ds,
\end{equation}
\begin{equation}  \label{a4}
0<\lambda <2\lambda _1,
\end{equation}
and $\lambda _1$ is a minimal characteristic number of the kernel $K\left(
x,s\right) $, converges in the mean to the solution of Eq. (\ref{a1}).

{\bf Proof}. Let us set in Eq. (\ref{a2}) 
\[
\varphi _n\left( x\right) =\varphi \left( x\right) +u_n\left( x\right) , 
\]
multiply both its parts by an eigenfunction of the kernel $v_i\left(
x\right) $ and integrate over $x$ from $a$ to $b$: 
\[
\alpha _i^n=\alpha _i^{n-1}-\int\limits_a^bv_i\left( x\right)
dx\int\limits_a^bK\left( x,s\right) u_{n-1}\left( s\right) ds, 
\]
where 
\[
\alpha _i^n=\int\limits_a^bu_n\left( x\right) v_i\left( x\right) dx. 
\]
As a result of the fact that $K\left( x,s\right) $ is symmetric and $%
v_i\left( x\right) $ satisfies the equation 
\[
v_i\left( x\right) -\lambda _i\int\limits_a^bK\left( x,s\right) v_i\left(
s\right) ds=0, 
\]
we have: 
\[
\int\limits_a^bv_i\left( x\right) dx\int\limits_a^bK\left( x,s\right)
u_{n-1}\left( s\right) ds=\int\limits_a^bu_{n-1}\left( x\right)
dx\int\limits_a^bK\left( x,s\right) v_i\left( s\right) ds 
\]
\[
=\frac 1{\lambda _i}\int\limits_a^bu_{n-1}\left( x\right) v_i\left( x\right)
dx=\frac{\alpha _i^{n-1}}{\lambda _i}. 
\]
Thus, 
\begin{equation}  \label{a5}
\alpha _i^n=\left( 1-\frac \lambda {\lambda _i}\right) \alpha
_i^{n-1}=\left( 1-\frac \lambda {\lambda _i}\right) ^n\alpha _i^0.
\end{equation}
By virtue of the completeness of the system of functions $v_i\left( x\right) 
$, 
\begin{equation}  \label{a6}
\int\limits_a^bu_n^2\left( x\right) dx=\sum_{i=1}^\infty \left( \alpha
_i^n\right) ^2,
\end{equation}
which is to be estimated.

The series $\sum_{i=k}^\infty \left( \alpha _i^n\right) ^2=\sum_{i=k}^\infty
\left( 1-\frac \lambda {\lambda _i}\right) ^{2n}\left( \alpha _i^0\right) ^2$
is majorized by the series $\sum_{i=k}^\infty \left( \alpha _i^0\right) ^2$,
because, by the inequality (\ref{a4}), 
\begin{equation}  \label{a7}
\left( 1-\frac \lambda {\lambda _i}\right) ^2\leq 1.
\end{equation}

Therefore, for any given small positive $\epsilon $, there exists $k>K\left(
\epsilon \right) $ that does not depend on $n$, such that 
\[
\sum_{i=k}^\infty \left( \alpha _i^n\right) ^2<\frac \epsilon 2. 
\]

At the same time, by choosing $n>N\left( \epsilon \right) $, we can make 
\[
\sum_{i=1}^k\left( \alpha _i^n\right) ^2=\sum_{i=1}^k\left( 1-\frac \lambda
{\lambda _i}\right) ^{2n}\left( \alpha _i^0\right) ^2<\frac \epsilon 2, 
\]
because for finite $i$ we have the inequality sign in the formula (\ref{a7}).

Thus, we arrive at the inequality 
\begin{equation}  \label{a8}
\int\limits_a^bu_n^2\left( x\right) dx<\epsilon ,
\end{equation}
which proves the Theorem.

\bigskip\ \ Submitted on March 29, 1954.

\end{document}